\documentclass[11pt,a4paper,english]{article}
\usepackage[pdftex]{graphicx}
\usepackage{amsmath}
\usepackage{epsf}
\usepackage[parfill]{parskip}
\usepackage[matrix,arrow]{xy}
\usepackage[usenames]{color}
\usepackage{hyperref}
\usepackage[vcentermath]{youngtab}
\usepackage[matrix,arrow]{xy}

\usepackage{epsfig}

\usepackage{amssymb,amsthm}
\usepackage{amscd}
\usepackage{amsfonts}
\usepackage{latexsym}
\usepackage{graphics}
\usepackage{color}
\usepackage{bbm}
\usepackage{bm}

\usepackage{xypic}
\usepackage{mathrsfs}

\input{xy}
\xyoption{all}

\input{epsf}

\makeatletter

\catcode`@=11
\renewcommand{\section}
{\@startsection{section}{1}{0pt}{\medskipamount}{\medskipamount}{\large\bf}}
\makeatletter\renewcommand{\subsection}
{\@startsection{subsection}{2}{\z@}{-3.25ex plus -1ex minus -.2ex}
{1.5ex plus .2ex}{\it }}

\numberwithin{equation}{section}
\catcode`@=12

\newcommand{\ba}{\begin{eqnarray*}}
\newcommand{\ea}{\end{eqnarray*}}
\newcommand{\ban}{\begin{eqnarray}}
\newcommand{\ean}{\end{eqnarray}}

\newcommand{\Tr}{{\rm Tr\,}}

\def\Id{{\rm id}}

\newcommand{\IZ}{\mathbb{Z}}
\newcommand{\IC}{\mathbb{C}}
\newcommand{\IP}{\mathbb{P}}

\newcommand{\IR}{\mathbb{R}}
\newcommand{\IQ}{\mathbb{Q}}
\newcommand{\IF}{\mathbb{F}}

\newcommand{\cI}{{\cal I}}
\newcommand{\cW}{{\cal W}}
\newcommand{\cN}{{\cal N}}

\newcommand{\cS}{{\cal S}}

\newcommand{\calH}{{\cal H}}

\newcommand{\cE}{{\cal E}}
\newcommand{\cO}{{\cal O}}

\newcommand{\cQ}{{\cal Q}}
\newcommand{\calR}{{\cal R}}
\newcommand{\cZ}{{\cal Z}}
\newcommand{\calL}{{\cal L}}
\newcommand{\cF}{{\cal F}}

\newcommand{\cT}{{\cal T}}
\newcommand{\cK}{{\cal K}}
\newcommand{\cP}{{\cal P}}

\newcommand{\frM}{{\mathfrak{M}}}

\newcommand{\frR}{{\mathfrak{R}}}
\newcommand{\frD}{{\mathfrak{D}}}

\newcommand{\sfA}{{\mathsf{A}}}
\newcommand{\sfe}{{\mathsf{e}}}

\newcommand{\sfP}{{\mathsf{P}}}

\newcommand{\sfQ}{{\mathsf{Q}}}
\newcommand{\sfD}{{\mathsf{D}}}
\newcommand{\sfS}{{\mathsf{S}}}

\newcommand{\sfR}{{\mathsf{R}}}

\newcommand{\sfW}{{\mathsf{W}}}
\newcommand{\sft}{{\mathsf{t}}}
\newcommand{\sfh}{{\mathsf{h}}}

\newcommand{\scrA}{{\mathscr{A}}}

\newcommand{\scrM}{{\mathscr{M}}}

\newcommand{\scrW}{{\mathscr{W}}}
\newcommand{\scrE}{{\mathscr{E}}}
\newcommand{\scrV}{{\mathscr{V}}}
\newcommand{\scrN}{{\mathscr{N}}}
\newcommand{\scrH}{{\mathscr{H}}}
\newcommand{\scrF}{{\mathscr{F}}}

\newcommand{\DT}{{\tt DT}}
\newcommand{\GW}{{\tt GW}}

\newcommand{\BPS}{{\tt BPS}}
\newcommand{\ttb}{{\tt b}}
\newcommand{\tta}{{\tt a}}

\newcommand{\Hilb}{{\rm Hilb}}

\newcommand{\mbf}[1]{{\boldsymbol {#1} }}
\newcommand{\complex}{{\mathbb C}} 
\newcommand{\zed}{{\mathbb Z}} 
\newcommand{\real}{{\mathbb R}} 
\newcommand{\torus}{{\mathbb T}}

\def\e{{\,\rm e}\,}

\newcommand{\Hcal}{{\cal H}}

\newcommand{\ch}{{\rm ch}}
\newcommand{\eul}{{\rm eul}}
\newcommand{\Ch}{{\sf ch}}

\def\ii{{\,{\rm i}\,}}
\def\dd{{\rm d}}

\newcommand{\Hom}{\mathrm{Hom}}

\newcommand{\End}{\mathrm{End}}

\newcommand{\bdiv}{\wp_{\infty}}
\newcommand{\Ext}{\mathrm{Ext}}
\newcommand{\Tor}{\mathrm{Tor}}

\newcommand{\gtop}{\lambda}

\def\beq{\begin{equation}}
\def\bee{\begin{equation}}
\def\eeq{\end{equation}}
\def\bea{\begin{eqnarray}}
\def\eea{\end{eqnarray}}
\def\bd{\begin{displaymath}}
\def\ed{\end{displaymath}}

\newcommand{\Cint}{\int\kern-10.5pt-\kern7pt}

\newcommand{\PP}{{\mathbb{P}}}

\newcommand{\be}{\begin{equation}}
\newcommand{\ee}{\end{equation}}

\newcommand\fverbit{\egroup\item[\fbox{\unhbox\pippobox}]}
\newbox\pippobox

{

\def\eps{\epsilon}

\def\w{\wedge}

\def\be{\begin{equation}}
\def\ee{\end{equation}}
\def\bea{\begin{eqnarray}}
\def\eea{\end{eqnarray}}

\begin{document}

\begin{titlepage}
\setcounter{page}{1}

\begin{flushright}
HWM--12--11\\
EMPG--12--19
\end{flushright}

\vskip 1.8cm

\begin{center}

{\Huge Curve counting, instantons \\[10pt] and McKay
  correspondences}\footnote{Invited contribution to the special issue
  ``Noncommutative Algebraic Geometry and its Applications to Physics''
of the {\sl Journal of  Geometry and Physics}, eds. G.~Cornelissen and
G.~Landi.}

\vspace{15mm}

{\large\bf Michele Cirafici}$^{(a)}$ \ and \ {\large\bf Richard J. Szabo}$^{(b)}$
\\[6mm]
\noindent{\em  $^{(a)}$  Centro de An\'{a}lise Matem\'{a}tica, Geometria e Sistemas
Din\^{a}micos\\ Departamento de Matem\'atica and LARSyS \\ Instituto Superior T\'ecnico\\
Av. Rovisco Pais, 1049-001 Lisboa, Portugal}\\ Email: \ {\tt cirafici@math.ist.utl.pt}
\\[6mm]
\noindent{\em $^{(b)}$ Department of Mathematics\\ Heriot--Watt
  University\\
Colin Maclaurin Building, Riccarton, Edinburgh EH14 4AS, UK}\\ and\\ 
{\em Maxwell Institute for Mathematical Sciences, Edinburgh, UK}\\ Email: {\tt
  R.J.Szabo@hw.ac.uk}

\vspace{15mm}


\begin{abstract}
\noindent
We survey some features of equivariant instanton partition functions
of topological gauge theories
on four and six dimensional toric K\"ahler varieties, and their
geometric and algebraic counterparts in the enumerative problem
of counting holomorphic curves. We discuss the relations of
instanton counting to representations of affine Lie algebras in
the four-dimensional case, and to Donaldson-Thomas theory for ideal sheaves on Calabi-Yau
threefolds. For resolutions of toric singularities, an algebraic structure induced by a quiver determines the instanton
moduli space through the McKay
correspondence and its generalizations. The correspondence elucidates
the realization of gauge theory partition functions as
quasi-modular forms, and reformulates the computation of
noncommutative Donaldson-Thomas invariants in terms of the enumeration
of generalized instantons. New results include a general presentation of the partition
functions on ALE spaces as affine characters, a rigorous treatment of equivariant
partition functions on Hirzebruch surfaces, and a putative connection
between the special McKay correspondence and instanton counting on
Hirzebruch-Jung spaces.

\end{abstract}

\end{center}
\end{titlepage}

\tableofcontents

\newpage

\section{Introduction}

The purpose of this survey is to outline some connections between
various topics in mathematical physics. Instantons, enumerative
invariants associated with holomorphic curves, quivers and the McKay
correspondence are recurrent themes which arise in different areas of physics
and mathematics, but are all common ingredients which enter into the
description of the BPS sectors of supersymmetric theories. Whether we
are talking about field theory or string theory, the study of the BPS
sector has unveiled an amazing series of insights and surprises. On
the physics side, the computation of quantities protected from quantum
corrections has led to a deeper understanding of the quantum physics
beyond the perturbative regime. Mathematically the computation of
enumerative invariants, in the guise of quantum correlators, has been
rewarded with the discovery of new structures within algebraic and
differential geometry. Thus Donaldson-Witten theory, Seiberg-Witten
invariants of four manifolds, the relation between Chern-Simons gauge
theory and knot invariants, mirror symmetry, Gromov-Witten invariants,
and the Kontsevich-Soibelman wall-crossing formula are currently subjects of intense investigation.

In this review we will focus on the problem of counting holomorphic curves on
toric varieties, instantons and Donaldson-Thomas invariants. While
these topics are well studied and several reviews are already
available in the literature (which we indicate throughout this paper), we wish to present them from a different angle. The
perspective we will take is that all of these topics can be understood
from constructions of instantons in suitable topological gauge theories, with certain variations, via techniques of equivariant localization. We will find that this perspective highlights the role played by quivers, and by the McKay correspondence and its generalizations.

Topological string theory on toric varieties is a prominent model to
study Gromov-Witten invariants since the toric symmetries make many
computations feasible. The enumerative invariants are the numbers of
worldsheet instantons which wrap holomorphic curves in the ambient toric
variety. A rather sophisticated mathematical theory allows one to make
very precise sense of the notion of  ``counting curves'', via the
intersection theory of the Deligne-Mumford moduli space of
curves. This problem has a combinatorial reformulation as the
statistical mechanics of a classical melting crystal. The underlying
mathematical reason is Donaldson-Thomas theory. In
physical parlance this represents a dual formulation of the curve
counting problem. Donaldson-Thomas invariants are associated with the
intersection theory of the moduli spaces of coherent sheaves on
Calabi-Yau threefolds. They are naturally related to a higher-dimensional generalization of the four-dimensional concept of instanton. This is directly transparent from a D-brane perspective, where the Donaldson-Thomas invariants count BPS bound states of D-branes wrapping holomorphic cycles in the Calabi-Yau geometry.  

Yet this is only part of the full story. A more complete and beautiful
picture has recently emerged. It has been known for some time that
\textit{stable} BPS states of D-branes on a Calabi-Yau manifold have a rather
intricate structure, dictated by Douglas' $\Pi$-stability conditions
on the derived category of coherent sheaves. But only in the last few
years has a combination of mathematical and physical insight paved the
way to high precision computations and direct evaluation of the
partition functions of BPS states in many cases. Thanks to the work of
Denef-Moore~\cite{Denef:2007vg} on enumeration of black hole
microstates and of Kontsevich-Soibelman~\cite{ks} on the mathematics
of generalized Donaldson-Thomas invariants, the study of BPS physics has
intensified into new and exciting directions.

As we explain more precisely later on, the counting of BPS states is physically encoded in the computation of the Witten index
\begin{equation}
\Omega \left( \gamma ; t \right) = \Tr_{\calH_{(\gamma ; t ) ,
    \mathrm{BPS}}}\, (-1)^F \ .
\end{equation}
This index has a hidden dependence on the value of the
background K\"ahler moduli $t = B + \ii J$. It can ``jump'' when the
K\"ahler parameters cross real codimension one walls in the moduli
space. This is at the core of the wall-crossing behaviour of BPS states:
The K\"ahler moduli space is divided into chambers by walls of
marginal stability and the index of BPS states is a piecewise constant
function inside each of the chambers, which however jumps as we cross
a wall of marginal stability. This means that as a wall is crossed
some BPS states may become unstable and decay into other states, or
novel bound states can form. Therefore the index can jump because the
\textit{single-particle} Hilbert space, over which we are taking the
trace, can lose or gain a sector. It is only in a certain chamber of
the K\"ahler moduli space that the corresponding generating function coincides with
the Donaldson-Thomas partition function. As we move along the moduli
space the partition function of BPS states defines
\textit{different} enumerative problems. All of these enumerative
problems are related by wall crossings, and a formula to
account precisely for this phenomenon was proposed by Kontsevich-Soibelman in~\cite{ks}.

To illustrate the wall crossing phenomenon, let us
consider a particular simple decay. Suppose a BPS particle with
charge $\gamma$ decays into two constituents with charges $\gamma_1$ and
$\gamma_2$ after crossing a wall of marginal stability. Charge conservation implies that $\gamma = \gamma_1 +
\gamma_2$. In particular the same relation must hold between the
central charges which are linear functions of the charge
vectors, and one has
\begin{equation} \label{ccharges}
Z (\gamma; t) = Z (\gamma_1 ; t) + Z (\gamma_2 ; t) \ .
\end{equation}
However, conservation of energy implies that at the location of the
wall given by $t=t_{ms}$ we have (see e.g.~\cite{Denef:2000ar} for a review)
\begin{equation} \label{energy}
\big|Z (\gamma ; t_{ms})\big| = \big|Z(\gamma_1 ; t_{ms}) + Z
(\gamma_2 ; t_{ms})\big| = \big|Z(\gamma_1 ;
t_{ms})\big|+\big|Z(\gamma_2 ; t_{ms})\big| \ .
\end{equation}
This implies that at the location of the wall of marginal stability the phases of the two central charges $Z(\gamma_1;t)$ and $Z (\gamma_2;t)$ align and therefore the equation of the wall is
\begin{equation}
\mathcal{W} (\gamma_1 ; \gamma_2) = \big\{  t \ \big\vert \ Z
(\gamma_1 ; t) = \lambda\, Z (\gamma_2 ; t) \  \text{for
some} \ \lambda \in \real_{\geq0}  \big\} \ .
\end{equation}

The problem now is to find the form of the walls of marginal stability
for a generic Calabi-Yau threefold and to compute the generating
function of stable BPS states in each chamber. This moreover hints
towards the existence of a generalized enumerative problem 
counting certain mathematical objects that also depend on a stability
parameter, which is identified with the physical notion of
stability. These enumerative invariants should reduce to the ordinary
Donaldson-Thomas invariants in a certain chamber but allow for walls
of marginal stability. Ordinary Donaldson-Thomas invariants are
automatically stable since they are represented geometrically by ideal
sheaves. There are currently far reaching proposals for theories of generalized Donaldson-Thomas invariants by Kontsevich-Soibelman \cite{ks}, Joyce-Song \cite{joycesong}, and a very explicit construction by Nagao-Nakajima for certain non-compact threefolds \cite{nakanagao,nagao}.

While a solution of the full problem is currently out of reach, there
are certain chambers where a direct evaluation of the invariants is
possible. An example is the case of noncommutative Donaldson-Thomas
invariants. These quantities still enumerate BPS bound states of D-branes, but
defined in a certain ``non-geometric'' chamber. This chamber is
called the noncommutative crepant resolution chamber. Here the classical
notions of geometry break down, and the problem is better described in
the language of quivers and their representations. It happens that a
certain noncommutative algebra associated with a quiver, the path
algebra, is itself the correct description of the geometry in this
chamber. Similarly the enumerative problem can be reformulated in
terms of the moduli space of quiver representations. It is precisely
this role played by quivers that allows one to reformulate the problem
\textit{again} as an instanton counting problem. In this chamber the
concept of instanton requires some slightly exotic gauge theory
construction, in terms of what we dub a \textit{stacky gauge theory}.

This leads to the exploration of how the corresponding enumeration
of BPS bound states of D-branes is
reflected in the more familiar world of four-dimensional gauge
theories, but now defined on toric varieties. The wrapped D-branes in
this case can be regarded as BPS particle states in a four-dimensional
supersymmetric gauge theory obtained by dimensional reduction over a
Calabi-Yau threefold. Flop transitions in a toric Calabi-Yau threefold
interpolate between instanton partition functions on different
non-compact toric four-cycles via wall-crossings. The same concepts, i.e. quivers, the
McKay correspondence and its generalizations, play a prominent role
and imply a relationship between the instanton counting problem and
the enumeration of holomorphic curves on complex toric surfaces. Understanding the associated curve counting
problems elucidates the
nonperturbative and geometric nature of supersymmetric gauge theories
in four dimensions, particularly those which arise by dimensional
reduction as low-energy effective field theories in superstring
compactifications. They moreover serve as toy models for their higher-dimensional
counterparts wherein the moduli spaces involved are much better
behaved and simpler in general, and computational progress is much
more feasible: By entering this realm we are able to transport a
wealth of instanton counting techniques in four dimensions over to the
six-dimensional cases. The enumerative
problems in four and six dimensions share many common features, and
their similarities and differences can shed light on each other; for
instance, one can in principle study the
connections of the six-dimensional generating functions with quasi-modular forms from the known modularity
properties of the instanton partition functions in four dimensions. From a six-dimensional perspective, these gauge
theories arise from dimensional reductions of the self-dual tensor field theory of
M5-branes over punctured Riemann surfaces; in this setting they are
expected to be related to
two-dimensional conformal field theory within the framework of the AGT
correspondence~\cite{AGT,Wyllard}, which conjectures that their
instanton moduli spaces carry natural geometric actions of certain
affine Lie algebras.

We have attempted to write this article in a way which we hope is
palatable to both mathematicians and physicists alike. While we do
make extensive use of jargon and concepts from string theory and gauge
theory, we have attempted to present them in formal mathematical terms
with the minimalist physical intuition required; likewise, many intricate
technical details of the mathematics we present have been streamlined
for brevity and readability. We consider various explicit examples of
the general formalism throughout. The next three sections deal exclusively
with the geometric problems of curve counting and their relations to
the enumeration of BPS states of D-branes. In \S\ref{sec:curveCY3} we
look at the problem on Calabi-Yau threefolds in the large radius limit
and indicate some reasons why a gauge theory description will become
relevant. Then we consider in \S\ref{sec:3CY} the analogous problems
in the noncommutative crepant resolution chamber where the D-branes
typically probe regions near conifold or orbifold singularities in the
Calabi-Yau moduli space. This inspires the much simpler problem of
curve counting on toric surfaces in \S\ref{sec:curvecountsurf}, which
we relate to instanton counting in four-dimensional maximally
supersymmetric gauge theory in \S\ref{sec:4dcohgt}; in particular, we
look in detail at the example of ALE spaces where the explicit construction of
the instanton moduli space is based on the McKay quiver, and the gauge
theory partition function is related to the representation theory of
affine Lie algebras through the McKay correspondence. In
\S\ref{sec:InstN2gt} we study $\cN=2$ gauge theories on resolutions of
toric singularities based on the equivariant cohomology of the framed
instanton moduli spaces; we give a new detailed rigorous analysis in the case
of $A_{p,1}$ singularities and discuss how the cohomology of the
instanton moduli space is related to more general affine Lie algebra
representations. In \S\ref{6dcohgt} we consider the extension of these
instanton counting techniques to the case of six-dimensional maximally
supersymmetric gauge theory, and how singular instanton solutions can be
used to reproduce Donaldson-Thomas invariants of toric Calabi-Yau
threefolds. In \S\ref{sec:Stacky} we discuss the extension to the
problem of computing orbifold Donaldson-Thomas invariants by means of
instanton solutions on toric Calabi-Yau
orbifolds, and discuss how the formalism of stacky gauge theory
naturally captures the noncommutative Donaldson-Thomas theory of the
orbifold singularity via the generalized McKay correspondence. Finally, we close in \S\ref{sec:specialmckay}
with a novel putative construction of instantons on generic
Hirzebruch-Jung resolutions by generalizing earlier constructions
using the special McKay correspondence; in particular, we propose how
one may generalize the ADHM construction to this general class. Although this project is far
from complete, we highlight the problems which need to be resolved and
why it may lead to novel descriptions of instantons in terms of
representation theoretic aspects of the instanton moduli spaces; in
particular, it could help elucidate the role of affine algebra
representations in this more general context.

For the reader's convenience, in \S\ref{EIsummary} we briefly summarise the various topological
invariants discussed in this paper, their mathematical and physical
definitions, and the relations among them.

\bigskip

\section{Counting curves in Calabi-Yau threefolds\label{sec:curveCY3}}

\subsection{BPS states and Hilbert schemes}

Type IIA string theory compactified on a 
Calabi-Yau threefold $X$ leads via dimensional reduction to a low-energy effective field theory
in four dimensions with $\cN=2$ supersymmetry. The BPS states are
those which preserve half of these supersymmetries. In the large
radius approximation they are labelled by a charge vector $\gamma$
which sits in a lattice $\Gamma$ given by
\begin{equation}
\gamma \ \in \ \Gamma = \Gamma^{\rm m} \oplus \Gamma^{\rm e} = \left( H^0 (X ,
  \zed) \oplus H^2 (X , \zed) \right) \ \oplus \ \left( H^4 (X , \zed)
  \oplus H^6 (X , \zed) \right) \ ,
\end{equation}
which we have separated into electric and magnetic charge sublattices
$\Gamma^{\rm e}$ and $\Gamma^{\rm m}$. The cohomology groups correspond to the
individual charges of the D$p$-branes wrapping $p$-cycles of $X$ which form the BPS states as
\begin{equation}
{\rm D}p \ \longleftrightarrow \ H^{6-p} (X , \zed) = H_{p} (X , \zed) \ , \qquad p = 0,2,4,6
\end{equation}
where we have used Poincar\'e duality (for non-compact Calabi-Yau
manifolds this whole discussion needs to be refined using cohomology
with compact support). The Dirac-Schwinger-Zwanziger intersection
product on $\Gamma$ is defined by
\beq
\big\langle\gamma\,,\,\gamma'\,\big\rangle_{\Gamma} = \int_X\, \gamma\wedge (-1)^{{\rm
    deg}/2}\, \gamma' \ .
\label{DSZint}\eeq
In the large radius limit these BPS states have central charge
\begin{equation}
Z_X (\gamma;t) = -\int_{X}\, \gamma \wedge \e^{-t}
\end{equation}
where $t = B+ \ii J$ is the K\"ahler modulus consisting of the
background supergravity two-form
$B$-field and the K\"ahler $(1,1)$-form $J$ of~$X$.

The {single-particle} Hilbert space of BPS states is graded
into sectors of fixed charge as
\begin{equation}
\calH_{\mathrm{BPS}}^X = \bigoplus_{\gamma \in \Gamma} \, \calH^X_{\gamma ,
  \mathrm{BPS}} \ .
\end{equation}
BPS states have an associated enumerative problem characterized by the Witten index
\begin{equation} \label{Windex}
\Omega_X \left( \gamma  \right) = \Tr_{\calH^X_{\gamma , \mathrm{BPS}}}\, (-1)^F
\end{equation}
which counts BPS states of a given charge $\gamma$ (up to a universal
contribution associated with the centre of mass of the BPS
particles); here $F$ is a suitable operator in the isometry group acting on one-particle
states of charge $\gamma$ in the four-dimensional $\cN=2$
supersymmetric gauge theory. Ordinary Donaldson-Thomas invariants correspond to a
specific choice of the charge vector $\gamma = (1 , 0 , - \beta , n)$
and are physically interpreted as the number of bound states of D2 and D0 branes with a single D6-brane. This information is encoded in the Donaldson-Thomas partition function
\begin{equation} \label{ZDTdef}
\cZ_{\rm BPS}^X (q , Q) = \sum_{\beta \in H_2 (X , \zed)}\, Q^{\beta}
\ Z_\beta^X(q) \qquad \mbox{with} \quad Z_\beta^X(q)= \sum_{n
  \in \zed} \, q^n  \ \Omega _X(n,\beta) \ ,
\end{equation}
where $Q^\beta:=\prod_i\, Q_i^{n_i}$ for an expansion $\beta=\sum_i\,
n_i\, S_i$, with $n_i\in\zed$, in a basis of two-cycles $S_i\in
H_2(X,\zed)$, and $Q_i=\e^{-t_i}$ with $t_i=\int_{S_i}\, t$.

The Donaldson-Thomas invariants $\Omega_X (n,\beta)$ have a more geometrical
definition via the introduction of a virtual fundamental class on the
moduli space of stable sheaves defined by Thomas in \cite{thomas}. In
string theory we are interested in a slightly specialized case of
Thomas' construction for ideal sheaves. An ideal sheaf $\cI$ on a
Calabi-Yau threefold $X$ is a torsion free sheaf
of rank one with trivial determinant. The torsion free condition means
that $\cI$ can be embedded in a bundle; we will loosely think
of $\cI$ as a ``singular bundle'', i.e. an entity which fails to be a
holomorphic line bundle only on a set of singularities. The
triviality of the determinant implies that the double dual of the
sheaf $\cI$ is isomorphic to the trivial bundle, i.e. $\cI^{\vee \vee}
\cong \cO_X$, and that $c_1(\cI)=0$. Note that in general the double dual of a sheaf is not the sheaf itself, contrary to what happens for holomorphic bundles.
Each ideal sheaf $\cI$ is associated with a subscheme
 $Y$ of $X$ via the short exact sequence
\begin{equation} \label{idealseq}
\xymatrix@1{
0 \ \ar[r] &  \ \cI \ \ar[r] & \ \cO_X \ \ar[r] & \ \cO_Y \ \ar[r] & \
0 \ ,
}
\end{equation}
which means that $\cI$ is the kernel of the
restriction map $\cO_X \rightarrow \cO_Y$ of structure sheaves.
In Donaldson-Thomas theory we are interested in the moduli space
$\scrM^{\rm BPS}_{n,\beta} (X)$ of ideal sheaves $\cI$ specified by the
topological data
\bea
\chi(\cI)=n \qquad \mbox{and} \qquad \ch_2(\cI)=-\beta \ .
\eea
Because of
(\ref{idealseq}), this moduli space can also be identified with the projective
Hilbert scheme $\mathrm{Hilb}_{n,\beta} (X)$ of points and curves on
$X$, i.e. subschemes $Y\subset X$ with no component of codimension
one such that
\begin{eqnarray}
n=\chi (\cO_Y) \qquad \mbox{and} \qquad
\beta = [Y] \ \in \ H_2 (X ,\zed) \ ,
\end{eqnarray}
where here $\chi$ denotes the holomorphic Euler characteristic. Then the Donaldson-Thomas invariants are defined by
\begin{equation} \label{DTdef}
\Omega_X(n,\beta)=\DT_{n,\beta}(X) := \int_{[\scrM^{\rm BPS}_{n,\beta} (X)]^{\mathrm{vir}}} \, 1 \ .
\end{equation}
The proper definition of how to integrate over this moduli scheme,
i.e. the definition of the length of the zero-dimensional virtual fundamental class $[\scrM^{\rm BPS}_{n,\beta}
(X)]^{\mathrm{vir}}$ within the framework of a symmetric perfect obstruction
theory, can be found in~\cite{thomas} (see~\cite{Szabo:2009vw} for a
description within the present context). Alternatively, we can use
Behrend's formulation~\cite{behrend} of Donaldson-Thomas invariants as the \emph{weighted} topological Euler characteristics
\begin{equation}
\DT_{n,\beta}(X) = \chi \big( \scrM^{\rm BPS}_{n,\beta} (X) \, , \, \nu_X \big) = \sum_{n\in\zed}\, n \
\chi\big(\nu_X^{-1}(n)\big) \ ,
\label{Behrend}\end{equation}
where $\nu_X:\scrM^{\rm BPS}_{n,\beta} (X) \to\zed$ is a canonical
constructible function; this equivalent definition avoids non-compactness issues
and is somewhat closer to the
physical intuition of the Witten index (\ref{Windex}) as the virtual
number of BPS particles on $X$. 

These invariants have a clear physical interpretation: They can be
thought of as the ``volume'' of the moduli space of BPS states
on $Y\subset X$ in that they
enumerate stable bound states that a \textit{single} D6-brane (since
ideal sheaves have rank one) wrapping the whole Calabi-Yau threefold $X$
can form with D2-branes wrapping rational curves $Y$ in class $\beta$ and
a number $n$ of
pointlike D0-branes; in this case the singularity sets of ideal
sheaves $\cI$ are the subschemes being counted. This identification essentially comes from the
interpretation of D-branes on a Calabi-Yau manifold as coherent sheaves. We
will provide direct evidence for this in the following
sections, where we study the moduli space of this D-brane system
and explicitly compute its partition function. 

\begin{figure}[h]
\centering
\includegraphics[width=3cm]{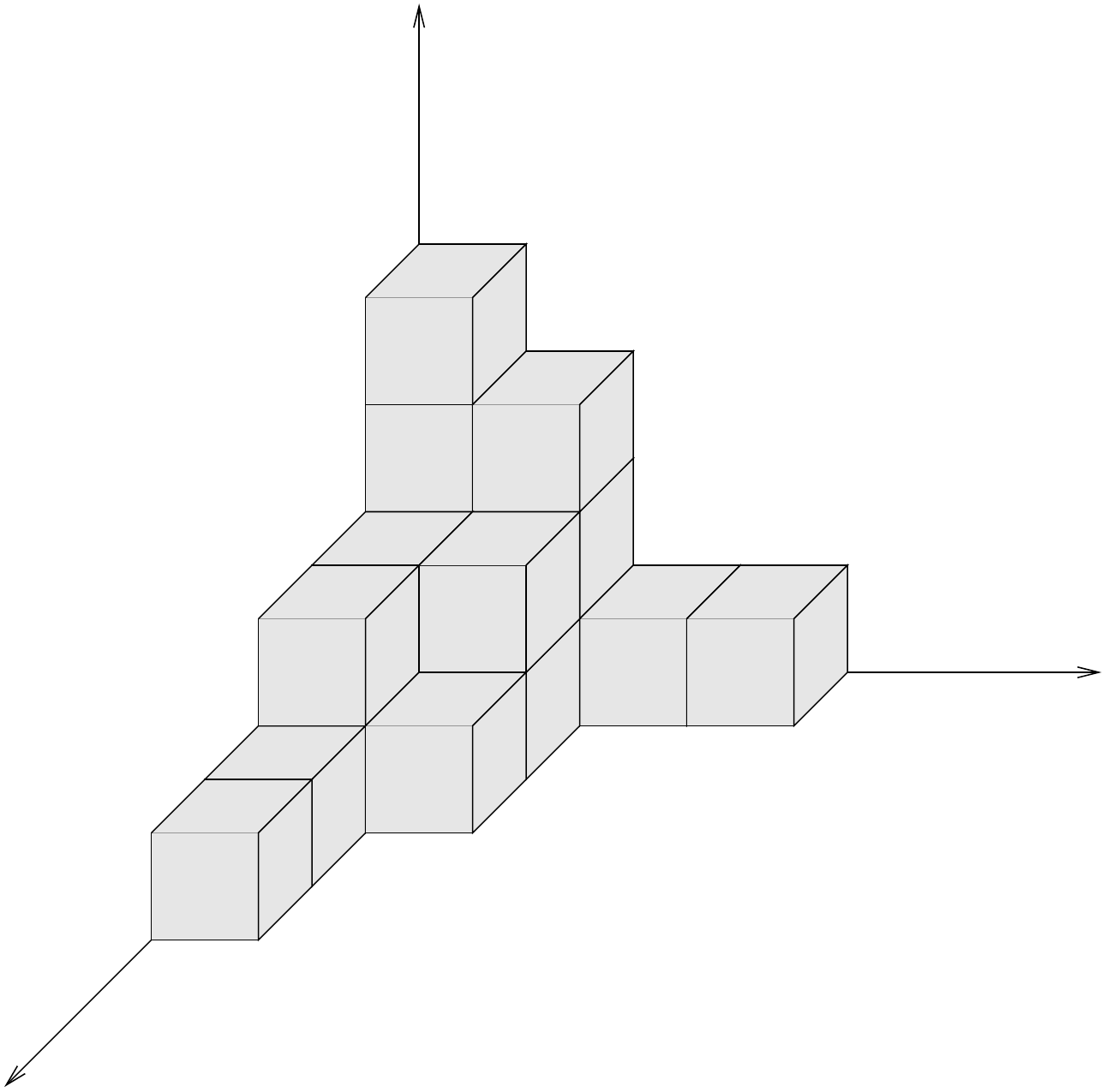}
\caption{\footnotesize{A three-dimensional Young diagram (plane partition)
  $\pi$. The corners of the upper boxes are located at lattice points
  $(i,j,\pi_{i,j})= (n_1,n_2,n_3)\in\IZ_{\geq0}^3$ with
  $\pi_{i,j}\geq\pi_{i+1,j}$ and $\pi_{i,j}\geq\pi_{i,j+1}$. The
  total number boxes of $\pi$ is $|\pi|=\sum_{i,j}\, \pi_{i,j}$.
}}
\label{3dYoungFig}\end{figure}

Using $\chi(\Hilb_n(X),\nu_X)=(-1)^n\, \chi(\Hilb_n(X))$ and Cheah's
formula for the generating function for the Euler characteristics of
Hilbert schemes $\Hilb_n(X)$ of zero-dimensional subschemes of
length $n$ in $X$, the degree zero contributions to the partition function (\ref{ZDTdef})
can be summed explicitly to give~\cite{fantechi}
\beq
Z_0^X(q) = \sum_{n=0}^\infty\, (-q)^n \ \chi\big(\Hilb_n(X)\big) = M(-q)^{\chi(X)} \ ,
\label{DTdeg0}\eeq
where $\chi(X)$ is the topological Euler characteristic of $X$ and 
\beq
M(q) = \prod_{n=1}^\infty\, \big(1-q^n\big)^{-n} = \sum_\pi\,
q^{|\pi|}
\label{MacMahon}\eeq
is the MacMahon function which enumerates plane partitions
(three-dimensional Young diagrams) $\pi$ with $n=|\pi|$ boxes (see
Fig.~\ref{3dYoungFig}).
We use the convention $\chi(X)=1$ when
$X=\complex^3$, so that
\beq
\cZ_{\rm BPS}^{\complex^3}(q) = M(-q) \ .
\eeq

\subsection{Topological string theory}

The Donaldson-Thomas partition function (\ref{ZDTdef}) is widely
believed (and in some cases rigorously proven) to be equivalent via
dualities to the partition function of the A-model topological string
theory on the same threefold $X$. The topological A-model
can be defined as a topological twist of an $\cN=(2,2)$ superconformal
sigma-model coupled to gravity. After localization onto classical
(BRST) minima, the result is a theory of holomorphic maps 
\begin{equation}
\phi \, :\, \Sigma_g  \ \longrightarrow \ X
\end{equation}
from a string worldsheet, which is a smooth projective curve
$\Sigma_g$ of arithmetic genus $g\geq0$, into the Calabi-Yau threefold $X$. This map
describes a worldsheet instanton wrapping a curve in the homology
class $\beta = \phi_* [ \Sigma_g ] \in H_2 (X , \zed)$. The string theory
path integral of the topological A-model localizes onto a sum of
integrals over the moduli spaces of these maps. Each (compactified) moduli space
$\overline{\scrM}_{g,\beta} (X) $ parametrizes stable maps of genus
$g$ wrapping a cycle $\beta$. The ``number'' of worldsheet instantons
in each topological sector is computed via the ``volume'' of this
moduli space as
\begin{equation}
\GW_{g , \beta} (X)= \int_{[\,\overline{\scrM}_{g,\beta} (X)\,]^{\rm vir} } \ 1 \ .
\end{equation}
This formula defines the $\IQ$-valued Gromov-Witten invariants of
$X$. The integration is properly defined using a symmetric perfect obstruction
theory in virtual dimension zero, which yields a virtual fundamental
cycle $[\,\overline{\scrM}_{g,\beta} (X)\,]^{\rm vir}$ in the
degree zero Chow group of subvarieties. 
The topological string amplitude has a genus expansion
\begin{equation}
F_{\rm top}^X (\gtop , Q ) = \sum_{g=0}^{\infty}\, F^X_g (Q) \
\gtop^{2g-2} \qquad \mbox{with} \quad F^X_g(Q) =  \sum_{\beta \in H_2 (X
  ,\zed)} \  \GW_{g,\beta}(X) \  Q^{\beta} \ ,
\end{equation}
where $\gtop=g_s$ is the string coupling, and the topological string
partition function is
\beq
Z_{\rm top}^X(\lambda,Q) = \e^{F_{\rm top}^X(\lambda,Q)} \ .
\eeq

For toric threefolds $X$ the duality is based on the
following observation. In the ``classical'' limit where the K\"ahler classes of $X$ are large, which is the limit $Q\to0$ wherein the
Calabi-Yau threefold
is well approximated by gluing together $\complex^3$ patches, the
individual genus $g\geq2$ amplitudes reduce to
\begin{equation} \label{FPintegrals}
\GW_{g,0}(X) = \lim_{Q\to0}\, F_g^X(Q) = \frac{\chi (X)}{2} \, \int_{\overline{\scrM}_g} \, c_{g-1}
\left( \scrH_g \right)^{\wedge 3} \ .
\end{equation}
The topological string amplitude can thus be compactly
expressed as an integral over Chern classes of the Hodge bundle
$\scrH_g$ over the Deligne-Mumford moduli
space $\overline{\scrM}_g$ of Riemann surfaces,
whose fibre at a point $\Sigma_g$ is the complex vector space of
holomorphic sections $H^0 (\Sigma_g , K_{\Sigma_g})$ of the canonical line bundle $K_{\Sigma_g}\to\Sigma_g$. This contribution comes from the
constant maps: As the K\"ahler classes tend to
infinity, every instanton contribution is suppressed and the path
integral is well approximated by a sum over constant maps
$\phi$ which take all of $\Sigma_g$ to a fixed point in $X$, with $\beta=0$. There are
additional terms in genera $g=0,1$ which are divergent in this
limit.
The integrals in (\ref{FPintegrals}) were explicitly computed by
Faber-Pandharipande~\cite{FaberPand} as
\beq
\int_{\overline{\scrM}_g} \, c_{g-1}
\left( \scrH_g \right)^{\wedge 3} = \frac{(-1)^g\, |B_{2g}\, B_{2g-2}|}{2g\, (2g-2)\,
  (2g-2)!}
\label{FPintegralscomp}\eeq
with $B_{2g}$ the Bernoulli numbers for $g\geq2$, and consequently~\cite{Gopakumar:1998jq}
\begin{equation}
\lim_{Q\to0}\, Z_{\rm top}^X(\lambda,Q) = M(q)^{\chi(X)/2}
\end{equation}
where $q=-\e^{\ii \gtop}$ and $M(q)$ is the MacMahon function (\ref{MacMahon}).
Heuristically, if we send the K\"ahler classes to infinity, the
topological string amplitude becomes a product of generating functions of
three-dimensional Young diagrams, each factor associated
with one of the copies of $\complex^3$ needed to cover the
Calabi-Yau manifold (in the sense of toric geometry), the total number of which is~$\chi(X)$.

\subsection{Topological
  vertex formalism\label{subsec:gaugestringduality}}

The proposed gauge/string theory duality can be stated precisely as the remarkable fact that the Gromov-Witten and
Donaldson-Thomas partition functions are actually the same in the
sense that~\cite{MNOP}
\begin{equation}
Z_{\rm top}^{\, X}(\gtop , Q )= M(q)^{-\chi(X)/2} \ {\cZ}_{\rm
  BPS}^{\,X} (q=-\e^{\ii\gtop} , Q) \ .
\label{gaugestringduality}\end{equation}
We interpret this equality as saying that the topological string
amplitude captures the degeneracies of the D6--D2--D0 BPS bound states
on a local threefold $X$. The duality can be extended to include open strings and
topological D-branes which wrap Lagrangian submanifolds of a
Calabi-Yau variety~$X$.

This equality can be proven for toric manifolds $X$, wherein the
partition functions can be evaluated combinatorially by applying virtual
equivariant localization techniques to the BPS indices (\ref{DTdef}) with respect to the induced action
of the three-torus $\torus^3$ on the Hilbert
scheme~\cite{MNOP}. Localization involving virtual fundamental classes
constructs a model for the virtual tangent space at the fixed points
of the torus action: One considers the infinitesimal deformations
around the fixed point and then \textit{subtracts} the obstructions to
these deformations. In this way one can regard the virtual tangent
space roughly as the difference between two cohomology groups, and the
relevant localization theorem gives the virtual Bott localization formula.

In the present case
we can choose an affine cover of $X$ whose local charts $U_{\alpha}\cong\IC^3$ are
centred at the fixed {points} of the toric action. The integral
(\ref{DTdef}) receives only two types of contributions, from the torus
fixed {points} and the torus fixed {lines} in $X$, which
correspond respectively to the vertices and edges of the toric diagram $\Delta$ of $X$. A torus fixed point $\pi_{\alpha}$ on $U_{\alpha}$ is represented by a  monomial ideal 
\begin{equation}
I_{\alpha} = \cI \big|_{U_{\alpha} } \ \subset \ \complex[z_1 , z_2 , z_3]
\end{equation}
which can be concretely represented by a three-dimensional Young diagram
\begin{equation}
\pi_{\alpha} = \big\{ (m_1 , m_2 , m_3) \in \zed^3_{\ge 0} \ \big| \
z_1^{m_1}\, z_2^{m_2}\, z_3^{m_3} \notin I_{\alpha} \big\} \ .
\label{pialpha}\end{equation}
The second type of contribution comes from overlaps of two open charts
$U_{\alpha}$ and $U_{\beta}$ which are glued along the line
corresponding to $z_1$ to give
\begin{equation}
I_{\alpha \beta} = \cI \big|_{U_{\alpha} \cap U_\beta} \ \subset \ \complex[z_1^{\pm\, 1} , z_2 , z_3]
\end{equation}
and are represented by two-dimensional Young diagrams
\begin{eqnarray}
\lambda_{\alpha \beta} =
\big\{(m_1 , m_2 , m_3) \ \big| \ z_2^{m_2}\, z_3^{m_3} \notin
I_{\alpha \beta} \big\} \ .
\label{lambdaalphabeta}\end{eqnarray}
A calculation in \v{C}ech cohomology then determines the contributions
to the Euler characteristic $n=\chi(\cI)$~\cite{MNOP}.

The combinatorial evaluation of the partition function (\ref{ZDTdef}) thus boils
down to decorating the trivalent planar toric graph $\Delta$ of $X$
with a three-dimensional Young diagram $\pi_v$ at each vertex $v$ and a
two-dimensional Young diagram $\lambda_e$ at each edge $e$
representing the asymptotics of the infinite plane partition
$\pi_v$. Symbolically, the partition function is then of the form
\beq
\cZ_{\rm BPS}^X(q,Q)= \sum_{\stackrel{\scriptstyle{\rm Young\
      diagrams}}{\scriptstyle\lambda_e}} \ \prod_{{\rm edges}\ e}\,
Q_e^{|\lambda_e|}~\prod_{\stackrel{\scriptstyle{\rm
      vertices}}{\scriptstyle v=(e_1,e_2,e_3)}}\,
M_{\lambda_{e_1},\lambda_{e_2},\lambda_{e_3}}(-q)
\label{ZDTcrystal}\eeq
where
\beq
M_{\lambda,\mu,\nu}(q)= \sum_{\pi\, :\, \partial\pi=(\lambda,\mu,\nu)}\, q^{|\pi|}
\label{topvertexcrystal}\eeq
is the generating function for plane partitions $\pi$ asymptotic to
$(\lambda,\mu,\nu)$ with $M_{\emptyset,\emptyset,\emptyset}(q)=M(q)$; we set $\lambda_e=\emptyset$ on all external
legs $e$ of the graph $\Delta$. The regularised box count $|\pi|$ of
an infinite plane partition $\pi$ with boundary is
computed using the degrees $(m_{e,1} , m_{e,2})$ specifying the normal bundle
$\cO_{\PP^1}(m_{e,1}) \oplus \cO_{\PP^1} (m_{e,2})$ over the rational curve in $X$ corresponding
to the given edge $e$ with the contribution
\beq
q^{\sum_{(i,j) \in \lambda_{e}}\, \left( m_{e,1} \, (i-1) + m_{e,2}\,
    (j-1) + 1 \right) + m_{e,1} \, | \lambda_{e} |} \ .
\label{framingfactor}\eeq
This gives the gluing rules for assembling
three-dimensional and two-dimensional Young diagrams together; the
``framing factors'' (\ref{framingfactor}) ensure that the gluing
procedure doesn't overcount or omit boxes when two plane partitions
are glued together along an edge $e$. The
localization integrals only contribute signs. It is
shown in~\cite{Okounkov:2003sp,MNOP} that 
(\ref{topvertexcrystal}) coincides (up to normalisation) with the topological
vertex~\cite{Aganagic:2003db} in the melting crystal
framework. Formally, the partition function (\ref{ZDTcrystal}) defines
the statistical mechanics of a ``Calabi-Yau crystal'' and reproduces
the topological string partition function within the formalism of the
topological vertex. In this setting the boxes of the Young diagrams correspond to sections
of the structure sheaves $\cO_Y$ of the associated closed subschemes
$Y\subset X$. A more general mathematical treatment of this algorithm,
including classes of non-toric local
Calabi-Yau threefolds involving non-planar graphs, can be found in~\cite{Li:2004uf}.

In the following we will interpret
the series (\ref{ZDTcrystal}) as a sum over generalized instantons in an
auxilliary topological gauge theory on the D6-brane. They
are states corresponding to ideal sheaves which can
be thought of as singular gauge fields. In this setting the large radius 
BPS index (\ref{Windex}) is regarded as the Witten index of the
field theory on the D-branes, and hence the BPS partition function
should be equivalent to the instanton partition function on the
D-branes in this limit.

\subsection{Conifold geometry\label{subsec:conifold1}}

Let us work through an explicit example. The
conifold singularity in six dimensions can be described as the locus $z_1\,z_2-z_3\, z_4=0$ in
$\IC^4$; the singularity at the origin can be removed by deforming the
conifold to the locus $z_1\,z_2-z_3\, z_4=t$ with the deformation parameter $t$ thought of
as the area of a projective line $\IP^1$ replacing the origin. The
crepant resolution of the conifold singularity is called the
resolved conifold and it is described geometrically as the total space of a rank two holomorphic
vector bundle $\cO_{\IP^1}(-1)\oplus\cO_{\IP^1}(-1)\to\IP^1$. The toric diagram
$\Delta$ consists of a single edge, representing the base $\IP^1$,
joining two trivalent vertices, representing the two $\IC^3$ patches
required to cover the resolved conifold. 

The large
radius partition function (\ref{ZDTcrystal}) reads
\bea
\cZ_{\rm BPS}^{\rm conifold}(q,Q) &=& \sum_{\lambda}\,
M_{\emptyset,\emptyset,\lambda}(-q)\,
M_{\emptyset,\emptyset,\lambda}(-q) \, Q^{|\lambda|} \nonumber \\[4pt] &=& \sum_{\pi_1,\pi_2,\lambda}\,
(-q)^{|\pi_1|+ |\pi_2|+\sum_{(i,j)\in\lambda} \, (i+j+1)} \,
Q^{|\lambda|} \ = \ M(-q)^2\, M(Q,-q)^{-1} \ ,
\label{largeRconifold}\eea
where the generating function
\beq
M(Q,q)= \prod_{n=1}^\infty\, \big(1-Q\,q^n\big)^{-n}
\label{genMacMahon}\eeq
counts weighted plane partitions with $M(1,q)=M(q)$. In the ``classical'' limit $Q\to0$
this reduces to $\big(\cZ_{\rm BPS}^{\IC^3}(q)\big)^2$, while the
``undeformed'' limit $Q\to1$ gives $\cZ_{\rm BPS}^{\IC^3}(q)$, as
expected. 

In the setting of topological string theory, the constant map
contributions $\GW_{g,0}({\rm conifold})$ for $g\geq2$ are given by (\ref{FPintegralscomp}),
while the two-homology of the resolved conifold is generated by its base
$\PP^1$ and the Gromov-Witten invariants which count
worldsheet instantons are given by~\cite{FaberPand}
\beq
\GW_{g,w}({\rm conifold}) = -w^{2g-3}\, \frac{|B_{2g}|}{2g\, (2g-2)!}
\eeq
for $g\geq2$ and curve class $\beta=w\, [\PP^1]$ with $w\in\IZ$. The
topological string amplitude is given by
\bea
F_{\rm top}^{\rm conifold}(\lambda,Q) &=& F^{\rm conifold}_0(Q)+F^{\rm
  conifold}_1(Q) \\ && +\, \sum_{g=2}^\infty\,\lambda^{2g-2}\,
\Big(\, \frac{(-1)^g\, |B_{2g}\, B_{2g-2}|}{2g\, (2g-2)\,
  (2g-2)!}- \frac{|B_{2g}|}{2g\, (2g-2)!}\, {\rm Li}_{3-2g}(Q)\, \Big)
\ , \nonumber
\eea
where the genus zero contribution contains the intersection numbers and the instanton factor ${\rm Li}_3 (Q) = \sum_{w\geq1}\, \frac{Q^w}{w^3}$, the
genus one part is given by the second Chern class of the tangent bundle of the
conifold, and the polylogarithm function ${\rm
  Li}_{3-2g}(Q)=\sum_{w\geq1}\, \frac{Q^w}{w^{3-2g}}$ sums the
contributions from genus $g\geq2$ worldsheet instantons.
The
gauge/string duality (\ref{gaugestringduality}) (with $\chi(X)=2$)
now follows from the exponential representation
\beq
M(Q,q)=\exp\Big(-\sum_{n=1}^\infty\, \frac{Q^n\, q^n}{n\,
  \big(1-q^n\big)^2}\, \Big)
\label{exprepMac}\eeq
of the generalized MacMahon function (\ref{genMacMahon}).

\bigskip

\section{Counting modules over 3-Calabi-Yau algebras\label{sec:3CY}}

\subsection{Quiver gauge theories}

So far we have been discussing ordinary Donaldson-Thomas invariants,
which are a particular case of a generalized set of invariants defined
over the whole Calabi-Yau moduli space that take into account the
chamber structure and wall-crossing phenomena. We will now describe a
different chamber, called the noncommutative crepant resolution
chamber, wherein the geometry is described via the path
algebra of a quiver, upon imposing a  set of relations which
can be derived from a superpotential. Physically this is the relevant
situation that one encounters when a D3-brane is used as a probe of a
singular Calabi-Yau threefold in a Type~IIB compactification. The
local geometry is reflected in the low-energy effective field theory
which has the form of a quiver gauge theory, i.e. a field theory whose
gauge and matter field content can be summarized in a
representation of a quiver. This supersymmetric gauge theory is
characterized by a superpotential which yields a set of relations as
F-term constraints. The path algebra associated to the quiver is
constrained by these relations; it encodes the geometry of the
Calabi-Yau singularity and can be used to set up an interesting
enumerative problem, where the roles of coherent sheaves are played by
finitely-generated modules over this algebra in the standard
dictionary of noncommutative algebraic geometry. Physically this corresponds to counting BPS states in the low-energy effective field theory; mathematically it goes under the name of {\it noncommutative Donaldson-Thomas theory}~\cite{szendroi,reineke}.

Recall that a quiver $\sf Q$ is a directed graph specified by a
set of vertices $v\in\sfQ_0$ and a set of arrows $\big(v\xrightarrow{ \ a
  \ }
w\big)\in \sfQ_1$ connecting vertices $v,w\in\sfQ_0$. Often a quiver comes with a set $\sfR$ of
relations among its arrows. A path in the quiver is a set of arrows
which compose; the relations $\sfR$ are realized as formal $\IC$-linear
combinations of paths. The paths modulo the ideal generated by the relations form an associative
$\IC$-algebra called the path algebra $\sfA=\complex \sf Q/\langle\sfR \rangle$, with product defined by concatenation of paths where possible and zero
otherwise. 

There is a particular class of quivers which come equipped with a
superpotential $\sfW:\sfQ_1\to \IC\sfQ$. In this case the mathematical
definition of relations descends directly from the physical one: The relations of
the quiver are the F-term equations derived from the superpotential, which is the ideal of relations generated by
\begin{equation}
\sfR = \langle \partial_a \sfW \ \vert \ a \in \sfQ_1 \rangle \ .
\end{equation}
Here we regard $\sf{W}$ as a sum of cyclic monomials (since
the gauge theory superpotential comes with a trace); the
differentiation with respect to the arrow $a$ is then formally taken by cyclically permuting the elements of a monomial until $a$ is in the first position and then deleting it.

A (linear) representation of a quiver with
relations $(\sfQ,\sfR)$ is a collection of complex vector spaces $V_v$ for
each vertex $v\in\sfQ_0$ together with a collection of linear transformations
$B_a:V_v\to V_w$ for each arrow $\big(v\xrightarrow{ \ a
  \ }
w\big)\in \sfQ_1$ which respect the ideal of relations $\sfR$. 
The representations of a quiver with
relations $(\sf Q,R)$ form a category ${\frR ep}(\sf Q , R)$ which is
equivalent to the category $\frM od(\sfA)$ of finitely-generated left $\sfA$-modules. In quiver gauge theories one
looks in general for representations of $(\sfQ,\sfR)$ in the category of complex vector
bundles (or better coherent sheaves), but in many cases the gauge theory is equivalently described
by a matrix quantum mechanics for which it suffices to study the module
category ${\frR ep}(\sf Q , R)$. We will be mostly interested in
isomorphism classes of quiver representations which are orbits under
the action of the gauge group $\prod_{v\in\sfQ_0} \, GL (V_v,\IC)$; they can be characterised using geometric invariant theory.
Equivalently, there is a more algebraic notion of stability of quiver
representations introduced by King, which is
closely related to the notion of stability for BPS states given by D-term
constraints in
supersymmetric gauge theories. 

To each vertex $v$ we can associate a one-dimensional simple module
$\sfD_v$ with $V_v = \complex$
and all other $V_w = 0$; in string theory these modules correspond to
fractional branes. Furthermore, if $\sfe_v$ is
the trivial path at $v$ of length zero, then $\sfP_v: =
\sfe_v\sfA$ is the subspace of the path algebra generated
by all paths that begin at vertex $v$; they are projective objects in
the category ${\frR ep}(\sf Q , R)$ which can be used to construct projective resolutions of the simple modules through
\begin{equation}
\begin{xy}
\xymatrix@C=8mm{  \cdots \ \ar[r] & \ \displaystyle{\bigoplus_{w\in
      {\sfQ_0}} \, \sfP_w{}^{\oplus d^{k}_{w,v}}} \ \ar[r] & \ \cdots \ \ar[r]
  & \ \displaystyle{\bigoplus_{w \in {\sfQ_0}} \, \sfP_w{}^{\oplus d^{1}_{w,v}}
  }  \ \ar[r] & \ar[r]  \ \sfP_v \ & \ \sfD_v \ \ar[r]  & \ 0
}
\end{xy}
\label{projressimple}\end{equation}
where
\begin{equation}
{d}^p_{w,v} = \dim \Ext^p_{\sf{A}} \left( \sfD_v , \sfD_w \right) \ .
\label{dpwvdimExt}\end{equation}
Note that $d^0_{w,v}=\delta_{w,v}$ since $\sfD_v$ are simple objects;
furthermore $d^1_{w,v}$ gives the number of arrows in $\sfQ_1$ from
vertex $w$ to vertex $v$ and $d^2_{w,v}$ is the number of relations,
while $d^3_{w,v}$ is the number of relations among the relations, and
so on.

\subsection{Noncommutative Donaldson-Thomas theory\label{subsec:NCDT}}

When a D-brane probes a singular
Calabi-Yau threefold, its low-energy effective field theory is
typically encoded in a quiver diagram. From the geometry of the
moduli space of the effective field theory one can reconstruct the
Calabi-Yau singularity; it often happens that the singularity is
\textit{resolved} by quantum effects. We can abstract these facts into
a mathematical statement: For Calabi-Yau singularities it is possible
to find a certain noncommutative algebra $\sfA$ whose centre is the
coordinate ring of the singularity and whose moduli spaces of
representations are resolutions of the singularity. We call $\sfA$ the
noncommutative crepant resolution of the singularity; the algebra
$\sfA$ is finitely generated as a module over its centre and has the
structure of a \emph{3-Calabi-Yau algebra}, see e.g.~\cite{Ginzburg-CY}. In the cases we are interested in, $\sfA$ is the path algebra of a quiver with relations.

On this noncommutative crepant resolution one can define Donaldson-Thomas invariants.
The starting point is the low-energy effective field theory of a
D2--D0 system on a toric Calabi-Yau threefold $X$. This
system of branes has a  ``leg'' outside $X$ in flat space $\real^4$;
the effective field theory is the dimensional reduction to one dimension of
four-dimensional $\cN=1$ supersymmetric Yang-Mills theory. The end
result is a supersymmetric quiver quantum mechanics with
superpotential. Several techniques are available for determining this
data. For example, Aspinwall-Katz have derived a fairly general
formalism to compute the superpotential
in~\cite{Aspinwall:2004bs}. Alternatively, the technology of brane
tilings gives an efficient algorithm which has also a neat physical
picture in terms of dualities~\cite{Feng:2005gw,Franco:2005rj}; see~\cite{Kennaway:2007tq,Yamazaki:2008bt} for reviews. A more geometric
perspective considers the toric geometry directly; then the quiver is
derived through the endomorphisms of a tilting object, which
guarantees that the derived category of representations of the quiver
contains all geometric information encoded in the derived category of
coherent sheaves on $X$. In the following we will assume that the relevant quiver $\sfQ$ and its superpotential $\sf W$ are known.

The path algebra $\sfA$ of this quiver with relations encoded in the
superpotential is a noncommutative crepant resolution of the
Calabi-Yau threefold $X$. However, the quiver so far
represents only the effective field theory of the D2--D0 system. To
incorporate the D6-brane we add an extra vertex $\{ \bullet \}$
together with an additional
arrow $a_\bullet$ from the new vertex to a reference vertex $v_0$ of $\sfQ$; this
operation is a \textit{framing} of the quiver and it ensures that the
moduli space of representations has nice properties. We will denote the
new quiver by $\hat \sfQ$; its vertex and arrow sets are given by
\begin{equation}
\hat \sfQ_0 = \sfQ_0 \cup \{ \bullet \} \qquad \mbox{and} \qquad \hat
\sfQ_1 = \sfQ_1 \cup \big\{ \bullet \xrightarrow{ \ a_{\bullet} \ } v_0 \big\} \ .
\end{equation}
This new quiver has its own path algebra $\hat \sfA$. Representations
are now constructed by specifying an $n_v$-dimensional vector space
$V_v$ on each node $v\in\sfQ_0$, while the framing node $\bullet$
always carries a one-dimensional vector space $\IC$. Then the moduli space
of stable representations $\scrM_{\mbf n}(\hat\sfQ,v_0)$ characterized by the
dimension vector ${\mbf n}=(n_v)_{v\in\sfQ_0}$ is compact and well
behaved~\cite{reineke}; an appropriate symmetric perfect obstruction theory is
developed in~\cite{szendroi,reineke}, which gives a sensible notion of integration. The index of BPS D6--D2--D0
states is then the noncommutative Donaldson-Thomas invariant which is given by the {weighted} Euler characteristic of this moduli space as
\begin{equation} \label{ncdt}
\Omega_{\sfA,v_0}(\mbf n)= \DT_{\mbf n , {v_0}}(\sfA) := \chi \big(\scrM_{\mbf n}(\hat\sfQ,v_0)
\,,\, \nu_\sfA
\big) \ ,
\end{equation}
which again counts the ``virtual'' number of points. We can therefore define a
partition function for the noncommutative Donaldson-Thomas invariants
obtained from the path algebra $\sfA$ as
\begin{equation}
\cZ_{\rm BPS}^{\sfA} (p , v_0) = \sum_{n_v \in \zed} \,
\Omega_{\sfA,v_0}(\mbf n) \ \prod_{v\in\sfQ_0}\, p_v^{n_v} \ .
\label{NCZBPS}\end{equation}

To set up an enumerative problem associated with the quiver, we begin
from the state containing a single D6-brane and describe it as the
space of paths on the quiver starting at the reference node $v_0$,
modulo the F-term relations. We assign a different kind of box to each
node of the quiver diagram by using different colours. This constructs
a sort of ``pyramid partition'', starting from the tip. In this
enumerative problem only the nodes of $\sf Q$ enter: The extra framing
node
$\{ \bullet \}$ plays only a passive role, in a certain sense fixing a
``boundary condition'' since it corresponds to the D6-brane which
extends to infinity. The combinatorial construction is
rendered more involved by the fact that at each step one has to impose
the relations derived from the superpotential. The best way to
rephrase the quiver relations into a practical rule is by the use of
brane tilings and related techniques, see~\cite{reineke,Ooguri:2008yb}. The end result
is that there is a direct correspondence between modules over the path
algebra and the coloured pyramid partitions which are built.

The
enumerative problem of noncommutative Donaldson-Thomas invariants is
obtained by a combinatorial algorithm derived via virtual localization on the
quiver representation moduli space with respect to a natural global
action of a torus $\torus$ which rescales the arrows and
preserves the superpotential; the fixed points define ideals of the
path algebra $\sfA$. Behrend-Fantechi prove in
\cite{fantechi} that for a symmetric perfect obstruction theory like the ones
constructed in \cite{szendroi,reineke}, which are compatible with the
torus action, the contribution of an isolated $\torus$-fixed point to the
Euler characteristic (\ref{ncdt})
is given by a sign determined by the parity of the dimension of the tangent space
{at the fixed point}. In particular, if all torus fixed points
$\pi$ are isolated then we obtain~\cite[Thm.~3.4]{fantechi}
\beq
\DT_{\mbf n , {v_0}}(\sfA) = \sum_{\pi\in\scrM_{\mbf n}(\hat\sfQ,v_0)^\torus} \, (-1)^{\dim T_{\pi}
  \scrM_{\mbf n}(\hat\sfQ,v_0)} \ .
\label{DTfantechi}\eeq

However, in constructing the low-energy
effective field theory one loses track of the precise relation between
box colours and D-brane charges. This change of variables can be
explicitly determined in some simple
examples, but no closed general formula has been found so far. This means that
the BPS states thus enumerated have charges that can be non-trivial
functions of the original D0 and D2 brane charges; in the
noncommutative crepant resolution chamber the coloured partitions are
the relevant dynamical variables and should be more properly thought
of as ``fractional branes'' pinned to the singularity.

\subsection{Conifold geometry}

Let us look at the noncommutative crepant resolution of the conifold
singularity discussed in \S\ref{subsec:conifold1}. 
In this case the quiver $\sfQ$ is the Klebanov-Witten
quiver~\cite{Klebanov:1998hh}, which contains two vertices $\sfQ_0 =
\{ 0, 1 \}$ and four arrows $\sfQ_1 = \{ a_1 , a_2 , b_1 , b_2 \}$
with the quiver diagram
\begin{equation}
\begin{xy}
\xymatrix@C=30mm{
  0 \ \ar@/_1pc/[r]|{ \ a_1 \ } \ar@/_2pc/[r]|{ \ a_2 \ } &  \
  \ar@/_1pc/[l]|{ \ b_1 \ } \ar@/_2pc/[l]|{ \ b_2 \ } \ 1 
}
\end{xy}
\end{equation}
and superpotential
\begin{equation}
{\sf W} = a_1 \, b_1 \,a_2 \,b_2 - a_1\, b_2 \,a_2\, b_1 \ .
\end{equation}
The path algebra is given by
\begin{equation}
{\sf A} = \complex [e_0  ,  e_1] \langle a_1 , a_2 , b_1 , b_2 \rangle
\, \big/ \, \langle b_1 \,a_i\, b_2 - b_2\, a_i\, b_1 \,,\, a_1\,
b_i\, a_2 - a_2\, b_i \,a_1 \ | \ i = 1,2 \rangle \ .
\end{equation}
The centre $Z(\sfA)$ of this algebra is generated by the elements
\begin{eqnarray}
z_1 &=& a_1\, b_1 + b_1\, a_1 \ , \nonumber \\[4pt]
z_2 &=& a_2 \,b_2 + b_2\, a_2 \ , \nonumber \\[4pt]
z_3 &=& a_1 \,b_2 + b_2\, a_1 \ , \nonumber \\[4pt]
z_4 &=& a_2 \,b_1 + b_1\, a_2 \ ,
\end{eqnarray}
and hence
\begin{equation}
{ Z(\sfA)} = \complex[z_1 , z_2 , z_3 , z_4] \, \big/ \, \left( z_1 \,z_2 - z_3 \,z_4 \right)
\end{equation}
which corresponds precisely to the nodal singularity of
the conifold. The path algebra $\sfA$ is thus a noncommutative crepant
resolution of the conifold singularity.

To construct noncommutative Donaldson-Thomas invariants, we consider
the framed conifold quiver
\begin{equation}
\begin{xy}
\xymatrix@C=30mm{
\bullet \ \ar@{.>}@//[dr]|< < < < < < < <{ \ a_{\bullet} \ } \\   & \ 0  \ \ar@/_1pc/[r]|{
  \ a_1 \ } \ar@/_2pc/[r]|{ \ a_2 \ } &  \ \ar@/_1pc/[l]|{ \ b_1 \ }
\ar@/_2pc/[l]|{ \ b_2 \ } \ 1 
}
\end{xy}
\end{equation}
We want to study the moduli space of finite-dimensional
representations of this quiver with relations that follow from the
superpotential $\sf W$, or equivalently finite-dimensional left $\sf
A$-modules. In particular, the moduli space of stable representations
is now equivalent to the moduli space of \textit{cyclic} modules. To construct this moduli
space, one considers the representation space
\begin{equation}
{\sf Rep} (\sfQ , v_0  ) = \bigoplus_{(v\longrightarrow w) \in
  {\sfQ}_1}\,  \Hom_\IC (V_{v} , V_{w}) \ \oplus \ \Hom_\IC(V_0 ,
\complex) \ ,
\end{equation}
where we have introduced a vector space $V_0$, $V_1$ for each of the
two nodes and the last summand corresponds to the framing arrow $a_{\bullet}$
to the reference node $v_0=0$. Now let ${\sf Rep} (\sfQ , v_0 ; \sfW )
$ be the subspace obtained upon imposing the F-term equations derived
from the superpotential $\sf W$ on ${\sf Rep} (\sfQ , v_0  )$. The
moduli space is obtained by factoring the natural action
of the gauge group by basis changes of the complex vector spaces $V_0$
and $V_1$ to get a smooth Artin stack
\begin{equation}
{\scrM}_{n_0,n_1}(\hat\sfQ) = \big[ {\sf Rep} (\sfQ , v_0 ; \sfW ) \,
\big/ \, GL(n_0,\IC)\times GL(n_1,\IC) \big] \ ,
\end{equation}
where we have dropped the reference vertex label for simplicity. After
suitably defining the quotient, the resulting space has nice
properties and carries a symmetric perfect obstruction theory~\cite{szendroi}.

\begin{figure}[h]
 \centering
  \includegraphics[width=6cm]{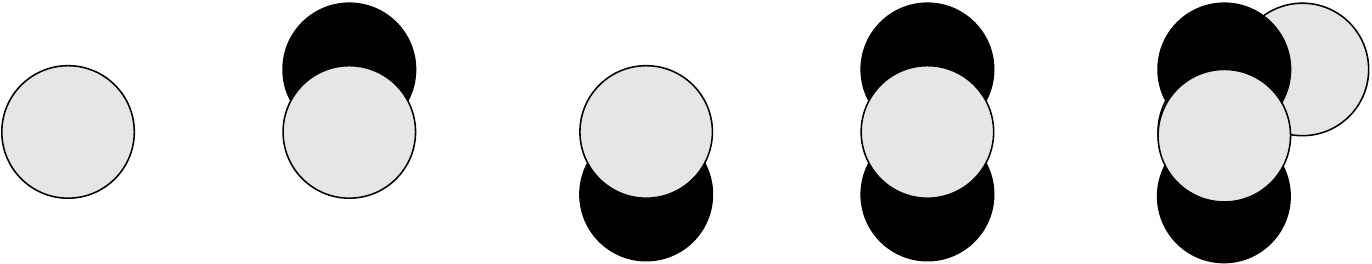}
 \caption{\footnotesize{A pyramid partition $\pi\in\mathcal{P}$.}}
 \label{toglipalle}
\end{figure}

Our task is now to compute the partition function for the noncommutative invariants
\begin{equation}
\cZ_{\rm BPS}^{\sf conifold} (p_0, p_1) = \sum_{ n_0, n_1 \in \zed} \,
\DT_{n_0,n_1}(\sfA) \ p_0^{n_0} \, p_1^{n_1}  \qquad \mbox{with}
\quad \DT_{n_0,n_1}(\sfA) =
\chi \big( {\scrM}_{n_0, n_1}(\hat\sfQ) \,,\, \nu_\sfA \big) \ .
\label{ZBPSNCconifold}\end{equation}
The invariants can be computed via equivariant localization with
respect to a natural torus action which rescales the arrows
diagonally; we keep only those transformations which leave
the superpotential invariant. The relevant torus is thus
\begin{equation}
\torus = \torus_{\sfW} \, \big/ \, S^1 \ ,
\end{equation}
where
\begin{equation}
\torus_\sfW = \big\{ ( \e^{\ii \epsilon_1} , \e^{\ii \epsilon_2} ,
\e^{\ii \epsilon_3} , \e^{\ii \epsilon_4} ) \ \big| \ \epsilon_1 +
\epsilon_2 + \epsilon_3 + \epsilon_4 = 0 \big\} \ ,
\end{equation}
while the subtorus
\begin{equation}
S^1 \cong \big\{ (\lambda , \lambda , \lambda^{-1} , \lambda^{-1}) \ \big|
\ \lambda\in S^1 \big\}
\end{equation}
is dictated by the charge vector which characterizes the resolved conifold as a
toric variety. To evaluate the invariants we therefore need to
classify the fixed points of this toric action on the moduli space and
compute the contributions of each fixed point. In~\cite{szendroi}
Szendr\H{o}i proves that there is a bijective correspondence between
the set of fixed points and the set $\mathcal{P}$ of plane
partitions $\pi$ such that if a box is present in $\pi$ then so are the
boxes immediately above it, which are of two different colours, say
black and white. Elements $\pi\in \cP$ are called
\textit{pyramid partitions} (see Fig.~\ref{toglipalle}).
These combinatorial arrangements are examples of those we discussed in
\S\ref{subsec:NCDT}; the enumeration of BPS states proceeds from the
framed conifold quiver via the
combinatorial algorithm that we sketched there. 

To assemble this data into the partition function
(\ref{ZBPSNCconifold}), all that is left to do is
compute the contribution of each fixed point $\pi
\in \mathcal{P}$, for which \cite{szendroi} proves
\begin{equation}
(-1)^{\dim T_\pi {\scrM}_{n_0, n_1}(\hat\sfQ)  } = (-1)^{|\pi_1|}
\end{equation}
where $n_1=|\pi_1|$ is the number of black boxes in the pyramid partition
$\pi$. If we similarly denote by $n_0=|\pi_0|$ the number of white boxes,
then the generating function of noncommutative Donaldson-Thomas invariants assumes the form
\begin{equation}
\cZ_{\rm BPS}^{\sf conifold}(p_0, p_1) = \sum_{ \pi \in \mathcal{P}} \,
(-1)^{|\pi_1|}  \ p_0^{|\pi_0|} \, p_1^{|\pi_1|} \ .
\end{equation}
This partition function has a nice infinite product representation \cite{young}: If we change variables to $q= p_0 \, p_1$ and $Q= p_1$ then
\begin{equation}
\cZ_{\rm BPS}^{\sf conifold}(q , Q) = M(-q)^2 \, M(Q,-q) \, M(Q^{-1},-q)  \ .
\label{NCcZBPSconifold}\end{equation}

It is interesting to compare the partition function
(\ref{NCcZBPSconifold}) with the
large radius generating function (\ref{largeRconifold}) of the
ordinary Donaldson-Thomas invariants. Szendr\H{o}i notes there is
a product formula that expresses the noncommutative invariants in
terms of the ordinary Donaldson-Thomas invariants of the \textit{two}
(commutative) crepant resolutions of the conifold singularity, which
are mutually related by a flop transition where the two-cycle shrinks
to zero size. If we label them by $\pm$, then
\begin{eqnarray}
\cZ_{\rm BPS}^{{\rm conifold}^-}(q , Q) = M(-q)^2 \, M(Q,-q) \qquad \mbox{and} \qquad
\cZ_{\rm BPS}^{{\rm conifold}^+}(q , Q) = M(-q)^2 \, M(Q^{-1},-q) \ ,
\end{eqnarray}
where now the parameter $Q$ is interpreted as the large radius
K\"ahler parameter $\e^{-t}$. Then one has the surprising equality
\begin{equation}
\cZ_{\rm BPS}^{\sf conifold} (q ,Q) = M(-q)^{-2} \ \cZ_{\rm BPS}^{{\rm
    conifold}^-}(q , Q) \, \cZ_{\rm BPS}^{{\rm conifold}^+}(q , Q) \ .
\end{equation}
This formula is interpreted as saying that the partition function of
noncommutative Donaldson-Thomas invariants can be obtained from the
topological string partition function from a number of wall crossings.

\bigskip

\section{Counting curves in toric surfaces\label{sec:curvecountsurf}}

\subsection{BPS states and Hilbert schemes}

In order to set up the gauge theory formulation of the curve counting
problems on Calabi-Yau threefolds, we shall start in a setting wherein
the gauge theory problem (instanton counting) has been more thoroughly
investigated and is much better understood.
In the next few sections we will study the relationship between curve
counting and gauge theory on a general toric surface $M$, in
particular with the enumeration of instantons in four dimensions. However, even in the
simplest $U(1)$ cases, surprisingly little is understood rigorously in
full generality; the best understood classes of toric surfaces are
ALE spaces and Hirzebruch surfaces. One of our goals in the following
will be an attempt to set up a rigorous framework that computes the gauge
theory partition functions on generic Hirzebruch-Jung spaces.
Geometrically, the respective problems correspond to counting
subschemes of dimension zero and one with compact support in a surface $M$,
and counting rank one torsion free sheaves $\cT$ on $M$ which now have the factorized form
\begin{equation} \label{factorization}
\cT = \calL \otimes \cI
\end{equation}
where $\calL=\cT^{\vee\vee}$ is a line bundle and $\cI$ is an ideal sheaf of points;
compared to the six-dimensional case, the extra factor $\calL$ is
necessary to include one-dimensional subschemes which in this case
occur as
components in codimension one.
From the gauge theory perspective, the main difference from
Donaldson-Thomas theory is now the presence of a non-trivial first
Chern class. This is essentially the reason why the two problems are
different. As we have defined them, Donaldson-Thomas invariants count
bound states of D6--D2--D0 branes. Adding D4-branes which wrap compact
four-cycles yields sheaves with non-trivial first Chern
class. In the following we will
enumerate BPS D2--D0 bound states in D4-branes which wrap a toric
surface $M$ inside a Calabi-Yau threefold~$X$ without D6-branes.

The pertinent moduli space is again the Hilbert scheme
$\scrM^{\rm BPS}_{n,\beta}(M)=\Hilb_{n,\beta}(M)$ of compact curves $Y\subset M$ with
\beq
n=\chi(\cO_Y) \qquad \mbox{and} \qquad \beta=[Y] \ \in \ H_2(M,\IZ) \
.
\eeq
For any smooth projective surface $M$, the structure of this scheme
simplifies tremendously, and it is a \emph{smooth manifold}. In this
case a codimension one subscheme of $M$ factorizes into divisors, and
sums of free and embedded points. This implies that the Hilbert scheme
factorizes into divisorial and punctual parts as
\beq
\scrM^{\rm BPS}_{n,\beta}(M) \cong {\rm Div}_{\beta}(M)\times \Hilb_{n-n_\beta}(M)
\ .
\label{Hilbfact}\eeq
The intersection number
\beq
n_\beta= -\mbox{$\frac12$} \, \langle\beta, \beta+K_M\rangle_{\Gamma}
\label{intnumbeta}\eeq
is the contribution of the
divisorial part $D$ of a subscheme $Y\in\scrM^{\rm BPS}_{n,\beta}(M)$
with $[D]=\beta$ to the
holomorphic Euler
characteristic $n=\chi(\cO_Y)$, with $K_M=-c_1(M)$ the canonical class of
$M$, and ${\rm Div}_\beta(M):= \Hilb_{n_\beta,\beta}(M)$ is the projective moduli space of
divisors in $M$. The remainder due to free and embedded points of $Y$
is contained in the Hilbert scheme $\Hilb_m(M)$ of $m=n-n_\beta$ points
on $M$, which in this case is non-singular of dimension $2m$.

Thus in this case the moduli space $\scrM^{\rm BPS}_{n,\beta}(M)$ is
smooth, and we can define generating functions using ordinary
fundamental classes, without recourse to virtual cycles as before. In
particular, we define the partition function for D4--D2--D0 BPS bound
states on $M$ as
\beq
\cZ_{\rm BPS}^M(q,Q) = \sum_{\beta \in H_2 (M , \zed)}\, Q^{\beta}
\ Z_\beta^M(q) \qquad \mbox{with} \quad Z_\beta^M(q)= \sum_{n
  \in \zed} \, q^n  \ \Omega _M(n,\beta) \ ,
\label{ZBPSsurface}\eeq
where now the index of BPS states is given by the topological Euler characteristic
\beq
\Omega _M(n,\beta) := \chi\big(\scrM^{\rm BPS}_{n,\beta}(M)\big) = \int_{\scrM^{\rm BPS}_{n,\beta}(M)}\,\eul\big(T
    \scrM^{\rm BPS}_{n,\beta}(M)\big)
\label{BPSEulerchar}\eeq
with $\eul\big(T
    \scrM^{\rm BPS}_{n,\beta}(M)\big)$ the Euler class of the \textit{stable} tangent
bundle on the smooth moduli space $\scrM^{\rm BPS}_{n,\beta}(M)$. The analog of
the formula (\ref{DTdeg0}) for the degree zero contributions are now
encoded by G\"ottsche's formula
\beq
Z_0^M(q) = \sum_{n=0}^\infty \, q^n \ 
  \chi\big(\Hilb_n(M) \big)= \hat\eta(q)^{-\chi(M)} 
\label{Gottsche}\eeq
for the generating function of the Euler characteristics of Hilbert
schemes of points on surfaces, where $\hat\eta(q)$ is the Euler
function whose inverse
\beq
\hat\eta(q)^{-1}= \prod_{n=1}^\infty\,\big(1-q^n\big)^{-1} =
\sum_\lambda\, q^{|\lambda|}
\eeq
enumerates two-dimensional Young diagrams (partitions)
$\lambda=(\lambda_1,\lambda_2,\dots)$, $\lambda_i\geq\lambda_{i+1}$
with $n=|\lambda|=\sum_i\,\lambda_i$ boxes.

\subsection{Vertex formalism}

\begin{figure}[h]
\centering
\includegraphics[width=3cm]{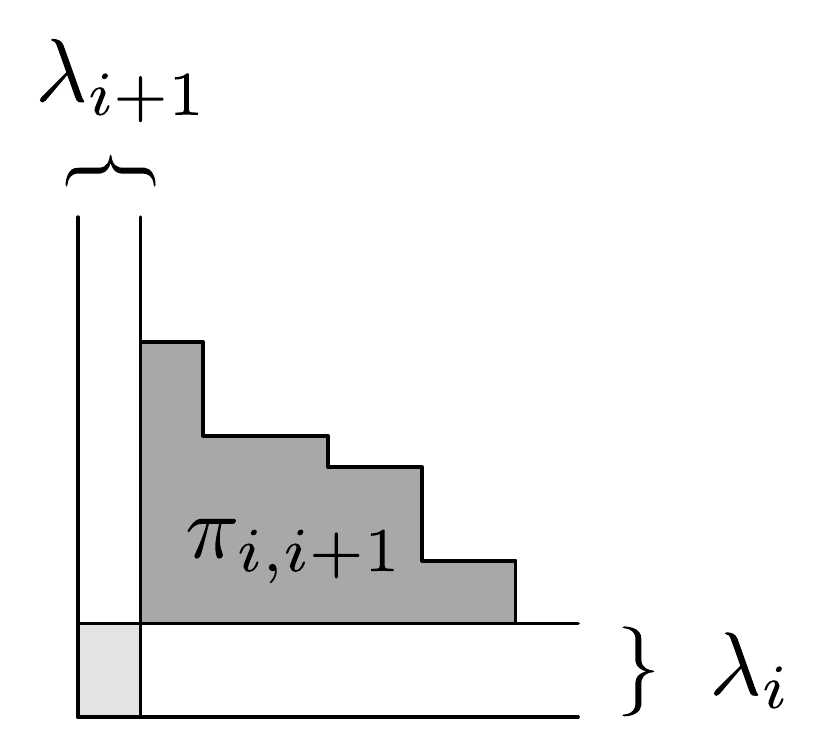}
\caption{\footnotesize{Factorization of an infinite two-dimensional Young diagram
  into a finite Young diagram $\pi_{i,i+1}$ and its boundaries along
  the two coordinate axes which are specified by integers
  $\lambda_i,\lambda_{i+1}\in\IZ_{\geq0}$. Here $i$ labels the
  one-cones of the toric fan of $M$, while $(i,i+1)$ labels the two
  bounding one-cones of each toric fixed point on $M$.
}}
\label{crystal2D}\end{figure}

The Euler characteristics (\ref{BPSEulerchar}) can be evaluated by
applying the localization theorem in equivariant Chow theory due to
Edidin-Graham~\cite{edidin}. In our case the localization formula is
simplified by the fact that the fixed point locus of the toric
action on the moduli space $\scrM^{\rm BPS}_{n,\beta}(M)$, induced by the action of the
two-torus $\torus^2$ on $M$, consists of isolated points and the
integrand is the Euler class of the tangent bundle, which cancels against
the tangent weights coming from the localization formula. The
collection of $\torus^2$-invariant ideal sheaves specifying the
one-dimensional subschemes $Y$ is in bijective correspondence with the
set of all (possibly infinite) Young
tableaux; in contrast to the case of plane partitions, there is a simple
factorization of infinite two-dimensional Young diagrams into a finite
part plus their asymptotic limits which we denote symbolically by
\beq
\{\mbox{infinite Young diagrams}\} ~ \cong ~ \IZ_{\geq0}^2
\ \times \ \{\mbox{finite Young diagrams}\} \ .
\eeq
This factorization is depicted in Fig.~\ref{crystal2D} and it
corresponds to the decomposition (\ref{Hilbfact}) of a subscheme $Y$ into a reduced and a zero-dimensional component.
The contribution of the free and embedded points to the holomorphic
Euler characteristic $n= \chi(\cO_Y)$ is given by the total box count of
the finite parts of Young diagrams, while for any compact toric
invariant divisor $D$ with an expansion
  $D=\sum_i\,\lambda_i\, D_i$, $\lambda_i\in\IZ_{\geq0}$ in a basis of
  $\torus^2$-invariant divisors $D_i$ with self-intersection numbers
  $\alpha_i= -\langle D_i,D_i\rangle_{\Gamma}$, a computation in \v{C}ech cohomology shows~\cite{Cirafici:2009ga}
\beq
\chi(\cO_D)=-\mbox{$\frac12$}\, \langle D,D+K_M\rangle_{\Gamma}= \sum_i\, \big(
\mbox{$\frac12$}\, \alpha_i\,
\lambda_i\, (\lambda_i-1)+\lambda_i - \lambda_i\,
\lambda_{i+1}\big) \ .
\label{chiOD}\eeq

Using this data one can combinatorially evaluate the BPS
partition function (\ref{ZBPSsurface}) via
a vertex formalism for toric surfaces $M$ which is analogous to the
topological vertex formalism for toric threefolds discussed in
\S\ref{subsec:gaugestringduality}. The partition function is
constructed on the bivalent
planar toric graph $\Delta$ which is the dual web diagram of the toric
fan of $M$. This formalism associates to each vertex $v=(e_1,e_2)$, where two edges
$e_1$ and $e_2$ of $\Delta$ meet, a factor
\begin{equation}
V_{\lambda_{e_1}, \lambda_{e_2}}(q) = \hat\eta(q)^{-1} ~ q^{-\lambda_{e_1} \,\lambda_{e_2}} \ ,
\end{equation}
where the inverse Euler function $\hat\eta(q)^{-1}$ counts two-dimensional
Young diagrams based at the vertex and the two non-negative integers
$\lambda_{e_1},\lambda_{e_2}$ label the asymptotics of the partitions
along the edges connecting two vertices with the product
$\lambda_{e_1} \,\lambda_{e_2}$ the contribution to the regularized
box count of an infinite Young diagram. Two vertices are glued
together along an edge $e$ carrying the common label $\lambda_e$ by a ``propagator''
\begin{equation}
G_{\lambda_e}(q,Q_e) = q^{\frac12\,\alpha_e \, \lambda_e
    \,(\lambda_e-1) + \lambda_e} \ Q_e^{\lambda_e} \ ,
\end{equation}
where $\alpha_e$ is the self-intersection number of the rational curve
represented by the edge $e$ and the parameter $Q_e$ weighs the
homology class of the edge. Then the partition function has the
symbolic form
\beq
\cZ_{\rm BPS}^M(q,Q) = \sum_{\lambda_e\in\IZ_{\geq0}} ~ \prod_{{\rm
    edges}\ e}\,
G_{\lambda_e}(q,Q_e)~\prod_{\stackrel{\scriptstyle{\rm
      vertices}}{\scriptstyle v=(e_1,e_2)}}\,
V_{\lambda_{e_1},\lambda_{e_2}}(q) \ ,
\label{ZBPSvertsurf}\eeq
where we set $\lambda_e=0$ on all external legs $e$ of the graph $\Delta$.
Since there is a factor $\hat\eta(q)^{-1}$ for each vertex, the
partition function (\ref{ZBPSvertsurf}) carries an overall factor
$\hat\eta(q)^{-\chi(M)}$ associated to the degree zero contributions
as in G\"ottsche's formula (\ref{Gottsche}), where $\chi(M)$ is the
topological Euler characteristic of $M$. It would be interesting to
find a four-dimensional version of topological string theory which
reproduces this counting, analogously to \S\ref{sec:curveCY3}.

\subsection{Hirzebruch-Jung surfaces\label{subsec:HJsurfaces}}

Our main examples of toric surfaces in this paper will fall into the
general class of Hirzebruch-Jung spaces $M=M_{p,p'}$, which are resolutions of
$A_{p,p'}$ singularities for a pair of coprime integers $p>p'>0$; they
represent the most general toric singularity in four
dimensions. In this case we may regard $M$ as a four-cycle in the
Calabi-Yau threefold $X=K_M$. Consider the quotient of $\complex^2$ by the action of the
cyclic group $G_{p,p'}\cong\IZ_p$ generated by
\begin{equation}
 G_{p,p'} = \mbox{$\frac1p$}\, \big(1\,,\,p'\, \big):= \left(
\begin{matrix}
\zeta & 0 \\
0 & \zeta^{p'} \\
\end{matrix}
\right)
\end{equation}
where $\zeta = \e^{2 \pi \ii /p}$. In the framework of toric geometry
this singular space is described by the two-cone spanned by two lattice
vectors ${\sf v}_0 = (1,0)$ and ${\sf v}_{m+1} = (-p',p)$ in $\IZ^2$. The minimal
resolution $\pi: M_{p,p'} \rightarrow
\complex^2 /  G_{p,p'}$ is the smooth connected moduli space
$\mathrm{Hilb}_{G_{p,p'}} (\complex^2)$ consisting of
\emph{$G_{p,p'}$-clusters}, i.e. $ G_{p,p'}$-invariant
zero-dimensional subschemes $Z$ of $\complex^2$ of length $| G_{p,p'}|$
such that $H^0(\cO_Z)$ is the regular representation of $ G_{p,p'}$. The toric diagram
of $M_{p,p'}=\mathrm{Hilb}_{G_{p,p'}} (\complex^2)$ is obtained by the subdivision with one-cones ${\sf v}_1, \ldots, {\sf v}_{m}$ such that
\begin{equation}
{\sf v}_{i-1} + {\sf v}_{i+1} = \alpha_i\, {\sf v}_i  \label{latrel}
\end{equation}
for $i=1, \ldots, m$, where the integers $m$ and $\alpha_i\geq2$ are determined by the continued fraction expansion
\begin{equation}
\frac{p}{p'} = [\alpha_1,\dots,\alpha_m]:= \alpha_1 - \frac{1}{\alpha_2 - \displaystyle{\frac{1}{\alpha_3-\displaystyle{\frac{1}{\ddots \,
    \displaystyle{\alpha_{m-1}-\frac1{\alpha_{m}}}}}}}} \ .
\label{contfrac}\end{equation}
Geometrically the singularity is resolved by a chain of $m$
exceptional divisors $D_i\cong \PP^1$ whose intersection matrix is  
\begin{equation}   \label{int_mat}
C= \big(\langle D_i, D_j\rangle_{\Gamma}\big) =
\begin{pmatrix} - \alpha_1 & 1 & 0 & \cdots &0\\
1 & -\alpha_2  & 1& \cdots &0\\
0 & 1 & - \alpha_3 &\cdots&0\\
\vdots &\vdots & \vdots &\ddots&\vdots\\0&0&0&\cdots&-\alpha_m
\end{pmatrix} \ .
\end{equation}
The exceptional curves $D_i$, $i=1,\dots,m$ generate the Mori cone in the rational
Chow group
$H_2^c(M_{p,p'},\IQ)$ consisting of linear combinations of
compact algebraic cycles with non-negative coefficients.

The vertex rules determine the BPS partition function
(\ref{ZBPSvertsurf}) as
\beq
\cZ_{\rm BPS}^{M_{p,p'}}(q,Q)=\frac1{\hat\eta(q)^{m+1}} \
  \sum_{\lambda_i \in \IZ_{\geq0}}\, q^{-\frac12\,
    \lambda\cdot C\lambda+\frac12\,\lambda\cdot C\delta + \frac12\,\lambda_1
    + \frac12\,\lambda_m} \ Q^\lambda
\eeq
where $\delta :=(1,\dots,1)$.
In particular, the spaces $M_{p+1,p}$ are the Calabi-Yau ALE
resolutions of $A_p$ singularities consisting of a chain of $m=p$
exceptional divisors $D_i$, for which $\alpha_i= 2$ for $i=1
, \ldots, p$, $K_{M_{p+1,p}}=0$, and (\ref{int_mat}) coincides with minus the Cartan matrix of the $A_{p}$
Dynkin diagram which represents the toric graph $\Delta$; the partition function in this case simplifies to
\beq
\cZ_{\rm BPS}^{A_p}(q,Q)=\frac1{\hat\eta(q)^{p+1}} \
  \sum_{\lambda_i \in \IZ_{\geq0}}\, q^{-\frac12\,
    \lambda\cdot C\lambda} \ Q^\lambda \ .
\label{ZBPSAn}\eeq
On the other hand, for $p'=1$ there is only one exceptional divisor with
self-intersection number $-\alpha_1=-p$ and the manifold $M_{p,1}$ can be
regarded as the total space of a holomorphic line bundle ${\cal
  O}_{\IP^1}(-p)$ of degree $p$ over the projective line $\IP^1$; the geometry is
the natural analog of the resolved conifold geometry from
\S\ref{subsec:conifold1}. This
space has a toric projectivization to the $p$-th Hirzebruch surface $\IF_p$
and the partition function is given by
\beq
\cZ_{\rm BPS}^{\IF_p}(q,Q)=\frac1{\hat\eta(q)^2}\,
\sum_{\lambda=0}^\infty\, q^{\frac p2\, \lambda^2} \ Q^\lambda \ .
\label{ZBPSFp}\eeq

\bigskip

\section{Four-dimensional cohomological gauge theory\label{sec:4dcohgt}}

\subsection{Vafa-Witten theory on toric surfaces\label{subsec:VafaWitten}}

A natural candidate gauge theory dual to the curve counting theory of
BPS states discussed in \S\ref{sec:curvecountsurf} is the topological
gauge theory studied by Vafa-Witten~\cite{Vafa:1994tf}, defined on
four-dimensional toric manifolds. This is the low-energy effective
field theory on D4-branes wrapping a holomorphic divisor $M$ in a
Calabi-Yau threefold~\cite{Bershadsky:1995qy,vafa}; the counting of instantons on
D4-branes is then equivalent to the enumeration of BPS D2--D0 staes in
the D4-branes, and hence we would expect the instanton partition function to be
equivalent to the BPS partition function for D4--D2--D0 bound states.
Under favourable circumstances, this gauge
theory captures the holomorphic sector
of $\cN=4$ supersymmetric Yang-Mills theory in four dimensions and can be used to understand its modular
properties under S-duality transformations (when viewed as a function
of the elliptic modulus $q$). This is particularly
interesting on non-compact toric manifolds, such as the $A_p$ series
of ALE spaces considered by Kronheimer-Nakajima~\cite{kronheimer}. On these spaces, by using celebrated results of Nakajima~\cite{nakaALE2}, Vafa-Witten relate the partition function of
the topological gauge theory with the characters of an affine Lie
algebra, thereby completely characterizing it as a quasi-modular form. In
subsequent sections we will extend these considerations to a
relation between Donaldson-Thomas theory and topological Yang-Mills theory in six dimensions. 

Here we consider the topologically
twisted version of $\cN=4$ supersymmetric Yang-Mills theory with gauge
group $G$ as
constructed in \cite{Vafa:1994tf} on a smooth connected K\"ahler
surface $(M,t)$; in the four-dimensional cases we will always set $B=0$.
When certain conditions are met (the Vafa-Witten vanishing theorems)
the corresponding partition function computes the Euler characteristic of the
instanton moduli space. The space of fields $\scrW$ is given by
\begin{equation} \label{N4fieldspace}
\scrW = \scrA(\cP) \times \Omega^0 \left( M , \mathrm{ad}\, \cP \right) \times
\Omega^{2,+} \left( M , \mathrm{ad}\,\cP \right)
\end{equation}
where $\scrA(\cP)$ denotes the affine space of connections on a
principal $G$-bundle $\cP \rightarrow M$,
$\mathrm{ad}\,\cP$ is the adjoint bundle of $\cP$, and the superscript
$+$ denotes the self-dual part; the superpartners of the fields in
(\ref{N4fieldspace}) are sections of the
tangent bundle over $\scrW$. For $(A,\phi,b^+)\in\scrW$, the twisted gauge theory corresponds to the moduli problem associated
with the field equations
\begin{eqnarray} 
\sigma&:=&F_A^+ + \mbox{$\frac{1}{4}$}\, t^{-1}\big(b^+\wedge b^+ \big) +
\mbox{$\frac{1}{2}$} \,
\phi \wedge b^+ \ = \ 0 \ , \nonumber \\[4pt]
\kappa&:=& t^{-1}\big(\nabla_Ab^+\big) + \nabla_A\phi \ = \ 0 \ ,
\label{VWeqs}\end{eqnarray}
where $F_A$ is the curvature of the gauge connection one-form $A$ and
$\nabla_A$ the associated covariant derivative, and the adjoint section
$\phi$ is called a Higgs field. Associated with these
field equations are a pair of doublets which are sections of the bundle $\scrF =  \Omega^{2,+} \left( M ,
\mathrm{ad}\,\cP \right) \oplus  \Omega^1 \left( M , \mathrm{ad}\,\cP
\right)$.

The path integral of the topological gauge theory localizes onto the
solutions of the field equations
(\ref{VWeqs}). The partition function can be interpreted geometrically
as a
Mathai-Quillen representative of the Thom class of the bundle
$\mathscr{V} = \mathscr{W} \times_{\cal{G}(\cP)} \mathscr{F}$, and its
pullback via the sections $(\sigma,\kappa)$ of (\ref{VWeqs}) gives the Euler class of
$\mathscr{V}$; here $\mathcal{G}(\cP)$ is the group of gauge
transformations, which are automorphisms of the $G$-bundle $\cP\to M$. Under
favourable conditions, appropriate vanishing theorems hold which
ensure that each solution of the system (\ref{VWeqs}) has
$\phi=b^+=0$ and corresponds to an \emph{instanton}, i.e. an
anti-self-dual connection; the space of gauge
equivalence classes of solutions to
the anti-self-duality equations $F_A^+=0$ is called the instanton moduli space $\mathscr{M}^{\rm inst}_\cP(M)$. Then the path integral localizes onto $\mathscr{M}^{\rm inst}_\cP(M)$ and reproduces a representative of the
Euler class of the tangent bundle $T \mathscr{M}^{\rm inst}_\cP(M)$. Therefore the
partition function computes volumes of the form
\beq \label{chiMPinst}
\chi \big(
\mathscr{M}^{\rm inst}_\cP(M) \big) = \int_{\mathscr{M}^{\rm inst}_\cP(M)}\, \eul \big( T
\mathscr{M}^{\rm inst}_\cP(M) \big)
\eeq
which give the Euler character of the
moduli space of $G$-instantons on $M$; it can be computed from the
index of the Atiyah-Hitchin-Singer instanton deformation complex~\cite{Labastida:1997vq}
\beq\label{AHScomplex}
\xymatrix{
  \Omega^0(M, \mathrm{ad}\,\cP)
   \quad \ar[r]^{ \hspace{-0.3cm} \nabla_A} & \quad
  \Omega^1(M, \mathrm{ad}\,\cP) \quad
  \ar[r]^{\nabla^+_A} & \quad
   \Omega^{2,+}(M, \mathrm{ad}\,\cP)
}
\eeq
associated with the equations (\ref{VWeqs}), where the first morphism is an infinitesimal gauge transformation while the second morphism corresponds to the linearization of the sections
$(\sigma, \kappa)$.
For unitary gauge bundles $\cE\to M$ of rank
$r$, the moduli space can be stratified into connected components $\scrM^{\rm inst}_{n,u;r}(M)$ with fixed instanton and monopole charges
\beq
n=\ch_2(\cE)\ \in \ H^4(M,\IZ)=\IZ \qquad \mbox{and} \qquad u=
c_1(\cE) \ \in \ 
H^2(M,\IZ) \ .
\eeq

Let us now look at the rank one case and embed the moduli space of $U(1)$
bundles with anti-self-dual connection into the space of (semi-stable)
torsion free sheaves $\cT$; this embedding provides a smooth Gieseker
compactification of the instanton moduli space which can be naturally
thought of as algebraic geometry data
parametrizing the extension of the moduli space to include noncommutative instantons. We can organize the
moduli space of isomorphism classes of these sheaves according to
their Chern classes as $\scrM^{\rm inst}_{n,\beta} (M)$ with
\begin{eqnarray}
n =\ch_2 (\cT) \qquad \mbox{and} \qquad 
\beta = c_1 (\cT) \ \in \ H_2 (M , \zed) \ ,
\end{eqnarray}
where we use Poincar\'e-Chow duality for surfaces to regard divisors of $M$
both as curves and as degree two cohomology
classes.
Note that for non-compact spaces the second Chern characteristic class
$n$ can be fractional. The factorization (\ref{factorization})
implies an analogous factorization of the moduli space strata
\beq
\scrM^{\rm inst}_{n,\beta}(M) \cong \mathrm{Pic}_\beta(M) \times \Hilb_{n-n_\beta}(M) \ ,
\label{instmodfact}\eeq
similarly to (\ref{Hilbfact}). The Picard lattice
$\mathrm{Pic}_\beta(M)$ parametrizes isomorphism classes of holomorphic line bundles $\calL=\cO_M(D)$ of
divisors $D$ on $M$ with $[D]=\beta$ which contribute the intersection number
(\ref{intnumbeta}) to the second Chern characteristic of a
torsion free sheaf and which are called ``fractional'' instantons,
while the Hilbert scheme of points $\Hilb_{m}(M)$ is the smooth
manifold of dimension $2m$ which
parametrizes ideal sheaves $\cI_Z$ with instanton number $m$ (and
$c_1(\cI_Z)=0$) corresponding to ``regular''
instantons supported on a zero-dimensional subscheme $Z\subset
M$ of length $m=n-n_\beta$. Using the Grothendieck-Riemann-Roch formula, the contribution of a
primitive element of (\ref{instmodfact}) to the instanton number $n$ is
found to be~\cite{Cirafici:2009ga}
\beq
\ch_2\big(\cO_M(D)\otimes\cI_Z\big) = \mbox{$\frac12$} \, \langle D,D\rangle_{\Gamma} -\chi(\cO_Z)
\eeq
with $\langle D,D\rangle_{\Gamma}=\int_M\, c_1(\cO_M(D))\wedge c_1(\cO_M(D))$.

The gauge theory partition function is given by 
\begin{equation}
Z_{\rm gauge}^M(q,Q) =
    \sum_{n,\beta} \, q^{n} \ Q^\beta \ \int_{\scrM^{\rm inst}_{n,\beta}(M)}\, \eul\big(T
    \scrM^{\rm inst}_{n,\beta}(M)\big) \ .
\label{Zgt}\end{equation}
By G\"ottsche's formula (\ref{Gottsche}), the regular instantons
contribute to (\ref{Zgt}) with a factor $\hat\eta(q)^{-\chi(M)}$ which
is identical to that of the BPS partition function
(\ref{ZBPSvertsurf}). On the other
hand, the discrete factor which comes from the sum over fractional
instantons $\calL\in{\rm Pic}_\beta(M)$ is different from that of (\ref{ZBPSvertsurf}); this series
is a
generalized theta-function which carries the relevant information
about the partition function as a quasi-modular form. Hence the curve counting
and instanton counting problems are related but not identical in four
dimensions; we shall see later on that the duality is however exact in
six dimensions.

For illustration, let us consider the class of Hirzebruch-Jung spaces
from \S\ref{subsec:HJsurfaces}. In this case, using linear equivalence
one can extend the complete set of torically invariant divisors,
including the non-compact ones, to an integral generating set for the
Picard group ${\rm Pic}(M_{p,p'})\cong H^2(M_{p,p'},\IZ)=\IZ^m$ of
line bundles given by~\cite{Cirafici:2009ga}
\beq
e^i=\sum_{j=1}^m \, \big(C^{-1}\big)^{ij}\,D_j \qquad \mbox{for} \quad
i=1,\dots, m \ ,
\label{Kahlerconegen}\eeq
where the matrix $C_{ij}= \langle D_i, D_j\rangle_{\Gamma}$ is the integer-valued
intersection form (\ref{int_mat}) for the compact two-cycles of the
resolution $M_{p,p'}$ whose inverse can be found in
e.g.~\cite[App.~A]{Griguolo}. These divisors satisfy
\begin{equation}
\big\langle D_i \,,\, e^j \,\big\rangle_{\Gamma} = \delta_{i}{}^{j} \ ,
\end{equation}
and hence generate the K\"ahler cone which is the
polyhedron in the rational Chow group
$H^2(M_{p,p'}, \IQ)$ dual to the Mori cone with respect to the
intersection pairing
\beq
\langle-,-\rangle_{\Gamma}\,:\, H_2^c(M_{p,p'},\IZ)\times H^2(M_{p,p'},\IZ) \ \longrightarrow \ H_{0}^c
(M_{p,p'} , \mathbb{Z}) = \mathbb{Z}
\label{intersection}\eeq
that includes the linear extension to non-compact divisors. 
They have intersection products $\langle e^i, e^j\, \rangle_{\Gamma} =(C^{-1})^{ij}$. Since $C$ is not unimodular, its inverse is
generally rational-valued and this change of basis naturally
incorporates the contributions from flat connections with non-trivial
holonomy at infinity. It corresponds to the basis of tautological line bundles
used by Kronheimer-Nakajima in~\cite{kronheimer}, and by
e.g.~\cite{Fucito:2006kn,Griguolo}, which we consider in \S\ref{subsec:ALEmodsp}. In this case the
partition function (\ref{Zgt}) evaluates to~\cite{Cirafici:2009ga}
\beq
Z_{\rm gauge}^{M_{p,p'}} (q,Q) = \frac{\Theta_{\Gamma}(q,Q)}{\hat\eta(q)^{m+1}} \ , 
\label{ZgtHJ}\eeq
where the Riemann theta-function on the magnetic charge
lattice $\Gamma= H^2(M_{p,p'},\IZ)\cong \IZ^m$ is given by
\beq
\Theta_{\Gamma}(q,Q) =
\sum_{u\in\Gamma} \,
q^{-\frac12\, u\cdot C^{-1} u} \ Q^u \ .
\eeq
This expression should be compared with e.g.~(\ref{ZBPSAn}) in the ALE
case, where the inverse of the Cartan matrix is given by
$(C^{-1})^{ij}=\frac{i\,j}{p+1}-{\rm min}(i,j)$ for $i,j=1,\dots,p$, or (\ref{ZBPSFp}) in the case of Hirzebruch surfaces where
\beq
Z_{\rm gauge}^{\IF_p} (q,Q) =
\frac{\theta_3(q^{1/p},Q)}{\hat\eta(q)^{2}} \ ,
\label{ZU1Fp}\eeq
with the Jacobi elliptic function
\beq
\theta_3(q,Q)=
\sum_{u\in\IZ} \, q^{\frac12\,u^2}\,Q^{u} \ .
\eeq
The formula (\ref{ZgtHJ}) agrees with the conjectural exact expression
of~\cite{Fucito:2006kn,Griguolo}; there is also a conjectured
factorization of the higher rank instanton partition function for
$r>1$ given by
\bea
Z^{M_{p,p'}}_{\rm gauge}(q,Q;r) = \Big(Z_{\rm gauge}^{M_{p,p'}}
(q,Q)\Big)^r = \frac1{\hat\eta(q)^{r\,(m+1)}}~\sum_{\vec u \in
  \Gamma^{r}} \,
q^{-\frac12\,\sum_{l=1}^r\, u_l\cdot C^{-1} u_l}~Q^{u} \ ,
\label{ZHJrfact}\eea
where $\vec u=(u_1,\dots,u_r)$ with $u_l\in\Gamma$, $l=1,\dots, r$, while
\beq
u=\sum_{l=1}^r\, u_l \qquad \mbox{and} \qquad
Q^{u} = \prod_{i=1}^{m} \,Q_i^{u_i} \ .
\eeq
This
factorized form can be derived from a localization calculation on the
ADHM parametrization of the instanton moduli space, which we
discuss in \S\ref{sec:InstN2gt}. The formula (\ref{ZHJrfact}) is
rigorously derived in~\cite{Fujii:2005dk} for ALE spaces using the
combinatorial formalism of quiver varieties, and
in~\cite{Bruzzo:2009uc} for Hirzebruch surfaces by a localization
calculation on the Coulomb branch of the gauge theory. Since $M_{p,p'}$ is
non-compact, one should also include properly the
boundary contributions that are known~\cite{Griguolo} in terms of
the induced Chern-Simons gauge theory on the three-dimensional
boundary of $M_{p,p'}$, which is a Lens space $L(p,p'\,)=S^3/G_{p,p'}$; such boundary
contributions will be incorporated in \S\ref{subsec:affinechar} for the case of the ALE spaces
$M_{p+1,p}$. 

\subsection{McKay correspondence\label{subsec:McKay}}

In the remainder of this section we will focus on the case of toric
Calabi-Yau twofolds $M$, i.e. ALE spaces, which have trivial canonical
class $K_M=0$ and are the natural
four-dimensional analogs of the spaces considered in
\S\ref{sec:curveCY3}; we will describe the instanton moduli spaces on these varieties. The
McKay correspondence allows a direct construction which is a straightforward
generalization of the construction of the instanton moduli space on
$\complex^2$: These moduli spaces are given by \textit{quiver
  varieties} based on the McKay quiver. In \S\ref{sec:Stacky} we will
see that this formalism can be extended to Calabi--Yau
threefolds; in particular, the generalized McKay correspondence will
enable us to
study the noncommutative enumerative invariants of
\S\ref{subsec:NCDT} as a generalized instanton counting
problem. 

We will now review the McKay correspondence for finite subgroups $G$
of $SL(2, \complex)$. This correspondence relates the quiver gauge
theory associated to a four-dimensional orbifold singularity with the
counting of D4--D2--D0 bound states on the minimal resolution of the
singularity.
We will only focus on cyclic groups $G=\IZ_{p+1}$,
but the line of reasoning can be extended to other groups of ADE
type. The $G$-action on $\complex[z_1,z_2]$ is given by
\begin{equation}
(z_1,z_2) \ \longrightarrow \ \big(\zeta\, z_1 \,,\,
\zeta^{p}\, z_2 \big)
\end{equation}
with $\zeta^{p+1}=1$. The quotient variety $\complex^2 / G$ has a
Kleinian or du~Val singularity of type $A_{p}$. It can be regarded
as an embedded hypersurface in $\complex^3$ via the defining equation
\begin{equation}
z_1^2 + z_2^2 + z_3^{p+1} = 0 \ .
\end{equation}
In the minimal crepant resolution of the singularity $\pi : M
\rightarrow \complex^2 / G$, the exceptional set and its intersection
indices can be encoded in a graph which has the form of a Dynkin
diagram of type~$A_{p}$.

Let $Q \cong \complex^2$ be the fundamental representation induced by
the inclusion $G \subset SL(2,\complex)$. If we denote the irreducible
representations of $G$ by $\rho_a$ with $a\in {\sf Q}_0 = \{
0,1,\dots,p\}$, where $\rho_0$ is the trivial representation and $\rho_a$ has character $\zeta^a$, then $Q = \rho_1 \oplus \rho_{p}$. Then the decomposition
\begin{equation}
Q \otimes \rho_a = \bigoplus_{b\in{\sf Q}_0}\,  {\tt a}_{ba}\, \rho_b
= \rho_{a+1} \oplus \rho_{a-1}  \qquad \text{with} \quad {\tt a}_{ab} = \dim \Hom_{G} \left( \rho_a ,  Q \otimes \rho_b \right)
\label{QotimesrhoC2}\end{equation}
can be used to define the McKay quiver $\sfQ_p$ as the graph with vertex set
$\sfQ_0$ and ${\tt a}_{ab}$ arrows from vertex $a$ to vertex $b$; only
${\tt a}_{a,a\pm1}=1$ are non-zero. This graph
has the form of an affine Dynkin diagram of type A and its subgraph
obtained by removing the affine vertex, which corresponds to the
trivial representation $\rho_0$, is a standard Dynkin diagram of type
A. We
will denote by $C$ minus the Cartan matrix corresponding to the Lie algebra
and by $\tilde{C}$ minus the Cartan matrix of its affine extension; both
matrices are non-negative and symmetric. The category of representations of the McKay quiver $\sfQ_p$ plays a crucial role in the McKay correspondence. 

The McKay
correspondence, in its simplest version, gives a bijection between the
set of irreducible representations (including the trivial one) and a
basis of $H_{\bullet} (M , \zed)$, which maps the exceptional curves
$D_a$ to the non-trivial representations $\rho_a$ and the homology
class of a point to the trivial representation $\rho_0$. To each
irreducible representation we associate a reflexive module $R_a:=\Hom_{G}
(\rho_a, \complex^2)$, and $\calR_a$ as the pullback sheaf on $M$
modulo torsion. The \textit{tautological sheaf} $\calR_a$ is locally
free and has rank one (the dimension of $\rho_a$). The collection of tautological sheaves has the property
\begin{equation}
\int_{D_b} \, c_1 (\calR_a) = \delta_{ab} 
\end{equation}
and the Chern classes $c_1 (\calR_a)$ form a basis of $H^2 (M, \zed)$;
this identifies $e^a=c_1(\calR_a)$ in (\ref{Kahlerconegen}), or
equivalently $\calR_a=\cO_M(e^a)$. Upon
including the trivial bundle $\mathcal{R}_0 = \mathcal{O}_{M}$,
which generates $H^0 (M , \mathbb{Z} )$, these bundles form the
canonical integral basis of the K-theory group $K(M)$ constructed
in~\cite{gonzalez}. Out of these locally free sheaves we construct the
tautological bundle~\cite{kronheimer}
\begin{equation}
\calR = \bigoplus_{a \in \widehat{G}}\, \calR_a \otimes \rho_a
\label{tautC2decomp}\end{equation}
where the sum runs over all irreducible representations of $G$.

For the McKay quiver the vertex set ${\sf Q}_0= \widehat{G}$ is
determined by the irreducible representations of the cyclic group $G$,
while the set of arrows is ${\sf Q}_1= \big\{ \, a_1^{(\rho)} , a_2^{(\rho)} \ |
\ \rho \in \widehat{G} \, \big\}$; the notation means that the arrow
$a_{1}^{(\rho)}$ (resp. $a_{2}^{(\rho)}$) goes from vertex $\rho_1
\otimes \rho$ (resp. $\rho_{p} \otimes \rho$) to vertex
$\rho$. With this notation understood the ideal of relations is
generated by
\begin{equation}
\sfR = \Big\{ \, a_2^{(\rho \otimes \rho_1)} \, a_1^{(\rho)} - a_1^{(\rho
  \otimes \rho_{p})} \, a_2^{(\rho)} \ \Big| \ \rho \in \widehat{G} \,
\Big\} \ .
\end{equation}
The fine moduli space $\scrM_\theta^{\delta} (\sfQ_p , \sfR)$  of
$\theta$-stable representations of the bound McKay quiver
$(\sfQ_p,\sfR)$, for the regular representation $R$ of $G$ where all vector spaces $V_a$ are one-dimensional, is 
smooth and the map $\pi : {\scrM^{\delta}_{\theta} (\sfQ_p , \sfR)}
\rightarrow {\scrM^{\delta}_{0} (\sfQ_p , \sfR)} \cong \complex^2 / G$ is
the minimal resolution of the orbifold singularity~$\complex^2 / G$.

We wish to now explain the relation between the intersection theory of
the exceptional set and the representation theory of the orbifold
group $ G$. We will show, following \cite{ItoNaka}, how the tensor
product decomposition of irreducible representations of $G$ appears in
the intersection matrix. Given the basis $\calR_a$ of tautological
bundles for $K(M)$, we introduce a dual basis ${\cS_a}$ for the
Grothendieck group $K^c(M)$ of
complexes of vector bundles which are exact outside the exceptional
locus $\pi^{-1}(0)$. Consider the complex on $M$ given by
\begin{equation}
\xymatrix@1{
 \calR \ \ar[r] & \ Q \otimes \calR \ \ar[r] & \ \bigwedge^2 Q \otimes
 \calR
} \ ,
\label{tautcomplex}\end{equation}
where the arrows are induced by multiplication with the coordinates
$(z_1,z_2)$ and here $Q$ is short-hand for the trivial bundle with
fibre $Q$. Since $ G
\subset SL(2 , \complex) $, the determinant
representation is trivial as a $ G$-module and we have $\bigwedge^2
Q \otimes \calR \cong \calR$. Then the transpose complex
\begin{equation}
\cS_a \ : \ \xymatrix@1{
 \calR^{\vee}_a \ \ar[r] & \
 \displaystyle{\bigoplus_{b\in\widehat{G}}\, {\tt a}_{ab} \,
 \calR_b^{\vee}} \ \ar[r] & \ 
 \calR_a^{\vee}
}
\label{McKaycSbasis}\end{equation}
gives the desired basis of $K^c (M)$. One
similarly defines the dual complexes 
\begin{equation}
\cS_a^{\vee} \ : \  - \Big[ \xymatrix@1{
 \calR_a \ \ar[r] & \ \displaystyle{\bigoplus_{b\in\widehat{G}}\, {\tt
     a}_{ab} \,
 \calR_b} \ \ar[r] & \ 
 \calR_a
 } \Big] \ .
\end{equation}
On $K^c (M)$ we can define a pairing
\begin{equation}
(\cS , \cT  )_{K^c} = \big\langle \Xi(\cS) \,,\, \cT \big\rangle_{K}
\end{equation}
where $\Xi :  K^c (M) \rightarrow K(M)$ is the map which takes a
complex of vector bundles to the corresponding element in $K(M)$,
i.e. the alternating sum of terms of the complex, and
$\langle-,-\rangle_{K}$ denotes the dual pairing.
It follows that
\begin{equation} \label{SvScartan}
\big(\cS_a^{\vee} \, , \, \cS_b\big)_{K^c} = \big\langle \Xi(\cS_a^{\vee}\,) \,,\,
\cS_b \big\rangle_{K} = \sum_{c\in\widehat{G}}\, ( \delta_{ac} - {\tt a}_{ac} +
\delta_{ac} )\, \langle \calR_c , \cS_b \rangle_{K} = 2 \delta_{ab} -
{\tt a}_{ab} =-\tilde C_{ab} \ .
\end{equation}
This result relates the tensor product decomposition (\ref{QotimesrhoC2}), and
therefore the extended Cartan matrix $-\tilde C$, with the intersection
pairing on $K^c (M)$; in particular, it naturally identifies the
resolved homology $H_2(M,\IZ)$ with the root lattice $\Gamma$ of the
$A_p$ Lie algebra and the
Grothendieck group $K^c(M)$ as the lattice of fractional instanton charges.

\subsection{Instanton moduli spaces\label{subsec:ALEmodsp}}

We review the construction of the framed moduli space of rank $r$ torsion free
sheaves on the orbifold compactification of the ALE
space $M$ given by $\overline{M} = M \cup \ell_{\infty}$ where
$\ell_{\infty} = \PP^1 /  G$~\cite{kronheimer,nakaALE1,nakaALE2,nakaALE3}. In a neighbourhood
of infinity we can regard $\overline{M}$ as the compact toric
orbifold $\PP^2 /
 G$ with the resolution of the singularity at the origin. More
precisely, the divisor $\ell_{\infty}$ is constructed by gluing together
the trivial bundle $\cO_M$ on $M$ with the line bundle $\cO_{[\PP^2/G]}(1)$ on $\PP^2 /
 G$. The latter bundle has a $ G$-equivariant structure such that
the restriction map of structure sheaves $\cO_M \rightarrow \cO_{[\PP^2/G]}(1)$ is $ G$-equivariant.

We begin by recalling the ADHM parametrization of the instanton moduli
space $\scrM^{\rm inst}_{n,r}(\IC^2)$ on affine space $\IC^2$; this parametrizes
stable D4--D0 bound states in the low-energy limit at large radius,
with BPS D0-branes regarded as instantons in the gauge theory on the
D4-branes~\cite{Douglas:1995bn,Douglas:1996sw}. It has a simple
description as a quotient by the action of the gauge group $GL(n,\IC)$ of the space spanned by the linear
operators
\be 
B_{1},B_{2} \in \End_\IC(V) \ , \qquad I \in \Hom_\IC(W,V) \
\qquad \mbox{and} \qquad J \in \Hom_\IC(V,W)
\ee
subject to the ADHM equation
\bea 
\label{adhm} [B_1 , B_2] + I \, J = 0
\eea
and constrained by a certain stability condition.
The vector spaces $V$ and $W$ have dimensions $n$ and $r$
respectively, and are given by cohomology groups characterizing the
instanton moduli space. For this, one has to parametrize holomorphic
bundles on $\complex^2$, or in a better compactified setting one works
with torsion free sheaves $\cE$ on the projective plane $\PP^2$. Then
using the Beilinson spectral sequence one can describe a generic
torsion free sheaf as the cohomology of a certain complex: Any torsion free sheaf $\cE$ on $\PP^2$ can be written as $\cE = p_{1 *} (p_2^* \cE \otimes \mathcal{O}_{\Delta})$ where
${\Delta}$ is the diagonal of $\PP^2 \times \PP^2$, and $p_1$ and $p_2$ are projections onto the first and second factors. The Koszul resolution of $\cO_{\Delta}$ gives the double complex
\begin{equation}
\mathbf{R}^{\bullet} p_{1*} (p_2^* \cE \otimes C^{\bullet}) \ ,
\label{KoszulP2}\end{equation}
where
\begin{equation}
C^i =  \mbox{$\bigwedge$}^{-i} \left( \cO_{\PP^2} (-\ell_\infty) \boxtimes \mathcal{Q}_{\PP^2}^{\vee} \right)
\label{CiP2}\end{equation} 
and the locally free sheaf $\cQ_{\PP^2}^{\vee}$ is defined by the exact sequence
\begin{equation}
\xymatrix@1{
  0 \ \ar[r] & \ \cO_{\PP^2} (-\ell_\infty) \ \ar[r] & \ \cO^{\oplus 3}_{\PP^2}
  \ \ar[r] & \ \cQ_{\PP^2} \ \ar[r] & \ 0 \ ,
}
\label{cQP2def}\end{equation}
with $\ell_\infty=\PP^2\setminus\IC^2\cong\PP^1$ a line at infinity. By Beilinson's theorem, the sheaf $\cE$ can be recovered from a spectral sequence whose first term is
\begin{equation}
 E_1^{p,q} =  \cO_{\PP^2}( p\,\ell_\infty ) \otimes H^q \big(\PP^2 \,,\, \cE (-\ell_{\infty}) \otimes \mbox{$\bigwedge$}^{-p} \mathcal{Q}_{\PP^2}^{\vee}\big)  
\end{equation}
This spectral sequence degenerates into a monad, after imposing the
framing condition that $\cE$ is trivial on $\ell_{\infty}$ and locally
free in a neighbourhood of $\ell_\infty$. After some homological
algebra one finds that all the relevant information is encoded into
two vector spaces $V$ and $W$ defined by $V = H^1 (\PP^2 , \cE (-2\ell_\infty))$
and $\cE\big|_{\ell_\infty}=W\otimes\cO_{\ell_\infty}$,
together with two linear maps 
\begin{equation} \label{compcomplex}
\xymatrix@1{
   V \otimes \mathcal{O}_{\PP^2}(-
  \ell_{\infty})
   \quad\ar[r] &\quad
   ( V \oplus V \oplus W) \otimes \cO_{\PP^2}
   \quad \ar[r] & \quad
  V \otimes \cO_{\PP^2} (\ell_{\infty}) \ .
}
\end{equation}
The condition that the two linear maps give a complex is precisely the ADHM equation (\ref{adhm}).

Consider now a generic torsion free sheaf $\cE$ on $\overline{M}$;
then $\cE = p_{1 *} (p_2^* \cE \otimes \mathcal{O}_{\Delta})$ where now
${\Delta}$ is the diagonal of $\overline{M} \times
\overline{M}$. The diagonal sheaf $\mathcal{O}_{\Delta}$ has a resolution which arises
from gluing the resolution of the diagonal of $M \times M$
constructed in \cite{kronheimer} from the complex of tautological
bundles (\ref{tautcomplex}), endowed with an appropriate
$G$-equivariant structure, with the resolution of the diagonal sheaf of
$\PP^2 \times \PP^2$. One arrives at a double complex of the same form
as (\ref{KoszulP2}) but with (\ref{CiP2}) replaced by
\begin{equation} \label{resolutionM} 
C^{k} = \Big( \mathcal{R} (k\, \ell_{\infty}) \boxtimes
\big(\mathcal{R}^\vee \otimes \mbox{$\bigwedge^{-k}$}
\mathcal{Q}^{\vee}\, \big)
\Big)^{ G} 
\ , 
\end{equation} 
where the line at infinity $\ell_\infty$ is $G$-invariant and the
locally free sheaf $\mathcal{Q}$ is defined as follows.
In the Kronheimer-Nakajima
construction~\cite{kronheimer}, one defines a trivial bundle with fibre $Q \cong
\complex^2$ on which the regular representation of $ G$ acts.
On the other hand, in the neighbourhood of the
divisor at infinity, $\overline{M}$ looks like $\PP^2$ (with the
$ G$-action). The sheaf $\mathcal{Q}_{\PP^2}$ on $\PP^2$ defined by (\ref{cQP2def}) has a natural $ G$-equivariant structure induced by the
identification $ \cO^{\oplus 3}_{\PP^2} \cong (Q \oplus R_0)
\otimes \cO_{\PP^2}$. We define the sheaf $\cQ$ by gluing together these
two sheaves, much in the same way that we can regard
$\overline{M}$ as obtained by gluing together $M$ and $\PP^2$ with
the appropriate equivariant structure (and $ G$-action): Since
$M$ is an ALE space, its compactification looks like
$\PP^2$ near the compactification divisor $\PP^1$ at infinity, and
the gluing is compatible with the action of the orbifold group $ G$ on
$\PP^1$.

We can now now apply Beilinson's theorem: For any torsion free
coherent sheaf $\cE$ on $\overline{M}$ there is a spectral sequence
with first term
\begin{equation}
 E_1^{p,q} = \Big( \mathcal{R}(p\, \ell_{\infty}) \otimes H^q \big(\,
 \overline{M} \,,\, \cE (-\ell_{\infty}) \otimes \mathcal{R}^\vee
 \otimes \mbox{$\bigwedge^{-p}$} \mathcal{Q}^{\vee}\, \big) \Big)^{ G}
\end{equation}
which converges to $\cE(-\ell_\infty)$. 
Upon imposing the framing condition that $\cE$ is trivial at infinity, the spectral sequence degenerates to a monad
with only non-vanishing middle cohomology given by
\begin{equation} \label{compcomplexG}
\xymatrix@1{
  (\mathcal{R}\otimes V)^G \otimes \mathcal{O}_{\overline{M}}(-
  \ell_{\infty})
   \quad\ar[r] &\quad
   {\begin{matrix} (\mathcal{R}\otimes Q \otimes V)^G
   \otimes \mathcal{O}_{\overline{M}}
   \\ \oplus \\
    (\mathcal{R}\otimes W)^G \otimes \mathcal{O}_{\overline{M}} \end{matrix}} \quad \ar[r] & \quad
   {\begin{matrix} (\mathcal{R}\otimes V )^G \otimes \mathcal{O}_{\overline{M}}(\ell_{\infty})
   \end{matrix}}
}
\end{equation}
where $V =H^1 \big(\, \overline{M} \,,\, \cE (-2 \ell_\infty)
\,\big)$ and $\cE\big|_{\ell_\infty}=(W\otimes\cO_{\ell_\infty})/G$.

This
construction realizes the instanton moduli space $\scrM^{\rm inst}_{n,\beta;r}(M)$
as a quiver
variety $\scrM (V,W)$, when a certain stability condition is
imposed: The middle cohomology of (\ref{compcomplexG}) is
the original sheaf $\cE$ by Beilinson's theorem. 
The condition that the sequence (\ref{compcomplexG}) is a complex is equivalent
to generalized ADHM equations which follow by decomposing (\ref{adhm})
under the action of the orbifold group~\cite{kronheimer}; this
describes the sheaf $\cE$ in terms of representations of the
\emph{framed} McKay
quiver, and provides a bijection between fractional 0-brane charges on
the orbifold and D2--D0 brane charges on the resolution which arises
from collapsing the two-cycles on $M$. For this,  we note that the two vector spaces $V$ and $W$
(regarded
as trivial vector bundles) have a natural grading under the action
of $ G$ into isotopical components
\begin{eqnarray}
V = \bigoplus_{a\in\widehat{G}}\, V_a \otimes \rho_a^{*} \qquad
\mbox{and} \qquad W
= \bigoplus_{a\in\widehat{G}}\, W_a \otimes \rho_a^{*} \ ,
\end{eqnarray}
with $V_a=\Hom_G(V,\rho_a)$ and $W_a=\Hom_G(W,\rho_a)$.
Let us assemble their dimensions $\dim V_a = n_a$ and
$\dim W_a = r_a$ into integer $p+1$-vectors $\mbf
n=(n_0, n_1,\dots,n_{p})$ and $\mbf r=(r_0,r_1,\dots,r_{p})$, where
$p+1$ is the number of vertices in the affine Dynkin graph associated with the ALE singularity (as well
as the number of irreducible representations of $ G$); this data
uniquely characterises the quiver variety $\scrM(\mbf n,\mbf r):=\scrM
(V,W) $. The McKay correspondence identifies these vectors as
expansions $\mbf n= \sum_a\, n_a \,\alpha_a$ and $\mbf r= \sum_a\, r_a
\,\lambda_a$ in the simple roots $\alpha_a$ and fundamental weights $\lambda_a$ of the associated affine ADE Lie algebra.
The equivariant maps $(B_1,B_2)\in\Hom_G(V,V\otimes Q)$, $I\in\Hom_G(W,V)$ and
$J\in\Hom_G(V,W)$ decompose accordingly by Schur's lemma into linear maps $B_1^{(a)}\in\Hom_\IC(V_a,V_{a-1})$,
$B^{(a)}_2\in\Hom_\IC(V_a, V_{a+1})$, $I^{(a)}\in\Hom_\IC(W_a,V_a)$ and
$J^{(a)}\in\Hom_\IC(V_a,W_a)$. Using the analogous decomposition of
the tautological bundle (\ref{tautC2decomp}), the generalized ADHM
equations are then
\begin{equation}
B_1^{(a-1)} \, B_2^{(a)} - B_2^{(a+1)} \, B_1^{(a)} + I^{(a)} \,
J^{(a)} = 0 \ .
\label{adhmgen}\end{equation}
This constructs the quiver variety $\scrM(V,W)$ as a suitable quotient by the action of the gauge group
$GL_G(V)$ on the subvariety of the representation space
\beq
{\sf Rep}(\sfQ) = \Hom_G(V,V\otimes Q) \ \oplus \ \Hom_G(W,V) \ \oplus
\ \Hom_G(V,W)
\eeq
cut out by the equations (\ref{adhmgen}).

In this setting the instanton moduli space can be regarded as the
$G$-invariant subspace of the moduli space $\scrM^{\rm inst}_{n,r}(\IC^2)$ of
framed torsion free sheaves $\cE$ on the projective plane $\PP^2$. The
$G$-action on $\PP^2$ lifts to a natural action of $G$ on $\scrM^{\rm
  inst}_{n,r}(\IC^2)$, once we fix a lift of the $G$-action to the
framing bundle $W\otimes\cO_{\ell_\infty}$. Then the $G$-fixed point
set of the moduli space has a stratification
\beq
\scrM^{\rm inst}_{n,r}(\IC^2)^G = \bigsqcup_{|\mbf n|=n} \, \scrM(\mbf
  n,\mbf r) \ ,
\label{quivervardecomp}\eeq
where $|\mbf n|:=\sum_a\, n_a$ and the connected components
$\scrM(\mbf n,\mbf r)$ are framed moduli spaces of
$G$-equivariant torsion free sheaves $\cE$ on $\PP^2$.

The instanton moduli space $\scrM(\mbf n,\mbf r)$, when non-empty, is smooth of dimension
\begin{equation}
\dim \scrM (\mbf n , \mbf r) = 2 \, \mbf n \cdot \mbf r + \mbf n\cdot
\tilde{C} \mbf n \ ,
\end{equation}
which follows by counting parameters and constraints.
It is also fine and the above construction determines a
universal sheaf $\scrE$ on $ \overline{M} \times \mathscr{M}^{\rm inst}(\mbf
n,\mbf r)$~\cite{nakaALE3}; the K-theory class of its fibre at a point
$[\cE]\in\scrM(\mbf n,\mbf r)$ is the virtual bundle defined by the
complex (\ref{compcomplexG}) which is given by
\begin{eqnarray}
\mathscr{E}\big|_{[\cE]} &=&
 \left( \mathcal{R} \otimes W \right)^{ G}
\otimes \mathcal{O}_{\overline{M}}  \ \oplus \ 
\left( \mathcal{R} \otimes Q \otimes V \right)^{ G} \otimes
\mathcal{O}_{\overline{M}} \nonumber \\ && \ \ominus \ 
\left( \mathcal{R} \otimes V \right)^{ G} \otimes
\mathcal{O}_{\overline{M}}(- \ell_{\infty}) \ \ominus \ 
\left( \mathcal{R} \otimes V \right)^{ G} \otimes \mathcal{O}_{\overline{M}}(
\ell_{\infty}) \ .
\end{eqnarray}
Evaluation of its Chern classes gives
\begin{eqnarray} \label{chernE}
c_1 (\cE) = \sum_{a\in\widehat{G}}\,  \Big( r_a +
\sum_{b\in\widehat{G}}\, \tilde{C}_{ab}\,
n_b \Big) \, c_1
\left( \mathcal{R}_a \right) = \sum_{a=0}^{p} \,
(r_a+n_{a-1}-2n_a+n_{a+1})\, c_1(\calR_a) 
\end{eqnarray}
and
\begin{eqnarray}\label{chernE2}
\mathrm{ch}_2 (\cE)
&=& \sum_{a\in\widehat{G}} \,
\Big( r_a+\sum_{b\in\widehat{G}}\, \tilde{C}_{ab} \, n_b \Big) \, \mathrm{ch}_2 \left(
\mathcal{R}_a \right) - 2\, \delta \cdot \mbf{n} \ \mathrm{ch}_2
\big(\mathcal{O}_{\overline{M}}( \ell_{\infty}) \big) \nonumber \\[4pt] &=& \sum_{a=0}^{p} \,
\Big((r_a+n_{a-1}-2n_a+n_{a+1})\, \ch_2(\calR_a) -\frac{n_a}{|G|} \,
\frac{t\wedge t}2 \Big) \ ,
\end{eqnarray}
where $|G|=p+1$ and we define $n_{-1}=0=n_{p+1}$, while the vector
$\delta\in\ker(\tilde C)$ is the unique vector in the kernel of the
extended Cartan matrix whose first component is~$1$,
i.e. $\delta=(1,\dots,1)$ is the vector of ranks of the tautological
bundles; geometrically it is related to the annihilator of the quadratic form which gives the intersection pairing. In the following we will often denote ${u} :=
\mbf{r} + \tilde{C} \mbf{n}$; by the McKay correspondence, it lies in
the weight lattice of the Kac-Moody algebra of type $A_p$.   

Hence there is a bijective
correspondence between the moduli space of framed torsion free
sheaves on $\overline{M}$ and the quiver variety $\scrM (\mbf n ,
\mbf r)$, where $\mbf r$ is determined by the asymptotic boundary
condition on the gauge field at infinity, while the vector $\mbf
n$ is determined by the Chern classes of $\cE$ (instanton charges) via
(\ref{chernE}); see~\cite{kronheimer,nakaALE3} for further details of this construction.

\subsection{Affine characters\label{subsec:affinechar}}

It is a celebrated result of~\cite{nakaALE2,Vafa:1994tf} that the rank $r$
gauge theory partition function (\ref{ZHJrfact}) on the $A_p$
ALE space $M=M_{p+1,p}$ coincides with a character of the affine
Kac-Moody algebra based on $u(p+1)$ at level $r$: Up to an overall
normalization one has
\beq
Z^{A_p}_{\rm gauge}(q,Q;r)= \sum_{n=0}^\infty ~ \sum_{\beta\in H_2(M,\IZ)}\, \Omega_{M}(n,\beta)~ q^{n-c/24}~Q^\beta =
\Tr\big(q^{L_0-c/24}~Q^{J_0}\big) \ ,
\label{ZApcharacter}\eeq
where the instanton zero-point energy $c=r\, \chi(M)$ is the central charge of a two-dimensional
conformal field theory of $r\,\chi(M)$ free chiral bosons with Virasoro
generator $L_0$, the sums run over characteristic classes $\beta=c_1(\cE)$ and $n=\ch_2(\cE)$ of an instanton gauge sheaf $\cE$ on the orbifold compactification $\overline{M}$, and $\Omega_{M}(n,\beta)$ are the degeneracies of
BPS states with the specified quantum numbers. The right-hand side of
this formula is the character of the integrable highest weight
representation of $\widehat{u}(p+1)_r$ corresponding to framing in the
trivial representation of the associated orbifold group $G=\IZ_{p+1}$,
i.e. $\mbf r=(r,0,\dots,0)$. The
operators $J_0$ come from the generators of the Cartan subalgebra $u(1)^{p+1}\subset
u(p+1)$. This connection can be strengthened in the realization
of the instanton moduli space as a quiver variety $\mathscr{M} (\mbf
n, \mbf r)$, wherein the gauge field (and consequently the partition
function) has a somewhat intricate dependence on boundary conditions
which we now describe in some detail.

For brevity we specialize the discussion here to the case of $U(1)$ gauge
theory, where $r=\dim W =1$. Then there are $p+1$ choices for the
framing vector given by $\mbf r= {\mbf e}_a$, where ${\mbf e}_a$ is
the vector with $1$ at entry $a$ and zeroes elsewhere for
$a=0,1,\dots p$. Physically this is a situation where the asymptotic
gauge field can be any of the $p+1$ flat connections which label
the boundary conditions. The
framing depends on the one-dimensional representation $\rho:G\to W$ of
$G=\IZ_{p+1}$~\cite{nakaALE3}. Near infinity $M$ looks like the Lens space
$L(p+1,p)=S^3/G$, so an anti-self-dual $U(1)$ gauge field can asymptote to a non-trivial
flat connection at infinity. Flat connections on $S^3/G$ are
classified by their holonomies which take values in the fundamental group
$\pi_1(S^3/G)=G$; whence a flat connection is labelled by a
one-dimensional representation $\rho\in\Hom_\IC(G,U(1))$. If $A^{(a)}$ denotes the flat connection at infinity
with holonomy $\zeta^a=\e^{2\pi\ii a/(p+1)}$, then the corresponding
framing shifts the second Chern class $\ch_2(\cE)$ by rational numbers
related to the Chern-Simons invariant~\cite{Griguolo}
\beq
h_a :=\frac1{4\pi^2}\, \int_{L(p+1,p)}\, A^{(a)}\wedge \dd
A^{(a)} =
\frac{p\, a^2}{p+1} \ .
\label{CSha}\eeq 
We can group sums over monopole charges $u\in\IZ^p$ into congruence classes modulo~$p+1$; the non-trivial congruence classes correspond to the
twisted sectors of the gauge theory with non-trivial holonomy at infinity.

For the trivial sector with $a=0$, the gauge theory partition function is given by
(\ref{ZgtHJ}). In
this case, the condition that the ideal sheaf $\cI$ in the
decomposition (\ref{factorization}) has vanishing first Chern class
(\ref{chernE}) is
equivalent to the condition that each entry $u_a$ of the vector $u=\mbf r+\tilde C\mbf n$ vanishes. For $U(1)$ instantons this
condition is uniquely solved by $\mbf{r} ={\mbf e}_0 =
(1,0,\dots,0)$ and $\mbf{n} = n\, {\delta}$, since $c_1 (\calR_0) = 0 $ and ${\delta}$
spans the kernel of the affine Cartan matrix. It follows from
(\ref{chernE2}) that the instanton number $n$ is then correctly
reproduced by the second Chern characteristic class $\ch_2(\cT)$ of
the gauge sheaf, and as we have explained the moduli space
\begin{equation}
\mathscr{M} (\mbf{n}=n\, \delta , \mbf{r}= {\mbf e}_0 ) =
\mathrm{Hilb}_n (M)
\end{equation}
is the Hilbert scheme of $n$ points on $M$. In this context, the
regular instantons on $M$ are associated with the regular
representation of the orbifold group $G$.

We now describe the modification of (\ref{ZgtHJ}) in twisted sectors. Since
there is a bijective correspondence between geometrical instanton moduli
spaces and quiver varieties, we can parametrize the sum over
first Chern classes in (\ref{ZgtHJ}) with a sum over the dimension vectors
$\mbf{n}$ and $\mbf{r}$ to write
\begin{equation} \label{partfct1}
Z_{\rm gauge}^{A_p} (q,Q;\mbf r) = \frac1{\hat\eta(q)^{p+1}} \,
\sum_{u} \,
q^{-\frac12\, u\cdot C^{-1} u} \ Q^{u}
\end{equation}
where the sum over ${u}$ runs over all values of $\mbf{n}$
and $\mbf{r}$ for which $\mathscr{M} (\mbf{n} , \mbf{r})$
is non-empty, or equivalently over integer vectors $\mbf n$ such that for
any fixed asymptotic boundary condition parametrized by $\mbf r$ the
difference vector $\mbf r - \mbf n$ belongs to an orbit of $\mbf r$
under the action of the affine Weyl group~\cite{nakaquiver}.

Let us rewrite the partition function (\ref{partfct1}) in a manner
which makes its relation to the characters of affine Lie algebras more
transparent. For this, we rewrite the sum over $u_1,\dots, u_p$ by
introducing new variables
\begin{eqnarray}
m_a = n_a-n_0
\end{eqnarray}
to get
\begin{equation}
u_a = r_a + (Cm)_{ab} \qquad \mbox{for} \quad
a=1,\dots,p \ ,
\end{equation}
which now involves the Cartan matrix $-C$. We prove that the new variables
$m_a$ span the integer magnetic charge lattice $\Gamma$ in the fractional instanton sector of the $U(1)$ gauge theory for each
{\it fixed} boundary condition $\mbf r={\mbf e}_a$. In this sector
the instanton numbers $n_0 , n_1, \dots , n_p$ are non-negative
integers constrained by the equation
\begin{equation} \label{eiconstraint}
0=n\cdot u= 2 n_a + n\cdot \tilde{C}n= 2 n_a + m\cdot {C}m \ .
\end{equation} 
By solving this constraint we can express the integers $n_a$ as
functions of the integers $m_a$ through
\begin{equation}
n_b=-\mbox{$\frac12$}\,m\cdot C m +m_b-m_a \qquad \mbox{for} \quad b=0,1,\dots , p \ ,
\end{equation}
which are non-negative since the Cartan matrix is positive
definite. Then the contribution to the partition function
(\ref{partfct1}) from the twisted sectors can be written
explicitly as
\beq
Z_{\rm gauge}^{A_p} (q,Q;{\mbf e}_a) = \frac{1}{\hat{\eta}(q)^{p+1}}\,
  \sum_{m}\, q^{\frac{a \,(p+1-a)}{2(p+1)} -m_a -\frac{1}{2}\, m\cdot
    Cm} \ Q_a \ Q^{Cm}
\label{bcZ}\eeq
for $a=0,1,\dots,p$, where we define $Q_0=1$. Similarly to~\cite{Dijkgraaf:2007sw,Dijkgraaf:2007fe}, this formula can be
expressed in terms of the characters $\chi_{\widehat{\lambda}}
(q,z)$ of $\widehat{su}(p+1)_1$ at a specialization point $z =
\sum_a\, x_a\, \alpha_a^{\vee} $ associated to a level one affine integrable
weight $\widehat{\lambda}$; they can be written using string-functions and theta-functions as
\begin{equation}
\chi_{\widehat{\lambda}} (q,z) = \sum_{\widehat{\lambda}\,'} \,
\Sigma_{\widehat{\lambda}}{}^{\widehat{\lambda}\,'}(q) \
\Theta_{\widehat{\lambda}\,' }(q,z) \ ,
\end{equation}
where
\begin{eqnarray}
\Sigma_{\widehat{\lambda}}{}^{\widehat{\lambda}\,'}(q) =
\delta_{\widehat{\lambda}}{}^{\widehat{\lambda}\,'} \ \eta (q)^{-p}
\qquad \mbox{and} \qquad
\Theta_{{\widehat{\lambda}}} (q , z) = \sum_{\alpha^{\vee} \in
  \Delta^{\vee}} \, q^{\frac{1}{2}\, \left( \alpha^{\vee} +
    \lambda\right)^2} \, \e^{2 \pi \ii z\cdot(\alpha^{\vee} + \lambda)}
\label{stringTheta}\end{eqnarray}
with $\lambda$ the finite part of $\widehat{\lambda}$ and
$\Delta^{\vee}$ the coroot lattice. For the fundamental weight
$\lambda = \lambda_a$ and the coroot $\alpha^{\vee} = \sum_a\, m_a\,
\alpha_a^{\vee}$ with $\lambda_a\cdot\alpha^\vee_b=\delta_{ab}$, we find
\begin{eqnarray}
\mbox{$\frac{1}{2}$} \, \left( \alpha^{\vee} + \lambda\right)^2 =
\sum_{a=1}^{p} \, \big(m_a^2 - m_a\, m_{a+1} \big) + m_a + \frac{a\,
  (p+1-a)}{2 p} \qquad \mbox{and} \qquad \e^{2 \pi \ii z\cdot
  (\alpha^{\vee} + \lambda)} = Q_a \ Q^{Cm} \nonumber \\
\end{eqnarray}
with $Q_a:=\e^{2\pi\ii x_a}$,
and hence the theta-function from (\ref{stringTheta}) reproduces the
series in (\ref{bcZ}) up to the redefinition $m \rightarrow - m$. The
remaining factor of the Dedekind function $\eta(q)=q^{1/24}\,
\hat\eta(q)$ comes from the Heisenberg algebra character associated to
the extra $\widehat{u}(1)$ part of $\widehat{u}(p+1)_1$; it arises
from integrating over the flat connections at infinity. This gives the
Fock space representation of $\widehat{u}(p+1)_1$ such that
(\ref{bcZ}) is its character. Moreover, via the McKay correspondence
the Chern-Simons invariant (\ref{CSha}) of the corresponding flat
connection maps to the conformal
dimension of the highest weight $\widehat{\lambda}_a$.

In the higher
rank cases, the analogous partition functions can be read off from the
Poincar\'e polynomial of the quiver variety $\scrM (\mbf v , \mbf w)$
which is computed in~\cite{hausel}, see also~\cite{Fujii:2005dk}, with
the result
\beq
Z_{\rm gauge}^{A_p} (q,Q;\mbf r) = \sum_{\mbf n} \,
\chi\big(\scrM(\mbf n,\mbf r)\big) \ 
  \prod_{a\in\widehat{G}} \, \xi_a^{n_a} = 
\frac1{\hat\eta(q)^{r\,(p+1)}} \ \prod_{a=0}^p\,
\Theta_\Gamma(q,Q_a)^{r_a} \ ,
\eeq
where the mapping between orbifold and resolution counting variables is given
by $q=\xi_0\,\xi_1\cdots \xi_p$ and
$Q_a=(\xi_{1+a},\dots,\xi_{p+a})$. 

This result underlies the consistency condition that requires these
partition functions to have the correct modular properties implied by
S-duality. Nakajima proves that there is a natural geometric action of
the affine Lie algebra
$\widehat{u}(p+1)_r$ on the middle cohomology of the instanton
moduli space; in particular the cohomology of (\ref{quivervardecomp})
is a direct sum of certain highest weight representations of
$\widehat{u}(p+1)_r$ determined by the framing vector $\mbf
r$. Setting $L_0=n$ and $J_0=\beta$ on $H^{\rm
  mid}(\scrM_{n,\beta;r}^{\rm inst}(M))$ gives the action of the Cartan
subalgebra of $\widehat{u}(p+1)_r$, while the rest of the action of
the affine Lie algebra is defined through operations of twisting
vector bundles along the exceptional divisors of the orbifold
resolution (corresponding to the action of conformal field theory
vertex operators). Consequently the generating
function of the Euler numbers of the instanton moduli space is
identified with a character of
$\widehat{u}(p+1)_r$~\cite{nakaALE2,nakaHeis}; see~\cite{nakaALE4} for
a review. The cohomology of the instanton moduli space
will be discussed in \S\ref{subsec:eqcohinst}.

Hence the $\cN=4$ gauge theory is described via a matrix quantum mechanics
which arises from the set of collective coordinates around an
instanton solution; it is described by the generalized ADHM equations
(\ref{adhmgen}) which are determined via representations of the McKay
quiver associated with the singularity. The instanton moduli space is
the associated quiver variety, with prescribed topological data
encoded in the representation theory variables. It
carries a natural action of the associated Kac-Moody algebras at a
certain level and its Euler characteristics are nicely organized into
characters of the affine Lie algebra.

\bigskip

\section{Equivariant gauge theories on toric surfaces\label{sec:InstN2gt}}

\subsection{Nekrasov functions\label{subsec:N2C2}}

We now turn to the problem of instanton counting in pure
$\mathcal{N}=2$ Yang-Mills theory on $M=\IC^2$
with gauge group $U(r)$, and some of its extensions, which include the
low-energy effective four-dimensional $\cN=2$ supersymmetric gauge
theories arising from dimensional reduction on a Calabi-Yau threefold;
we follow the formalism devised by Nekrasov
in~\cite{Nekrasov:2002qd}. These gauge
theory partition functions are not quite the natural counterparts of the BPS partition
functions we constructed in \S\ref{sec:curvecountsurf}, but they are
natural relatives of the six-dimensional $\cN=2$ gauge theory that we will
study in \S\ref{6dcohgt} in the sense that the instanton moduli
space integrals are defined as ``volumes''; the difference is that
here the (stable) fundamental cycle has positive dimension, so its
volume is defined in the sense of equivariant integration. On the
other hand, as a maximally supersymmetric gauge theory the partition
functions of \S\ref{6dcohgt} compute the virtual Euler
characteristics of the instanton moduli schemes; its counterpart in
four dimensions is the $\cN=4$ supersymmetric Yang-Mills theory which
was treated in \S\ref{sec:4dcohgt}. In a sense the
six-dimensional cohomological gauge theory is a hybrid of the
four-dimensional $\cN=2$
and $\cN=4$ gauge theories, and we will export techniques used to
study both of them. From the six-dimensional perspective, the
low-energy sectors of
$\cN=2$ gauge theories on a toric surface $M$ arise as the tensor field theories of an M5-brane on the six-dimensional product
manifold $M\times \Sigma$, where $\Sigma$ is the Seiberg-Witten curve;
see e.g.~\cite{Bonelli:2012ny} for a recent account. More generally,
there is a large class of gauge theories which arise by compactifying
coincident M5-branes on $M\times C$, where $C$ is a punctured Riemann
surface and the Seiberg-Witten curve $\Sigma$ is a branched cover of
$C$; some of these theories are studied by~\cite{Gaiotto:2009hg}. This feature
underlies the connection between these four-dimensional gauge theories
and natural geometric representations of Heisenberg algebras and $W$-algebras on the equivariant cohomology of
the instanton moduli spaces within the framework of the AGT correspondence~\cite{AGT,Wyllard}.

As before, the set of
observables that enter into the instanton counting problem are captured by
the topologically twisted $\cN=2$ gauge theory; these observables compute the intersection theory of the (compactified) instanton moduli
space. For computational purposes, one should further deform the gauge
theory by defining it on a noncommutative space and in an appropriate
supergravity background called the ``$\Omega$-background''. The
noncommutative deformation can be regarded as modifying the moment map equations (\ref{adhm}) that define the
instanton moduli space; see~\cite{Szabo:2011mj} for an explicit description in the present context. The effect of the $\Omega$-deformation is
that the new observables are equivariant differential
forms with respect to the isometry group of $\IC^2$ and the new BRST
operator, whose cohomology determines physical observables, can be interpreted as an equivariant
differential on the
space of fields; this deformation thereby mixes gauge transformations
with rotations.

As in \S\ref{subsec:VafaWitten}, the relevant fields in the bosonic sector of the twisted gauge theory
comprise a connection one-form $A$ on a rank $r$ vector bundle $\cE$, and a complex Higgs field
$\phi$ which is a local section of the adjoint bundle of $\cE$. One
can study the gauge theory equivariantly; the equivariant gauge theory is
topological and localizes onto the moduli space of instantons
$\scrM^{\rm inst}_{n,r}(\IC^2)$. The instanton partition function is called a
\emph{Nekrasov function}. It has an explicit
dependence on the eigenvalues $a=(a_1,\dots,a_r)$ of the Higgs field $\phi$
which determine a framing of $\cE$ by a representation of the maximal
torus $U(1)^r$ of the gauge group, and the equivariant parameters
$\epsilon=(\epsilon_1,\epsilon_2)$ for the natural scaling action of the
two-torus $\torus^2$ on $\IC^2$; it can be written as 
\be 
Z_{\mathrm{inst}}^{\IC^2}
(\epsilon,a;q;r) = \sum_{n=0}^{\infty} \, q^n \ \cZ_{n}^{\IC^2}
(\epsilon,a;r) \label{partfunct} \ee
where the
counting parameter $q$ again weighs the instanton number and is
determined by the gauge coupling. The quantity $\cZ_{n}^{\IC^2}
({a},\epsilon;r)$
can be interpreted geometrically as a representative of a particular
$\torus^2\times U(1)^r$ equivariant characteristic class of a bundle over the instanton
moduli space; for the pure $\cN=2$ gauge theory it is just the
equivariant integral
\beq
\cZ_{n}^{\IC^2}
(\epsilon,a;r)^{\rm gauge} =\oint_{\scrM^{\rm inst}_{n,r}(\IC^2)}\, 1
\label{ZnC2eqvol}\eeq
defined by the pushforward to a point in equivariant cohomology;
throughout we use the symbol $\oint$ to distinguish equivariant
integration from ordinary integration. In contrast to the $\cN=4$ gauge theory, where the instanton partition function gives
the complete answer, here the full partition function $Z_{\mathrm{gauge}}^{\IC^2}
(\epsilon,a;q;r)$ should also include a
perturbative contribution which we will insert later on.

By using the ADHM parametrization of the instanton moduli space from
\S\ref{subsec:ALEmodsp}, the evaluation of the partition function of the deformed gauge theory is
reduced to the computation of the equivariant ``volumes''
(\ref{ZnC2eqvol}) of the instanton
moduli spaces \cite{Moore:1997dj,Moore:1998et}; these volumes are computed via
equivariant localization. The instanton moduli scheme
$\scrM^{\rm inst}_{n,r}(\IC^2)$ is a fine moduli space; it is moreover smooth of
dimension $2\,r\,n$ and the tangent space at a point
corresponding to a torsion free sheaf $\cE$ on $\PP^2$ is given by
\beq
T_{[\cE]}\scrM^{\rm inst}_{n,r}\big(\IC^2 \big) =
\Ext^1_{\cO_{\PP^2}}\big(\cE\,,\,\cE(-\ell_\infty) \big) \ .
\label{tangentspExt1}\eeq
We may
thus compute the volumes (\ref{ZnC2eqvol}) using the Atiyah-Bott localization theorem in
the equivariant cohomology $H^\bullet_{\torus}(\scrM^{\rm inst}_{n,r}(\IC^2))$,
whose coefficient ring is $H^\bullet_{\torus}({\rm
  pt})\cong\IC[\epsilon_1,\epsilon_2,a_1,\dots,a_r]$; here $\torus:= \torus^2\times
  U(1)^r$. This reduces the evaluation of the instanton
measure $\cZ_{n}^{\IC^2}(\epsilon,a;r)$ to two steps: a classification
of the fixed points of the $\torus$-action on the moduli space, and the computation of
the weights of the toric action on the tangent space to the instanton
moduli space at each fixed point.

The classification of the fixed points was given by
Nakajima-Yoshioka~\cite{nakajima,NakYosh}; it follows from the identification of the
rank one
instanton moduli space $\scrM^{\rm inst}_{n,1}(\IC^2)$ with the Hilbert scheme
of points ${\rm Hilb}_n(\mathbb{C}^2)$. The fixed points in this case
are isolated point-like instantons which are in bijective correspondence
with Young diagrams $\lambda=(\lambda_1,\lambda_2,\dots)$ having
$|\lambda|=n$ boxes. For the $U(r)$ gauge theory in the Coulomb branch
where the eigenvalues $a_1,\dots,a_r$ are all distinct, the problem
reduces to $r$ copies of the rank one case and the fixed points are
parametrized by the finite set of length $r$ sequences
$\vec\lambda=(\lambda^1,\dots,\lambda^r)$ of Young diagrams of size
$|\vec\lambda\,|=\sum_{l=1}^r\, |\lambda^l|=n$. The localization
formula then evaluates the equivariant integral (\ref{ZnC2eqvol}) as
\beq
\cZ_{n}^{\IC^2}
(\epsilon,a;r)^{\rm gauge} = \sum_{\vec\lambda\,:\, |\vec\lambda\,|=n}
\ \frac1{\eul_\torus\big(T_{\vec\lambda}\scrM^{\rm inst}_{n,r}(\IC^2) \big)} \ .
\label{ZnC2euler}\eeq

To compute the equivariant Euler class of the tangent bundle at the fixed points
$\vec\lambda$ in (\ref{ZnC2euler}), following Nakajima-Yoshioka
\cite{nakajima,NakYosh} one introduces a two-dimensional $\torus^2$-module $Q$
to keep track of the geometric torus
action, with weights
$t_{i} = \e^{\ii \epsilon_{i}}$ for $i=1,2$. At a fixed point
$\vec\lambda$ of the $\torus^2\times U(1)^r$
action we can decompose the vector spaces $V$ and $W$ introduced in the ADHM
construction as elements of
the representation ring of $\torus$ to get
\begin{equation}
V_{\vec \lambda} = \sum_{l=1}^r \,e_l~ \sum_{(i,j)\in \lambda^l }\, t_1^{i-1}
\,t_2^{j-1}  \qquad  \mbox{and} \qquad W_{\vec \lambda} = \sum_{l=1}^r\,e_l
\label{decompos}
\end{equation}
where $e_l = \e^{\ii a_l}$. With this notation the fields $(B_1 ,
B_2 , I , J )$ corresponding to a fixed point configuration $[\cE]$ are
elements $(B_1,B_2)\in \End_\IC(V_{\vec \lambda}) \otimes Q$, $I\in
\Hom_\IC(W_{\vec \lambda},V_{\vec \lambda})$, and $J \in \Hom_\IC(V_{\vec \lambda},W_{\vec \lambda}) \otimes \bigwedge^2 Q$.
Using (\ref{tangentspExt1}) we may describe the local structure of the instanton moduli space
by the complex
\begin{equation} \label{adhmdefcomplex}
\xymatrix@1{
  \Hom_\IC(V_{\vec \lambda} , V_{\vec \lambda})
   \quad\ar[r]&\quad
   {\begin{matrix} \Hom_\IC(V_{\vec \lambda} , V_{\vec \lambda}) \otimes Q
   \\ \oplus \\
   \Hom_\IC(W_{\vec \lambda} , V_{\vec \lambda}) \\ \oplus  \\ \Hom_\IC(V_{\vec \lambda} ,
   W_{\vec \lambda}) \otimes \bigwedge^2
   Q \end{matrix}}\quad \ar[r] & \quad
   {\begin{matrix} \Hom_\IC(V_{\vec \lambda} , V_{\vec \lambda}) \otimes \bigwedge^2
       Q
   \end{matrix}}
}
\end{equation}
which is a finite-dimensional version of the
Atiyah-Hitchin-Singer instanton deformation complex (\ref{AHScomplex}). The first
map corresponds to infinitesimal (complex) gauge transformations
while the second map is the linearization of the ADHM equation
(\ref{adhm}). In general the complex (\ref{adhmdefcomplex}) has three
non-vanishing cohomology groups; we can safely assume for our purposes
that they vanish in both degree zero and two. Then the only
non-vanishing degree one cohomology
describes fields that obey the linearized version of the ADHM
equation (\ref{adhm}) but are not gauge variations; it is thus a
local model for the tangent space $T_{[\cE]}\scrM^{\rm inst}_{n,r}(\IC^2)$. The
weights of the toric action on the tangent space are given by the equivariant
index of the complex (\ref{adhmdefcomplex}), which can be expressed
in terms of the characters of the representations as
\begin{eqnarray} \label{character4d}
\Ch_\torus\big(T_{\vec\lambda}\scrM^{\rm inst}_{n,r}(\IC^2)\big) = W^*_{\vec \lambda } \otimes
V_{\vec \lambda } + \frac{V^*_{\vec \lambda } \otimes W_{\vec \lambda }}{ t_1\,
t_2} - V^*_{\vec \lambda } \otimes V_{\vec \lambda } \
\frac{ (1-t_1)\, (1-t_2)}{t_1\, t_2} \ ,
\end{eqnarray}
where the dual involution acts on the weights as $t_i^*=t_i^{-1}$ and
$e_l^*=e_l^{-1}$. To state the final result, let
us recall some definitions which will be used throughout this section in combinatorial
expansions of partition functions. Let $\lambda$ be a Young
diagram. Define the arm and leg lengths of a box $s=(i,j)\in \lambda $
respectively by
\beq
A_\lambda (s)= \lambda_i-j \qquad \mbox{and} \qquad L_\lambda (s)= \lambda_j^t-i \
, 
\eeq
where $\lambda_i$ is the length of the $i$-th column of $\lambda $ and
$\lambda_j^t$ is the length of the $j$-th row of $\lambda $. Define the arm
and leg colength of $s=(i,j)$ respectively by
\beq
A_\lambda ^t(s)= j-1 \qquad \mbox{and} \qquad L_\lambda ^t(s)= i-1 \ .
\eeq
Then the character (\ref{character4d}) can be expressed in the
form~\cite{Flume:2002az,NakYosh}
\beq
\Ch_\torus\big(T_{\vec\lambda}\scrM^{\rm inst}_{n,r}(\IC^2)\big) =
\sum_{l,l'=1}^r\, e_{l'}\, e_l^{-1} \
M^{\IC^2}_{\lambda^l,\lambda^{l'}}(t_1,t_2) \ ,
\eeq
where
\beq\label{MYC2}
M^{\IC^2}_{\lambda,\lambda'}(t_1,t_2) = \sum_{s\in\lambda}\,
t_1^{-L_{\lambda'}(s)}\, t_2^{A_\lambda(s)+1} + \sum_{s'\in\lambda'}\,
t_1^{L_\lambda(s'\,)+1}\, t_2^{-A_{\lambda'}(s'\,)}
\eeq
for a pair of Young diagrams $\lambda,\lambda'$.
The corresponding top Chern polynomial then yields the desired product
of weights that enters the localization formula (\ref{ZnC2euler}). One can thereby write down an explicit expression for the partition function of the matrix
quantum mechanics that corresponds to the instanton factor (\ref{partfunct}), with
topological charge $n$ that corresponds to the total number of boxes of
$\vec \lambda$~\cite{Bruzzo:2002xf,nakajima}. 

Finally the instanton partition function of the equivariant
$\cN=2$ gauge theory on $\IC^2$ can be written as
\beq
Z_{\mathrm{inst}}^{\IC^2}
(\epsilon,a;q;r)=
\sum_{\vec \lambda}\,q^{|\vec
  \lambda\,|}~\cZ^{\IC^2}_{\vec\lambda}(\epsilon,a;r)^{\rm gauge} \ ,
\label{N2inst}\eeq
where the Euler classes computed via localization are
\bea
\cZ^{\IC^2}_{\vec\lambda}(\epsilon,a;r)^{\rm gauge} &=&
\prod_{l,l'=1}^r \ \prod_{s\in \lambda^l}\,\Big(
a_{l'}-a_l-L_{\lambda^{l'}}(s)\,\epsilon_1+\big(
A_{\lambda^l}(s)+1\big)\,\epsilon_2 \Big)^{-1} \nonumber\\ && \qquad
\qquad \times~
\prod_{s'\in \lambda^{l'}}\,\Big(a_{l'} -a_l +
\big(L_{\lambda^l}(s'\,)+1\big)\,\epsilon_1- A_{\lambda^{l'}}(s'\, )\,\epsilon_2
\Big)^{-1} \ .
\label{weights}\eea
In the rank one case, the sum over Young diagrams $\lambda$ can be
done explicitly with the remarkably simple
result~\cite[\S4]{NakYosh}
\beq
Z_{\mathrm{inst}}^{\IC^2}
(\epsilon;q;1) = \e^{q/\epsilon_1\,\epsilon_2} \ ,
\label{ZC2N2U1}\eeq
in which we observe explicitly the dependence on the equivariant
volume $\oint_{\IC^2}\,1=\frac1{\epsilon_1\,\epsilon_2}$ provided by
the $\Omega$-deformation that
regularizes the volume of the instanton moduli space.

One can rephrase this computation in a fashion that
is more suitable for extension to generic toric surfaces. Define the universal sheaf $\scrE$ on $\complex^2 \times
\mathscr{M}^{\rm inst}_{n,r}(\IC^2) $ with fibres~\cite{nakajima2}
\begin{equation}
\scrE\big|_{[\mathcal{E}]} = W \ \oplus \ V \otimes \big(S^- \ominus S^+\big)
\end{equation}
where $S^{\pm}$ are the positive and negative chirality spinor
bundles. Its character at a fixed point is~\cite{nakajima2}
\begin{equation} \label{C2chern}
\Ch_\torus \big(\scrE\big|_{[\mathcal{E}]}\big) = W_{\vec \lambda} - (1 -
t_1)\, (1 - t_2) \, V_{\vec \lambda} \ ,
\end{equation}
and with a
slight abuse of notation one can formally write \cite{Losev:2003py}
\begin{eqnarray} \label{indexC2}
\Ch_\torus\big(T_{\vec\lambda}\scrM^{\rm inst}_{n,r}(\IC^2)\big) &=& - \int_{\complex^2} \,
\Ch_\torus \big(\scrE\big|_{[\mathcal{E}]}\big)\wedge \Ch_\torus \big(\scrE^\vee\,
\big|_{[\mathcal{E}]}\big) \wedge \mathrm{td}_{\torus^2}(\complex^2)
\nonumber \\[4pt] &=& -
\frac{\Big[\Ch_\torus \big(\scrE\big|_{[\mathcal{E}]}\big)\wedge \Ch_\torus \big(\scrE^\vee\,
\big|_{[\mathcal{E}]}\big)\Big]_0}{(1-t_1)\, (1-t_2)}
\end{eqnarray}
where $\mathrm{td}_{\torus^2}(\IC^2)$ is the equivariant Todd class. The
integral in (\ref{indexC2}) is evaluated by localization with respect to the
action of the torus $\torus^2$ on $\complex^2$ given by $z_i
\rightarrow t_i\, z_i$ for $i=1,2$. The only fixed point is the origin
$z_i=0$, and so in the second equality we are left with the character of the
universal sheaf computed at the point $0 \times
\mathscr{M}^{\rm inst}_{n,r}(\IC^2)$ which using equivariant integration can be expressed as a sum over the fixed points
of the action of the torus $\torus^2\times U(1)^r$ on the moduli
space, given by Young
diagrams. Substituting (\ref{C2chern}) we easily find
\begin{eqnarray}
\Ch_\torus\big(T_{\vec\lambda}\scrM^{\rm inst}_{n,r}(\IC^2)\big) = - \frac{W_{\vec
    \lambda } \otimes W^*_{\vec \lambda }}{(1-t_1)\, (1-t_2)} -
V^*_{\vec \lambda } \otimes V_{\vec \lambda } \ \frac{(1-t_1)\,
  (1-t_2)} {t_1 \,
t_2} + W^*_{\vec \lambda } \otimes V_{\vec \lambda } +
\frac{V^*_{\vec \lambda } \otimes W_{\vec \lambda }}{t_1 \, t_2} \ ,
\nonumber \\
\label{Chpert}\end{eqnarray}
which reproduces exactly the character (\ref{character4d}) up to the first
term which is proportional to $W_{\vec\lambda} \otimes
W_{\vec\lambda}^*$ and hence
is independent of the partition vector $\vec
\lambda$.

The first term in (\ref{Chpert}) is
interpreted as the perturbative contribution. Expanding it as
a power series in $t_1,t_2$ and extracting the top Chern form gives a
contribution to the partition function in the form of
products of functional determinants
\beq
\prod_{i,j=0}^\infty\ \frac1{x-i\,\epsilon_1-j\,\epsilon_2} \ ,
\eeq
in the Higgs field eigenvalues $x=a_l-a_{l'}$. Double zeta-function
regularization of this infinite product in the proper time
representation gives
the Barnes double gamma-function
\beq
\Gamma_2(x|-\epsilon_1,-\epsilon_2)= \exp\big(\gamma_{\epsilon_1,\epsilon_2}(x)\big)
\eeq
with
\beq
\gamma_{\epsilon_1,\epsilon_2}(x):=\left. \frac\dd{\dd
    s}\,\right|_{s=0} ~\frac1{\Gamma(s)}\, \int_0^\infty\, \frac{\dd
  t}{t}~t^s~\frac{\e^{-t\,x}}{\big(\e^{\epsilon_1\,t}-1\big)\,
  \big(\e^{\epsilon_2\,t}-1\big)} \ .
\eeq
The leading behaviour for small $\epsilon_1,\epsilon_2$ can be
extracted from
\beq
\lim_{\epsilon_1,\epsilon_2\to0}\,\epsilon_1\,\epsilon_2\,
\gamma_{\epsilon_1,\epsilon_2}(x) = -\mbox{$\frac12$}\,\log
x+\mbox{$\frac34$}\,x^2 \ .
\eeq
We define the perturbative partition function as
\begin{equation}
Z_{\mathrm{pert}}^{\IC^2} (\epsilon, a;r)=
\prod_{l,l'=1}^r\,\exp \big( - \gamma_{\epsilon_1 , \epsilon_2} (a_l -
a_{l'}) \big) \ .
\label{N2pert}\end{equation}

Finally, one also has a classical contribution
\begin{equation}
Z_{\mathrm{class}}^{\rm \IC^2} (\epsilon, a;q;r)=
\prod_{l=1}^r \, q^{-{a^2_l}/{2 \, \epsilon_1 \, \epsilon_2}} \ .
\label{N2class}\end{equation}

Altogether, one can define the full partition function $Z_{\rm
  gauge}^{\IC^2} (\epsilon, a;q;r)$ as the product of the classical part
(\ref{N2class}), the perturbative part (\ref{N2pert}), and the
instanton piece (\ref{N2inst}); in the limit $\epsilon_1
, \epsilon_2 \rightarrow 0 $, the quantity $\epsilon_1\,\epsilon_2\, 
\log Z_{\rm
  gauge}^{\IC^2} (\epsilon, a;q;r)$ is the Seiberg-Witten prepotential of
$\cN=2$ supersymmetric Yang-Mills theory~\cite{Nekrasov:2002qd}.

\subsection{${\mathcal{N}=2^*}$ gauge theory\label{subsec:N2star}}

We will now describe the structure of other equivariant $\cN=2$ gauge theory partition
functions on more general toric surfaces, defering the technical details of their
evaluation to \S\ref{subsec:N2quiver}. We follow the treatment of
Gasparim-Liu~\cite{Gasparim} where partition functions of general
$\cN=2$ gauge theories are rigorously derived for a certain class of
toric surfaces using equivariant localization techniques on the
instanton moduli spaces, and suitably modify them to correctly take into
account contributions from non-compact divisors as explained in \S\ref{subsec:VafaWitten}.

Consider first the deformation of the $\mathcal{N}=4$ gauge theory on
a toric surface $M$ obtained by turning on a mass parameter for the adjoint field
hypermultiplet; this is called the $\mathcal{N}=2^*$ supersymmetric Yang-Mills theory on $M$. In this case, toric localization also involves the
maximal torus $\torus_{\mu}$ of the $U(1)$ flavour symmetry group
whose equivariant parameter $\mu$ is interpreted as the mass of the
adjoint matter field. In the context of the AGT correspondence, it should
be related to some two-dimensional conformal field theory on a
one-punctured torus.

The classical contributions on any Hirzebruch-Jung surface can be obtained from (\ref{N2class}) by
orbifold projection with respect to the action of the cyclic group
$G_{p,p'}\cong\IZ_p$, yielding
\beq
Z_{\rm class}^{M_{p,p'}}(\epsilon, a;q;r)=\prod_{l=1}^r \,
q^{-{a^2_l}/{2 \, p \, \epsilon_1 \, \epsilon_2}} \ .
\eeq

The perturbative contributions to the partition function are also straightforward to write down; in the
notation of \S\ref{subsec:N2C2} they are given by
\beq
Z_{\rm pert}^{M}(\epsilon, a,\mu;r)=
Z_{\rm pert}^{M}(\epsilon, a;r)^{\rm gauge}~
Z_{\rm pert}^{M}(\epsilon, a,\mu;r)^{\rm adj}
\label{ZpertMN2star}\eeq
where
\beq
Z_{\rm pert}^{M}(\epsilon, a;r)^{\rm gauge}=
\prod_{l,l'=1}^r\,\exp\big(-\gamma_{-w,u}(a_l-a_{l'})-\gamma_{w,u-k\,w}(a_l-a_{l'})
\big)
\eeq
and 
\beq
Z_{\rm pert}^{M}(\epsilon, a,\mu;r)^{\rm adj}=
\prod_{l,l'=1}^r \,\exp\big(\gamma_{-w,u}(a_l-a_{l'}+\mu)+\gamma_{w,u-k\,w}(a_l-a_{l'}
+\mu)
\big) \ ,
\eeq
where as before the vector $a=(a_1,\dots,a_r)$ contains
the Higgs eigenvalues in the Lie algebra $u(r)$. The parameters $w$
and $u$ are the weights of the tangent and normal bundles under the
toric action at a fixed point on a distinguished compactification
divisor $\ell_\infty$ for $M$, disjoint from the other compact
two-cycles of $M$, with $k:=\langle\ell_\infty,\ell_\infty\rangle_{\Gamma} >0$.

In our prototypical
cases of Hirzebruch-Jung surfaces, this class of examples
encompasses the total spaces of the holomorphic line bundles
$\mathcal{O}_{\mathbb{P}^1}(-p)$, $p>0$, i.e. the minimal resolutions
of $A_{p,p'}$ singularities with $p'=1$ whose projectivization
$\IF_p$ is the $p$-th Hirzebruch surface; the case $p=2$
is the $A_1$ ALE space. In this case one has~\cite{Gasparim}
\beq
w = \epsilon_1 \ , \qquad u= -\epsilon_2 \qquad \mbox{and} \qquad
k=p \ .
\eeq
Partition functions similar to those of this section are given in~\cite{Bonelli:2011kv}.

Let us now write down the instanton contributions to the
$\mathcal{N}=2^*$ gauge theory partition function. The corresponding Nekrasov instanton partition functions on $\IC^2$ appear as the
building blocks of those on $M$. They are given by 
\beq
Z_{{\rm inst}}^{\IC^2}(\epsilon, a,\mu;q;r) =
\sum_{\vec \lambda}\,q^{|\vec \lambda\,|}~\cZ_{\vec
\lambda}^{\IC^2}(\epsilon, a;r)^{\rm gauge} ~
\cZ_{\vec \lambda}^{\IC^2}(\epsilon, a,\mu;r)^{\rm adj} \ ,
\eeq
where
\bea
\cZ^{\IC^2}_{\vec\lambda}(\epsilon,a,\mu;r)^{\rm adj} &=&
\prod_{l,l'=1}^r \ \prod_{s\in \lambda^l}\,\Big(
a_{l'}-a_l-L_{\lambda^{l'}}(s)\,\epsilon_1+\big(
A_{\lambda^l}(s)+1\big)\,\epsilon_2 +\mu\Big) \nonumber\\ && \qquad
\qquad \times~
\prod_{s'\in \lambda^{l'}}\,\Big(a_{l'} -a_l +
\big(L_{\lambda^l}(s'\,)+1\big)\,\epsilon_1- A_{\lambda^{l'}}(s'\, )\,\epsilon_2
+\mu \Big)
 \label{weightsM}\eea
represents the equivariant Euler class
$\eul_{\torus\times\torus_\mu}\big(T_{\vec\lambda}\scrM^{\rm inst}_{n,r}(\IC^2) \big)$ at the
fixed point $\vec\lambda$. Note that
\beq
\cZ_{\vec\lambda}^{\IC^2}(\epsilon, a;r)^{\rm gauge} =
\frac1{\cZ_{\vec\lambda}^{\IC^2}(\epsilon, a,\mu=0;r)^{\rm adj}} \ .
\eeq
As before the sum runs over $r$-vectors of Young tableaux $\vec
\lambda=(\lambda^1,\dots,\lambda^r)$ having $|\vec \lambda\,|=\sum_l\,|\lambda^l|$ boxes in
total, which parametrize the regular instantons on $M$. The matter field
contribution is written for the adjoint representation of
the $U(r)$ gauge group; the adjoint matter fields are local sections of
the adjoint bundle of the tangent bundle on the
instanton moduli space.

For the $A_{p,1}$ spaces, the instanton partition function can
be expressed as a magnetic
lattice sum
\bea
Z_{\rm inst}^{\IF_p}(\epsilon, a,\mu; q,Q; r) &=&
\sum_{\vec u\in\IZ^r}\, q^{\frac1{2p} \,\sum_{l=1}^r\, u_l^2} \ Q^{u}~
\prod_{l\neq l'}\,\frac{\ell^{\vec
  u}_{l,l'}(\epsilon, a,\mu)}{\ell^{\vec
  u}_{l,l'}(\epsilon, a)} ~
Z_{{\rm inst}}^{\IC^2}\big(\epsilon_1\,,\epsilon_2\,,\,
a+ \epsilon_1\,\vec u\,,\,\mu\, ;\, q\, ;\,r \big) \nonumber \\ &&
\qquad \qquad \times~
Z_{{\rm
    inst}}^{\IC^2}\big(-\epsilon_1\,,\,\epsilon_2+p\,\epsilon_1\,,\,
a+ (\epsilon_2+p\,\epsilon_1)\,\vec
u\,,\,\mu\,;\,q\,; \,r\big) \ ,
\label{N2starinst}\eea
where the leg factors are given by
\beq
\ell^{\vec u}_{l,l'}(\epsilon, a,\mu)=\left\{
  \begin{array}{l} \displaystyle
~ \prod_{j=0}^{u_{l} -u_{l'}-1}~\prod_{i=0}^{p\,j}\,
\big(a_{l'} -a_l-i\,\epsilon_1-j
\,\epsilon_2 +\mu \big) \ , \qquad
u_l>u_{l'} \ , \\ ~~ \\ \displaystyle
~ \prod_{j=1}^{u_{'}-u_{l}}~\prod_{i=1}^{p\,j-1}\,
\big(a_{l'}-a_l+i \,\epsilon_1+j
\,\epsilon_2+\mu \big) \ , \qquad
u_l<u_{l'} \ , \\ ~~ \\ \displaystyle
~ 1 \ , \qquad u_l=u_{l'}
\end{array} \right.
\eeq
and 
\beq
\ell^{\vec u}_{l,l'}(\epsilon, a)=
\ell^{\vec u}_{l,l'}(\epsilon, a,\mu
  =0) \ .
\eeq
Here $\vec u=(u_1,\dots,u_r)$, with $u=\sum_l\, u_l$, parametrize the
contributions from fractional instantons. When $\mu=0$, the
perturbative contribution (\ref{ZpertMN2star}) is unity and the
formula (\ref{N2starinst}) reproduces $r$ powers of the anticipated
$\cN=4$ result (\ref{ZU1Fp}). On the other hand, the $\mu\to\infty$
limit $Z_{\rm inst}^{\IF_p}(\epsilon, a,\mu=\infty; q=0,Q;
r)\big|_{\mu^2\,q=\Lambda}$ gives partition functions for the
pure $\cN=2$ gauge theory on $M_{p,1}$ (with regular counting
parameter $\Lambda$) which amounts to dropping all numerator products.

In the rank one case $r=1$, the $\cN=2^*$ instanton partition function (\ref{N2starinst})
reduces to
\bea
Z_{\rm inst}^{\IF_p}(\epsilon_1,\epsilon_2,\mu; q,Q;1) =
\theta_3(q^{1/p},Q)~ \Big(\,
Z_{{\rm inst}}^{\IC^2}(\epsilon_1,\epsilon_2, \mu;q;1) ~
Z_{{\rm
    inst}}^{\IC^2}(-\epsilon_1,\epsilon_2+p\, \epsilon_1,\mu;q;1) \, \Big)
\label{ZU1FpN2star}\eea
where
\beq
Z_{{\rm inst}}^{\IC^2}(\epsilon_1,\epsilon_2,\mu;q;1) =
\sum_{\lambda}\,q^{|\lambda|}~\cZ_{\lambda}^{\IC^2}(
\epsilon_1,\epsilon_2;1)^{\rm gauge} ~
\cZ_{\lambda}^{\IC^2}(
\epsilon_1,\epsilon_2,\mu;1)^{\rm adj}
\label{ZinstC2U1N2star}\eeq
with
\beq
\cZ_{\lambda}^{\IC^2}(
\epsilon_1,\epsilon_2,\mu;1)^{\rm adj} =
\prod_{s\in \lambda}\,\Big( -L_{\lambda}(s)\,\epsilon_1+\big(
A_{\lambda}(s)+1\big)\,\epsilon_2 +\mu \Big) \,\Big(
\big(L_{\lambda}(s)+1\big)\,\epsilon_1- A_{\lambda}(s)\,\epsilon_2
+\mu \Big) \ ,
\eeq
and
\beq
\cZ_{\lambda}^{\IC^2}(
\epsilon_1,\epsilon_2;1)^{\rm gauge} = \frac1{\cZ_{\lambda}^{\IC^2}(
\epsilon_1,\epsilon_2,\mu=0;1)^{\rm adj}} \ .
\eeq
The factor involving the Jacobi elliptic function
is proportional to the rank one $\cN=4$ partition function
(\ref{ZU1Fp}), with the
factor $\hat\eta(q)^{-2}$ replaced by a product of two Nekrasov
functions on $\IC^2$. Using (\ref{ZC2N2U1}), the pure $\cN=2$ gauge theory partition function obtained from
(\ref{ZU1FpN2star}) is given by the simple expression
\beq
Z_{\rm inst}^{\IF_p}(\epsilon_1,\epsilon_2,\mu=\infty; q=0,Q;1)\big|_{\mu^2\,q=\Lambda} =
\theta_3(\Lambda^{1/p},Q)~ \exp\Big(\, \frac{p\, \Lambda}{\epsilon_2\,
  (\epsilon_2+p\, \epsilon_1)} \, \Big) \ .
\eeq
At the Calabi-Yau specialization
$\epsilon_1+\epsilon_2=0$ of the
$\Omega$-deformation, one has the remarkable combinatorial
identity~\cite[eq.~(6.12)]{Nekrasov:2003rj}
\beq
Z_{{\rm inst}}^{\IC^2}(-\varepsilon,\varepsilon,\mu;q;1) =
\hat\eta(q)^{\tilde\mu^2-1} \qquad \mbox{with} \quad
\tilde\mu=\frac\mu{\varepsilon} \ ,
\eeq
and hence for the $A_1$ ALE space the partition function
(\ref{ZU1FpN2star}) with $p=2$ at this locus
takes the simple form
\beq
Z_{\rm inst}^{A_1}(-\varepsilon,\varepsilon,\mu; q,Q;1)
= \hat\eta(q)^{2(\tilde\mu^2-1)} \ 
\theta_3(q^{1/2},Q)
\eeq
generalizing (\ref{ZU1Fp}).

The structure of the formula (\ref{N2starinst}) generalizes to give the
$\cN=2^*$ partition functions on the more complicated
toric resolutions of $A_{p,p'}$ singularities with $p'>1$, which can
be written as combinatorial expansions based on the toric diagram
$\Delta$ of $M_{p,p'}$ as in the case of the maximally supersymmetric
gauge theory. There are $\chi(M_{p,p'})=m+1$ copies of the
Nekrasov partition function on $\IC^2$ for each of the regular instantons
(``vertex'' contributions),
plus $b_2(M_{p,p'})=m$ contributions from the fractional instantons (``edge''
contributions) on each exceptional divisor of the orbifold
resolution, see~\cite[Prop.~5.11]{Gasparim} for the general symbolic formula; such ``blow-up formulas'' have been developed recently
in~\cite{Bonelli:2012ny} based on physical considerations from
M-theory within the context of the AGT correspondence. In contrast
to the pure $\cN=4$ gauge theory, the regular and fractional
instanton contributions do not completely decouple in this case. This
will be the generic situation for all (quiver) gauge theory partition
functions computed below. The
denominator factors arise from a localization integral over the
Euler class of the tangent bundle on the instanton moduli space,
while the numerator factors come from the Euler class of the vector bundles
of which the matter fields are local sections. In the rank one cases, the instanton
partition functions are always independent of the Higgs eigenvalues $a$,
and the sum over monopole numbers $u\in\IZ$ always factorizes out as
in~(\ref{ZU1FpN2star}) when the matter field content involves only adjoint sections.

\subsection{${\mathcal{N}=2}$,  ${{N}_f=2r}$ gauge theory\label{subsec:fundmatter}}

We will now compute
the instanton contributions to the rank $r$ partition function of
$\mathcal{N}=2$ supersymmetric Yang-Mills theory on $M$ with $N_f$
flavours in the fundamental matter field hypermultiplet; the value
$N_f=2r$ is
selected by the requirement of superconformal invariance of the $U(r)$
gauge theory. In this case, toric localization also involves the
maximal torus $\torus_{\vec\mu}$ of the $U(2r)$ flavour symmetry group
with equivariant parameters
$\vec\mu:=(\mu_1,\dots,\mu_{2r})$, where $\mu_f$, $f=1,\dots,2r$
are interpreted as
masses for fundamental matter fields. The gauge theory in this
instance is conjecturally dual to some two-dimensional field theory of conformal blocks on a
sphere with $2r$ punctures.

The Nekrasov instanton partition function on $\IC^2$ will appear as the
building blocks of those on $M$. They are given by 
\beq
Z_{{\rm inst}}^{\IC^2}(\epsilon, a,\vec\mu;q;r) =
\sum_{\vec \lambda}\,q^{|\vec
  \lambda\,|}~\cZ_{\vec\lambda}^{\IC^2}(\epsilon, a;r)^{\rm gauge}~
\prod_{f=1}^{2r} \,
\cZ_{\vec\lambda}^{\IC^2}(\epsilon,a,\mu_f;r)^{\rm fund} \ ,
\label{ZinstC2flav}\eeq
where
\beq
\cZ_{\vec\lambda}^{\IC^2}(\epsilon, a,\mu_f;r)^{\rm fund}=\prod_{l=1}^r~
\prod_{s\in \lambda^l}\,\big(a_l-L_{\lambda^l}^t(s)\,\epsilon_1
-A_{\lambda^l}^t(s)\,\epsilon_2+\mu_f \big)
\label{ZlambdaC2fund}\eeq
and the product in (\ref{ZinstC2flav}) is a representative for the
equivariant Euler class
$\eul_{\torus\times\torus_{\vec\mu}}\big(\scrV_{n,r}\otimes
V_{2r} \big)_{\vec\lambda}$ at the fixed point $\vec\lambda$, where
$\scrV_{n,r}$ is the Dirac bundle over the instanton moduli space
$\scrM^{\rm inst}_{n,r}(\IC^2)$ and $V_{2r}$ denotes the fundamental vector
representation of $U(2r)$.
The matter field
contributions are all written for the fundamental representation of
the $U(r)$ gauge group. If we wish the $f$-th
matter field to instead transform in the anti-fundamental
representation, then we should use the complex conjugate of the Dirac
bundle $\scrV_{n,r}$ of which the matter fields are sections. This amounts to
a shift $\mu_f\to\epsilon_1+\epsilon_2-\mu_f$ so that
\beq
\cZ_{\vec\lambda}^{\IC^2}(\epsilon, a,\mu_f;r)^{\overline{\rm fund}} =
\cZ_{\vec\lambda}^{\IC^2}(\epsilon, a,\epsilon_1+\epsilon_2-\mu_f;r)^{\rm fund}
\ .
\eeq

For the $A_{p,1}$ spaces, the instanton partition function can
again be expressed as a magnetic
lattice sum
\bea
Z_{\rm inst}^{\IF_p}(\epsilon, a,\vec \mu; q,Q;r) &=&
\sum_{\vec u\in\IZ^r}\, q^{\frac 1{2p}\,\sum_{l=1}^r\, u_l^2} \ Q^{u}~
\prod_{l\neq l'}\,\frac1{\ell^{\vec
  u}_{l,l'}(\epsilon,a)}~
\prod_{f=1}^{2r}~\prod_{l=1}^r\,\ell_l^{\vec
  u}(\epsilon, a,\mu_f) \nonumber \\ && \qquad \qquad \times~
Z_{{\rm inst}}^{\IC^2}\big(\epsilon_1\,,\,\epsilon_2\,,\,
a+ \epsilon_1\,\vec u\,,\,\vec\mu\,;q\,;\,r \big) \\ && \qquad \qquad \times~
Z_{{\rm
    inst}}^{\IC^2}\big(-\epsilon_1\,,\,\epsilon_2+p\,\epsilon_1\,,\,
a+ (\epsilon_2+p\,\epsilon_1)\,\vec
u\,,\,\vec\mu\,;\,q\,;\,r \big) \nonumber
\eea
where
\beq
\ell^{\vec u}_{l}(\epsilon,a, \mu_f) =\left\{
  \begin{array}{l} \displaystyle
~ \prod_{j=0}^{-u_l-1}~\prod_{i=0}^{p\, j}\,
\big(a_l-i\,\epsilon_1-j \,\epsilon_2+\mu_f\big) \ , \qquad
u_l< 0 \ , \\ {~~} \\ \displaystyle
~ \prod_{j=1}^{u_l}~\prod_{i=1}^{p\, j-1}\,
\big(a_l+i \,\epsilon_1+j
\,\epsilon_2+\mu_f \big) \ , \qquad
u_l>0 \ , \\ ~~ \\ \displaystyle
~ 1 \ , \qquad u_l =0 \ .
\end{array} \right.
\eeq
The
denominator factors above arise from a localization integral over the
Euler class of the tangent bundle on the instanton moduli space,
while the numerator factors come from the Euler classes of the Dirac
bundles for the matter fields.

Again in the rank one case $r=1$, everything is independent of the
Higgs equivariant parameters~\cite{Gasparim}, but in this case the partition
functions still mix contributions
from regular and fractional instantons as the monopole numbers shift
the mass parameters in (\ref{ZlambdaC2fund}). At the Calabi-Yau
specialization locus, one has the combinatorial
identity~\cite{Marshakov:2009gs}
\beq
Z_{\rm inst}^{\IC^2}(-\varepsilon,\varepsilon,\mu_1,\mu_2;q;1)=
(1-q)^{-\tilde\mu_1\, \tilde\mu_2} \qquad \mbox{with} \quad
\tilde\mu_f=\frac{\mu_f}{\varepsilon} \ ,
\eeq
and the rank one partition function for the $A_1$ ALE space can be
expressed in the simplified form
\bea
Z_{\rm inst}^{A_1}(-\varepsilon,\varepsilon,\mu_1,\mu_2; q,Q;1) &=&
\frac1{(1-q)^{2\,\tilde\mu_1\, \tilde\mu_2}} \ \sum_{u\in\IZ}\,
  \Big(\, \frac{\, \varepsilon^2\,q^{1/4} }{(1-q)^2} \, \Big)^{u^2} \ Q^u
  \\ && \qquad \qquad \qquad \qquad \times \ \prod_{j=1}^{|u|} \
  \prod_{i=1}^{2j-1}\, \big(i-j-\tilde\mu_1\big)\,
  \big(i-j-\tilde\mu_2 \big) \nonumber
\eea
where the products are defined to be~$1$ at $u=0$.

\subsection{Quiver gauge theories\label{subsec:N2quiver}}

We will now explain how to derive the partition functions of this
section. For this, we will study the moduli spaces of framed
instantons in some detail; an ADHM-type parametrization of this space
can be found in~\cite{rava}.
Let $\scrM^{\rm inst}_{n,\beta;r}(M)$ be the moduli space of
torsion free sheaves
$\cE$ on
$M=M_{p,1}$ of rank $r$, first Chern class (monopole number) $c_1(\cE)=\beta$, and second Chern
character (instanton number) ${\rm ch}_2(\cE)=n$, which are trivialized on a
line at infinity $\ell_\infty\cong \PP^1$ in $M$; it is proven
in~\cite{Gasparim,Bruzzo:2009uc,rava} that $\scrM^{\rm inst}_{n,\beta;r}(M)$ is a
smooth quasi-projective variety which gives a fine moduli space of
framed sheaves on $\IF_p$. There is a natural
action of the torus $\torus=\torus_\epsilon\times \torus_{a}$ on this moduli
space, where $\torus_\epsilon\cong \torus^2$ with equivariant parameters
$\epsilon= (\epsilon_1,\epsilon_2)$ is induced by the toric action on the
surface $M$, and $\torus_{a}\cong U(1)^r$ is the maximal torus of the
$U(r)$ gauge group with equivariant parameters the Higgs eigenvalues $a=(a_1,\dots,a_r)$. The $\torus$-fixed points $\cE\in \big(\scrM^{\rm inst}_{n,\beta;r}(M)\big)^\torus$ are given by sums of ideal sheaves of
divisors $D_l\subset M$ as
\beq
\cE = \cI_1(D_1)\oplus\cdots\oplus \cI_r(D_r) \ ,
\eeq
such that $\cI_l(D_l)=\cI_l\otimes\cO_M(D_l)$ for
some ideal sheaf $\cI_l$ of a zero-dimensional subscheme $Z_l$
of $M$; fractional instantons are
supported on $D_l$ with their monopole charges while regular
instantons are supported on $Z_l$. The $\torus_\epsilon$-invariant
zero-dimensional subschemes $Z_l$ correspond to vertices and the
$\torus_\epsilon$-invariant divisors $D_l$ to edges in the underlying
toric diagram $\Delta$ of $M$; for $M=M_{p,1}$ there are two vertices connected
together by a single edge. Such a fixed point is
therefore parametrized by a triple of vectors 
\beq
x=\big(\, \vec\lambda_1\,,\,\vec \lambda_2\,,\,\vec u\,\big) \ ,
\eeq
where $\vec \lambda_a=(\lambda^1_a,\dots,\lambda_a^r)$ for $a=1,2$ are Young diagrams corresponding
to the $Z_l$ and $\vec u=(u_1,\dots,u_r)\in\IZ^r$ with
$u_l=c_1(\cO_M(D_l))$ for $l=1,\dots,r$. These vectors are related to the
instanton charges through~\cite{Cirafici:2009ga}
\beq
\beta= u \qquad \mbox{and} \qquad n =
|\vec \lambda_1| +|\vec \lambda_2|
-\frac{1}{2p}\, \sum_{l=1}^r\, u_l^2
\eeq
with $|\vec \lambda_a|:=\sum_l \, |\lambda^l_a|$ and $u :=\sum_l\,
u_l$.

We shall derive the basic building blocks for all instanton
partition functions, which are called \emph{bifundamental weights}.
Following~\cite{Carlsson,Feigin}, we consider the virtual bundle
$\scrE_{n,\beta;r}$ over $\scrM^{\rm inst}_{n,\beta;r}(M)\times\scrM^{\rm inst}_{n,\beta;r}(M)$ with fibre over a
pair of torsion free sheaves $(\cE,\cE'\,)$ given by the cohomology group
\beq
\scrE_{n,\beta;r}\big|_{(\cE,\cE'\,)}=\Ext^1_{\cO_M}\big(\cE\,,
\,\cE'(-\ell_\infty)\big) \ ,
\eeq
where $\cE'(-\ell_\infty)=\cE'\otimes \cO_M(-\ell_\infty)$. By the
Kodaira-Spencer theorem, the restriction $\scrE_{n,\beta;r}\big|_\Delta$ to the
diagonal subspace $\Delta\subset \scrM^{\rm inst}_{n,\beta;r}(M)\times\scrM^{\rm inst}_{n,\beta;r}(M)$
coincides with the tangent bundle $T\scrM^{\rm inst}_{n,\beta;r}(M)$, the sections of
which are gauge and adjoint matter fields. If
$p_1:\scrM^{\rm inst}_{n,\beta;r}(M)\times\scrM^{\rm inst}_{n,\beta;r}(M)\to \scrM^{\rm inst}_{n,\beta;r}(M)$
denotes projection onto the first factor, then the push-forward
$p_{1*}\scrE_{n,\beta;r}$ coincides with the Dirac bundle $\scrV_{n,\beta;r}$, the sections of
which are fundamental matter fields.

We claim that $\scrE_{n,\beta;r}$ is a vector bundle of rank
$\dim(\scrM^{\rm inst}_{n,\beta;r}(M))$ on
$\scrM^{\rm inst}_{n,\beta;r}(M)\times\scrM^{\rm inst}_{n,\beta;r}(M)$. This follows from the
following facts, which generalize the vanishing theorems
of~\cite[Prop.~1]{Gasparim}:
\begin{itemize}
\item \ Since $\cE,\cE'$ are sheaves on a surface all of their $\Ext_{\cO_M}^i$
  groups vanish for $i>2$;
\item \ Since $\cE,\cE'$ have the same trivialization at $\ell_\infty$, their
  charges are the same: $\ch(\cE)=\ch(\cE'\,)$; in particular
  $c_1(\cE)=c_1(\cE'\,)$ and whence $\Hom_{\cO_M}(\cE,\cE'(-\ell_\infty))=0$;
\item \ Since $M=M_{2,1}$ is Calabi-Yau, by Serre duality one has
\beq
\Ext^2_{\cO_M}\big(\cE\,,\,\cE'(-\ell_\infty)\big)\cong
  \Hom_{\cO_M}\big(\cE'\,,\,\cE(-\ell_\infty) \big) =0 \ ,
\eeq
and similarly when $K_M\neq0$, as in~\cite{Gasparim,Bruzzo:2009uc}.
\end{itemize}
It follows that $\scrE_{n,\beta;r}$ is a vector bundle on
$\scrM^{\rm inst}_{n,\beta;r}(M)\times\scrM^{\rm inst}_{n,\beta;r}(M)$ of rank
\beq
\dim \Ext^1_{\cO_M}\big(\cE\,,
\,\cE'(-\ell_\infty)\big) = - \chi\big(\,\overline{\cE}\otimes
\cE'\otimes\cO_M(-\ell_\infty) \big) \ .
\eeq
The Euler characteristic here can be computed by the
Hirzebruch-Riemann-Roch theorem. Since $\ch(\cE)=\ch(\cE'\,)$, the
computation is exactly as in~\cite{Gasparim}, yielding
\beq
\ch_0(\scrE) = \dim \scrM^{\rm inst}_{n,\beta;r}(M) = 2\, r\,n+(r-1)\, \beta^2
\eeq
where $\beta^2=\int_M\,c_1(\cE)\wedge c_1(\cE)$ for $\cE\in \scrM^{\rm inst}_{n,\beta;r}(M)$.

On the ``double'' of the instanton moduli space
$\scrM^{\rm inst}_{n,\beta;r}(M)\times\scrM^{\rm inst}_{n,\beta;r}(M)$ there is a natural action of
the extended torus $\tilde \torus=\torus_\epsilon\times \torus_{a}\times \torus_{a'}$, which acts as $\torus_\epsilon\times \torus_{a}$ on the first
factor ($\torus_{a'}$ acting trivially) and as $\torus_\epsilon\times \torus_{a'}$ on the second
factor ($\torus_{a}$ acting trivially). We want to compute the
character of $\tilde \torus$ in $\scrE_{n,\beta;r}\big|_{(\cE,\cE'\,)}$ at a fixed point
$(\cE,\cE'\,)\in \big(\scrM^{\rm inst}_{n,\beta;r}(M)\times\scrM^{\rm inst}_{n,\beta;r}(M))^{\tilde
  \torus}$, corresponding to a pair of triples of vectors
\beq
x=\big(\, \vec \lambda_1\,,\,\vec \lambda_2\,,\,\vec u\,\big) \qquad \mbox{and}
\qquad x'=\big(\, \vec \lambda^{\,\prime}_1 \,,\,\vec \lambda^{\,\prime}_2\,,\,\vec
u\,' \,\big) \ .
\eeq
Recalling that $\cE=\cI_1(D_1)\oplus\cdots\oplus \cI_r(D_r)$ and
$\cE'=\cI_1'(D_1'\, )\oplus\cdots\oplus \cI_r(D_r'\,)$, we have
\bea
\Ch_{\tilde \torus}\, \scrE_{n,\beta;r}\big|_{(\cE,\cE'\,)} &=& \Ch_{\tilde \torus}\, \Ext^1_{\cO_M}\big(\cE\,,
\,\cE'(-\ell_\infty)\big) \nonumber\\[4pt] &=& 
-\sum_{l,l'=1}^r\, \Ch_{\tilde \torus}\,
\Ext^\bullet_{\cO_M}\big(\cI_l(D_l)\,,\,
\cI_{l'}'(D_{l'}'-\ell_\infty) \big) \nonumber\\[4pt] &=&
-\sum_{l,l'=1}^r\, e'_{l'}\, e_l^{-1} \ \Ch_{\torus_\epsilon}\,
\Ext^\bullet_{\cO_M}\big(\cI_l(D_l)\,,\,
\cI_{l'}'(D_{l'}'-\ell_\infty) \big) \ ,
\eea
where $e_l=\e^{\ii a_l}$ and $e_{l'}'=\e^{\ii a'_{l'}}$ for $l,l'=1,\dots,r$.

The computation of the $\torus_\epsilon$ character $\Ch_{\torus_\epsilon}\,
\Ext^\bullet_{\cO_M}\big(\cI_l(D_l)\,,\,
\cI_{l'}'(D_{l'}'-\ell_\infty) \big)$ is no different from the
calculations of~\cite{Gasparim,Bruzzo:2009uc} for the tangent bundle which has
$\cE'=\cE$: We only need to keep track of prime labels on all quantities
indexed by $l'$ here. Proceeding in this way, we arrive at
\beq
\Ch_{\tilde \torus}\, \scrE_{n,\beta;r}\big|_{(\cE,\cE'\,)} = \sum_{l,l'=1}^r\,
e_{l'}'\, e_l^{-1} \ \Big( M^{\IF_p}_{l,l'}(t_1,t_2)
+ L^{\IF_p}_{l,l'}(t_1,t_2)\Big) \ .
\eeq
The vertex contribution (by a calculation in \v{C}ech
cohomology) is given by~\cite[Prop.~5.1]{Gasparim}
\bea
M^{\IF_p}_{l,l'}(t_1,t_2) = t_2^{u_{l'}'-u_l} \
M^{\IC^2}_{\lambda_1^l,\lambda_1'{}^{l'}} (t_1,t_2)
+ t_1^{p\,(u_{l'}'-u_l)} \, t_2^{u_{l'}'-u_l}
   \ M^{\IC^2}_{\lambda_2^l,\lambda_2'{}^{l'}}
   \big(t_1^{-1}\,,\,t_1^p\, t_2 \big)
\eea
where $M^{\IC^2}_{\lambda,\lambda'}(t_1,t_2)$ is the weight
decomposition (\ref{MYC2}) of the equivariant character on $\IC^2$. The edge contribution (by the Grothendieck-Riemann-Roch
theorem) is given by~\cite[Prop.~5.5]{Gasparim}
\beq
L^{\IF_p}_{l,l'}(t_1,t_2) =\left\{
  \begin{array}{l} \displaystyle
~ \sum_{j=0}^{u_l-u_{l'}'-1}~\sum_{i=0}^{p\,j}\,
t_1^{-i}\, t_2^{-j} \ , \qquad
u_l>u_{l'}' \ , \\ ~~ \\ \displaystyle
~ \sum_{j=1}^{u_{l'}'-u_l}~\sum_{i=1}^{p\,j-1}\,
t_1^{i}\, t_2^{j} \ , \qquad
u_l<u_{l'}' \ , \\ ~~ \\ \displaystyle
~ 0 \ , \qquad u_l=u_{l'}' \ .
\end{array} \right.
\eeq

From these formulas we can read off the weights for the action of the
torus $\tilde
\torus\times \torus_\mu$, with $\torus_\mu\cong U(1)$ the flavour
symmetry group with equivariant parameter $\mu$, which defines the equivariant Euler class of
the bundle $\scrE_{n,\beta;r}$; we refer to it as a ``bifundamental
weight''. We incorporate the orbifold compactification on
$\ell_\infty/G_{p,1}$ discussed in~\cite{Bruzzo:2009uc} which is
analogous to that described for ALE spaces in \S\ref{subsec:ALEmodsp};
it multiplies divisors by the intersection number $p$ under the linear equivalences
of~\cite{Cirafici:2009ga} which incorporate the contributions from
non-compact divisors (see~\cite[Lem.~4.1]{Bruzzo:2009uc}). In this way
we arrive at the bifundamental weight
\bea
&& \cZ^{\IF_p}\Big[{\scriptsize\begin{matrix} \vec \lambda_1 & \vec \lambda_2 & \vec
  u\\ \vec \lambda_1^{\,\prime} & \vec \lambda^{\,\prime}_2 & \vec
  u\,'\end{matrix}}\Big](\epsilon;a,a';\mu)^{\rm bif} \nonumber \\ &&
\qquad\qquad \ = \ \prod_{l,l'=1}^r\,
\ell_{l,l'}^{\vec u,\vec u\,'}(\epsilon;a,
a';\mu) \
\cZ^{\IC^2|l,l'}_{\vec \lambda_1,\vec
  \lambda^{\,\prime}_1}\big(\epsilon_1\,,\,\epsilon_2\,;\, a+ \epsilon_1\, \vec u\,, \, 
a'+ \epsilon_1\,\vec u\,'\,;\, \mu \big)^{\rm bif}
\label{ZbifFp} \\ && \qquad\qquad\qquad\qquad\times \
\cZ^{\IC^2|l,l'}_{\vec \lambda_2,\vec
  \lambda^{\,\prime}_2}\big(-\epsilon_1\,,\,\epsilon_2+p\,\epsilon_1 \,;\, 
a+ (\epsilon_2+p\,\epsilon_1) \, \vec u\, ,\, 
a'+(\epsilon_2+p\, \epsilon_1)\,\vec u\,'\,;\, \mu)^{\rm bif} \ , \nonumber
\eea
where
\beq
\ell^{\vec u,\vec u\,'}_{l,l'}(\epsilon;
a,a';\mu)=\left\{
  \begin{array}{l} \displaystyle
~ \prod_{j=0}^{u_l-u'_{l'}-1}~\prod_{i=0}^{p\,j}\,
\big(a_{l'}'-a_l-i\,\epsilon_1-j\,\epsilon_2 +\mu \big) \ , \qquad
u_l>u'_{l'} \ , \\ ~~ \\ \displaystyle
~ \prod_{j=1}^{u'_{l'}-u_l}~\prod_{i=1}^{p\,j-1}\,
\big(a'_{l'}-a_l+i\,\epsilon_1+j\,\epsilon_2+\mu \big) \ , \qquad
u_l<u'_{l'} \ , \\ ~~ \\ \displaystyle
~ 1 \ , \qquad u_l=u'_{l'} \ ,
\end{array} \right.
\eeq
and
\bea
\cZ^{\IC^2|l,l'}_{\vec \lambda,\vec
  \lambda^{\,\prime}}(\epsilon;a,
a';\mu)^{\rm bif} &=& \prod_{s\in \lambda^l}\,\Big(
a'_{l'}-a_l-L_{\lambda'\,^{l'}}(s)\,\epsilon_1+\big(
A_{\lambda^l}(s)+1\big)\,\epsilon_2 +\mu \Big) \label{ZbifC2} \\ && \times~
\prod_{s'\in \lambda'\,^{l'}}\,\Big(a'_{l'}-a_l +
\big(L_{\lambda^l}(s'\,)+1\big)\,\epsilon_1- A_{\lambda'\,^{l'}}(s'\, )\,\epsilon_2
+\mu \Big) \ . \nonumber 
\eea

The bifundamental weights give the contributions of a matter field of mass
$\mu$ in the
bifundamental hypermultiplet which is a section of the bundle
$\scrE_{n,\beta;r}$, and hence is charged under the group $U(r)\times
U(r)$. They are the building blocks for all
instanton partition functions. The basic weights that we use are given by
\bea
Z^{\IF_p}_{\rm adj}[\vec \lambda_1,\vec \lambda_2,\vec u\, ](\epsilon; a;\mu) &=& \cZ^{\IF_p}\Big[{\scriptsize\begin{matrix} \vec \lambda_1 & \vec \lambda_2 & \vec
  u\\ \vec \lambda_1 & \vec \lambda_2 & \vec u\end{matrix}}\Big](\epsilon; a,
a;\mu)^{\rm bif} \ , \nonumber\\[4pt]
Z^{\IF_p}_{\rm gauge}[\vec \lambda_1,\vec \lambda_2,\vec u\, ](\epsilon; a) &=& Z^{\IF_p}_{\rm adj}[\vec
\lambda_1,\vec \lambda_2,\vec u\, ](\epsilon; a;0) \ , \nonumber\\[4pt]
Z^{\IF_p}_{\rm fund}[\vec \lambda_1,\vec \lambda_2,\vec u\, ](\epsilon; a;\mu) &=& \cZ^{\IF_p}\Big[{\scriptsize\begin{matrix} \vec \lambda_1 & \vec \lambda_2 & \vec
  u\\ \emptyset & \emptyset & \vec 0 \end{matrix}}\Big](\epsilon; a,
0;\mu)^{\rm bif} \ , \nonumber\\[4pt]
Z^{\IF_p}_{\overline{\rm fund}}[\vec \lambda_1,\vec \lambda_2,\vec
  u\, ](\epsilon; a;\mu) &=& Z^{\IF_p}_{\rm fund}[\vec \lambda_1,\vec
  \lambda_2,\vec u\, ](\epsilon; a;\epsilon_1+\epsilon_2-\mu)
  \ .
\eea
The instanton partition functions on $M$ are generalizations
of (\ref{partfunct}) given by generating functions for the
stratification $\scrM^{\rm inst}_r(M)=\bigsqcup_{n,\beta}\, \scrM^{\rm inst}_{n,\beta;r}(M)$
of the symbolic form
\beq
Z_{\rm inst}^{M}(\epsilon,a,\vec \mu;q,Q;r)= \sum_{n,\beta}\, q^n
\ Q^\beta \ \oint_{\scrM^{\rm inst}_{n,\beta;r}(M)} \, {\rm
  char}_{\torus\times\torus_{\vec\mu}} \ ,
\eeq
which involve equivariant integrals over characteristic classes ${\rm
  char}_{\torus\times\torus_{\vec\mu}}$ obtained by suitable
restrictions of the Euler classes
$\eul_{\tilde\torus\times\torus_{\vec\mu}}(\scrE_{n,\beta;r})$ on the
instanton moduli space. After application of the
Atiyah-Bott localization formula for
$M=M_{p,1}$, the partition functions take the generic forms
\beq
Z_{\rm inst}^{\IF_p}(\epsilon,a,\vec \mu;q,Q;r)=
\sum_{\vec \lambda_1,\vec \lambda_2}\, q^{|\vec \lambda_1|+|\vec \lambda_2|}~ \sum_{\vec
  u\in\IZ^r}\, q^{\frac 1{2p}\,\sum_{l=1}^r\,u_l^2} \ Q^{u} \
\cW_\cN^{\IF_p}(\epsilon, a,\vec \mu\, )[\vec \lambda_1,\vec \lambda_2,\vec u\,] \ ,
\eeq
where the weights $\cW_\cN^{\IF_p}(\epsilon, a,\vec \mu\, )[\vec \lambda_1,\vec \lambda_2,\vec u\,]$ arise
from localization integrals of ratios of Euler classes of the tangent
and Dirac bundles over the instanton moduli space at the fixed points,
and they depend on the particular quiver gauge theory in question
which has
$\cN$ supersymmetries. The basic
examples we have considered include
\bea
\cW^{\IF_p}_{\cN=4}&=&1 \ , \nonumber\\[4pt]
\cW^{\IF_p}_{\cN=2^*}(\epsilon, a,\mu)[\vec \lambda_1,\vec
\lambda_2,\vec u\,] &=& \frac{Z_{\rm adj}^{\IF_p}[\vec \lambda_1,\vec \lambda_2,\vec
  u\,](\epsilon; a;\mu)}{Z_{\rm gauge}^{\IF_p}[\vec \lambda_1,\vec
  \lambda_2,\vec u\,](\epsilon; a)} \ , \nonumber\\[4pt]
\cW^{\IF_p}_{\cN=2,N_f\leq r}(\epsilon, a,\vec \mu\, )[\vec
\lambda_1,\vec \lambda_2,\vec u\,] &=& \cZ^{\IF_p}\Big[{\scriptsize\begin{matrix} \vec \lambda_1 & \vec \lambda_2 & \vec
  u\\ \emptyset & \emptyset & \vec 0
\end{matrix}}\Big](\epsilon; a,-\vec \mu;0)^{\rm bif} \ ,
\eea
which respectively reproduce the partition functions of
\S\ref{subsec:VafaWitten}, \S\ref{subsec:N2star} and
\S\ref{subsec:fundmatter}, and
\beq
\cW^{\IF_p}_{\cN=2}(\epsilon,a)[\vec
\lambda_1,\vec \lambda_2,\vec u\,] = \lim_{\mu\to\infty} \,
\frac1{\mu^{2( |\vec\lambda_1| + |\vec\lambda_2|)}} \, \cW^{\IF_p}_{\cN=2^*}(\epsilon, a,\mu)[\vec \lambda_1,\vec
\lambda_2,\vec u\,] = \frac1{Z_{\rm gauge}^{\IF_p}[\vec \lambda_1,\vec
  \lambda_2,\vec u\,](\epsilon; a)}
\eeq
for the pure $\cN=2$ supersymmetric Yang-Mills theory on $M_{p,1}$.

\subsection{Representations of affine algebras\label{subsec:eqcohinst}}

Our equivariant partition functions may be formulated in a purely algebraic
fashion that could help to elucidate the existence of natural
geometric representations of infinite-dimensional Lie algebras on the
cohomology of the instanton moduli space. Such representations were already discussed in
\S\ref{subsec:affinechar} in the case of ALE spaces, and we will now
sketch how they may arise in the classes of toric surfaces
considered in this section.

Let us begin by recalling some facts from equivariant localization theory that we will
use in the following. Let $\scrM$ be a non-compact manifold acted upon
by the torus $\torus$ with finitely many isolated fixed points
$\scrM^\torus$, and let
$H_\torus^\bullet(\scrM)=H_\torus^\bullet(\scrM,\IC)$ be its
$\torus$-equivariant cohomology.
Let $H=H_\torus^\bullet(\scrM)\otimes_{H_\torus^\bullet({\rm
    pt})}H_\torus^\bullet({\rm pt})_{\rm frac}$, where $H_\torus^\bullet({\rm pt})\cong
\IC[\epsilon_1,\epsilon_2,a_1,\dots,a_r]$ and $H_\torus^\bullet({\rm
  pt})_{\rm frac}$ is the localization of the ring $H_\torus^\bullet({\rm
  pt})$ to its field of
fractions, i.e. at the maximal ideal generated by
$\epsilon_1,\epsilon_2,a_1,\dots,a_r$. Then the localization theorem
in equivariant cohomology
states that the restriction map
\beq
H_\torus^\bullet(\scrM)\otimes_{H_\torus^\bullet({\rm
    pt})}H_\torus^\bullet({\rm pt})_{\rm frac} \ \longrightarrow \
H^\bullet(\scrM^\torus\,)\otimes_{\IC }H_\torus^\bullet({\rm pt})_{\rm frac}
\eeq
is an isomorphism. Let $\iota_x:x \hookrightarrow \scrM$ be the
inclusion of a fixed point $x\in\scrM^\torus$. Then the push-forwards
$|x\rangle\!\rangle_{a}:=(\iota_x)_*1$ form a basis for the equivariant cohomology group
$H$ as a vector space over the field of fractions $H_\torus^\bullet({\rm
  pt})_{\rm frac}$. The complex vector space $H$ becomes a Hilbert space upon
introducing an inner product $\langle-,-\rangle_\torus: H\otimes H\to
H_\torus^\bullet({\rm pt})_{\rm frac}$ defined by equivariant integration
\beq
\langle \alpha,\beta\rangle_\torus = \oint_\scrM\, \alpha\wedge\beta:= (-1)^m\, \sum_{x\in\scrM^\torus}\
\frac{\iota_x^*(\alpha\wedge \beta)}{\eul_\torus(T_x\scrM)} \ ,
\label{Tinnerprod}\eeq
where $2m=\dim(\scrM)$ and the equivariant Euler class of the tangent
bundle of $\scrM$ can be computed in terms of the weights of the torus action on
$T_x\scrM$ as $\eul_\torus(T_x\scrM)= \prod_{w\in T_x\scrM}\, w$. The middle-dimensional (non-localized)
equivariant cohomology $H_\torus^{\rm
  mid}(\scrM)$ is a vector space of dimension equal to the number of
fixed points $|\scrM^\torus|$, and the restriction of $\langle-,-\rangle_\torus$ to
$H_\torus^{\rm
  mid}(\scrM)$ is
non-degenerate and $\IC$-valued. Moreover, by the localization theorem
and standard properties of equivariant
integration, the classes $|x\rangle\!\rangle_{a}$ for
$x\in\scrM^\torus$ form an orthogonal basis for $H_\torus^{\rm
  mid}(\scrM)$ with
\beq
{}_{a}\langle\!\langle x|y\rangle\!\rangle_{a} = (-1)^m\,
\sum_{x'\in\scrM^\torus}\, \frac{\iota_{x'}^*\big((\iota_x)_*1\wedge
(\iota_y)_*1\big)}{\eul_\torus(T_{x'}\scrM)} = \delta_{x,y}~ \eul_\torus(T_x\scrM)^{-1} \ .
\eeq

We apply these facts to the instanton moduli space $\scrM=\scrM^{\rm inst}_r(M)= \bigsqcup_{n,\beta}\,
\scrM^{\rm inst}_{n,\beta;r}(M)$ and the Hilbert space
\beq
\calH_{ a}=\bigoplus_{n,\beta}\,
H_\torus^\bullet\big(\scrM^{\rm inst}_{n,\beta;r}(M) \big)
\otimes_{\IC}
H_\torus^\bullet({\rm pt})_{\rm frac} \ .
\eeq
Fixed points $x\in\scrM^\torus$ are then parametrized by \emph{all} triples
$(\vec \lambda_1,\vec \lambda_2,\vec u\,)$ from
\S\ref{subsec:N2quiver}, and the inner products of fixed point
states $|x\rangle\!\rangle_{a}=|\vec \lambda_1,\vec \lambda_2,\vec
u\,\rangle\!\rangle_{a}$ are given by
\beq
{}_{a}\langle\!\langle \vec \lambda_1,\vec \lambda_2,\vec
u\,|\vec \lambda_1^{\,\prime},\vec \lambda_2^{\,\prime},\vec
u\,'\,\rangle\!\rangle_{a} = Z^{\IF_p}_{\rm gauge}[\vec \lambda_1,\vec \lambda_2,\vec
u\,](\epsilon; a)^{-1} \ \delta_{\vec \lambda_1,\vec \lambda_1^{\,\prime}}\,
\delta_{\vec \lambda_2, \vec \lambda_2^{\,\prime}}\, \delta_{\vec
u, \vec u\,'} \ .
\eeq
We also introduce operators on $\calH_{a}$ whose eigenvalues give the grading into
instanton numbers $n$ and first Chern classes $\beta$ through
\beq
L_0|\vec \lambda_1,\vec \lambda_2,\vec
u\,\rangle\!\rangle_{a} = \Big(|\vec \lambda_1|+|\vec
\lambda_2|+\frac1{2p}\, \sum_{l=1}^r\, u_l^2\Big)\, 
|\vec \lambda_1,\vec \lambda_2,\vec
u\,\rangle\!\rangle_{a} \qquad \mbox{and} \qquad J_0|\vec \lambda_1,\vec \lambda_2,\vec
u\,\rangle\!\rangle_{a} = u \, |\vec \lambda_1,\vec \lambda_2,\vec
u\,\rangle\!\rangle_{a} \ .
\label{gradingops}\eeq
Finally, following~\cite{Carlsson,Alday} we define intertwining operators determined by the bifundamental
hypermultiplet of fields through
\bea \nonumber 
\Phi_{a,a'}^\mu & : &  \calH_{a'}\ \longrightarrow \ \calH_{a} \ ,
\\[4pt] 
\Phi_{a,a'}^\mu |\vec \lambda_1,\vec \lambda_2,\vec
u\,\rangle\!\rangle_{a'} &=& \sum_{\vec
  \lambda_1^{\,\prime},\vec \lambda_2^{\, \prime},\vec
u\, '}\, \frac{\cZ^{\IF_p}\Big[{\scriptsize\begin{matrix} \vec \lambda_1 & \vec \lambda_2 & \vec
  u\\ \vec \lambda_1^{\,\prime} & \vec \lambda_2^{\,\prime} & \vec u\,'\end{matrix}}\Big](\epsilon; a,
a';\mu)^{\rm bif}}{Z^{\IF_p}_{\rm gauge}[\vec \lambda_1,\vec \lambda_2,\vec
u\, ](\epsilon; a'\,)} \ |\vec \lambda_1^{\,\prime},\vec \lambda_2^{\,\prime},\vec
u\,'\,\rangle\!\rangle_{a} \ .
\label{bifundop}\eea
Then any quiver gauge theory partition function can be expressed in
terms of suitable matrix elements of combinations of all these operators. For
example, one has the trace formulas
\beq
Z_{\rm gauge}^{\IF_p}(q,Q;r) = \Tr_{\calH_{a}}\big(q^{L_0} \ Q^{J_0} \big) \qquad \mbox{and} \qquad 
Z_{\rm inst}^{\IF_p}(\epsilon, a,\mu;q,Q;r) =
\Tr_{\calH_{a}}\big(\Phi^\mu_{a,a}~q^{L_0} \ Q^{J_0} \big)
\eeq
for the instanton partition functions of the $\cN=4$ and $\cN=2^*$
gauge theories, respectively.

The simplifications of the rank one partition functions can also be
understood from the ensuing simplifications of the equivariant
cohomology of the instanton moduli space. For $r=1$
the localization torus is $\torus=\torus_\epsilon=\torus^2$ and the
natural factorization of the strata for rank one torsion free sheaves
in (\ref{instmodfact}) implies that
$\scrM=\Gamma\times\bigsqcup_{n\geq0}\, \Hilb_n(M)$, where the
magnetic charge lattice $\Gamma$ is the Picard group of line bundles ${\rm Pic}(M)=H^2(M,\IZ) \cong\IZ$ on
$M=M_{p,1}$. The
corresponding equivariant cohomology group is the Hilbert space
\beq
\calH= \cF_M \otimes\IC[\Gamma] \qquad \mbox{with} \quad
\cF_M=\bigoplus_{n=0}^\infty\, \cF_n := \bigoplus_{n=0}^\infty\,
H_\torus^\bullet\big(\Hilb_n(M) \big)\otimes_{\IC[\epsilon_1,\epsilon_2]}
\IC(\epsilon_1,\epsilon_2) \ ,
\label{rank1Hilbert}\eeq
whose fixed point basis states factorize according to $|\lambda_1,\lambda_2,
u\rangle\!\rangle= |\lambda_1,\lambda_2\rangle \otimes |u\rangle$ with
$|\lambda_1,\lambda_2\rangle\in H_\torus^{4n}\big(\Hilb_n(M)\big)$
when $|\lambda_1|+|\lambda_2|=n$.

We will now elucidate the geometrical and algebraic significance of
the bifundamental operators (\ref{bifundop}) in this case by regarding
(\ref{rank1Hilbert}) as the Hilbert space of a free boson representation. For this,
recall that there is a natural geometric action on the vector space
$\cF_{M}$ of the affine $u(1)\oplus u(1)$
algebra associated to the cohomology lattice $H_c^\bullet(M,\IZ)\cong\IZ^2$, with
the intersection product~\cite{grojnowski,nakaHeis}.
To any cohomology class $\gamma\in H_c^{\bullet}(M,\IZ)$ one can
associate a geometrically defined Nakajima operator
\beq
\alpha_{-m}(\gamma) \,:\, H_\torus^\bullet\big(\Hilb_n(M)\big) \ \longrightarrow \
H_\torus^{\bullet+2m+{\rm deg}(\gamma)-2}\big(\Hilb_{n+m}(M)\big)
\eeq
for $m>0$; this operator is defined by its matrix elements with
respect to the inner product (\ref{Tinnerprod}) as
\beq
\big\langle\alpha_{-m}(\gamma) \eta\,,\,\xi\big\rangle_\torus := \int_{\mathscr{Z}^{n,n+m}(\gamma)}\, \eta\wedge \xi
\eeq
for $\xi\in\cF_n$ and $\eta\in\cF_{n+m}$, where $\mathscr{Z}^{n,n+m}(\gamma)\subset \Hilb_n(M)\times \Hilb_{n+m}(M)$ is the incidence variety of relative ideal sheaves supported at a single point of $M$ lying on the Poincar\'e dual cycle to $\gamma$ in $M$. We also define
\beq
\alpha_m(\gamma)=(-1)^{m+1}\, \alpha_{-m}(\gamma)^\dag
\eeq
where the adjoint operator on $\cF_{M}$ is defined with respect to the
inner product (\ref{Tinnerprod}). 
The celebrated result of Nakajima~\cite{nakaHeis} is that these operators satisfy the commutation relations of the Heisenberg
algebra
\beq
\big[\alpha_m(\gamma)\,,\,\alpha_{m'}(\gamma'\,)\big]=m \
\big\langle\gamma \,,\, \gamma'\, \big\rangle_{\Gamma} \
\delta_{m+m',0} \ .
\eeq
By G\"ottsche's formula for the Poincar\'e polynomials of the Hilbert
schemes of points $\Hilb_n(M)$, this representation of
$\widehat{u}(1)\oplus \widehat{u}(1)$
is irreducible; hence $\cF_{M}$ is the bosonic Fock space representation, with
vacuum vector
\beq
|0\rangle \ := \ 1\in H_\torus^0\big(\Hilb_0(M)\big) \ .
\eeq

Since points and divisors on $M$ have vanishing intersection product, 
there is a natural splitting of the cohomology lattice $H_c^\bullet(M,\IZ)= H^0(M,\IZ)\oplus \Gamma$. For the generator of the degree zero cohomology we can take any of the two torus fixed points $v_1,v_2\in M$; we will choose $H^0(M,\IZ)=\IZ[v_2]$ and denote
\beq
\alpha_{-m}^2:=\alpha_{-m}(v_2) \,:\, H_\torus^\bullet\big(\Hilb_n(M)\big) \ \longrightarrow \
H_\torus^{\bullet+2m-2}\big(\Hilb_{n+m}(M)\big) \ .
\eeq
These operators satisfy
\beq
\big[\alpha_m^2\,,\,\alpha^2_{m'} \big]= m \ \delta_{m+m',0} \ .
\eeq

The operators generated by divisors $[D] \in H_c^{\rm mid}(M,\IZ)={\rm Pic}(M)$ act on the middle-dimensional cohomology
\beq
\cF_{M}^{\rm mid} := \bigoplus_{n=0}^\infty\,
H_\torus^{2n}\big(\Hilb_n(M) \big) \qquad \mbox{with} \quad
\alpha_{-m}(D)\,:\, H_\torus^{2n}\big(\Hilb_n(M) \big) \ \longrightarrow \
H_\torus^{2(n+m)}\big(\Hilb_{n+m}(M) \big) \ .
\eeq
Denote the Nakajima operators associated to the $\torus$-invariant integral generator
$e^1$ of the Picard group ${\rm Pic}(M)$ by
$\alpha_m^1:= \alpha_m(e^1)$; they
satisfy
\beq
\big[\alpha_m^1\,,\,\alpha_{m'}^1\big]=- \mbox{$\frac mp$} \
\delta_{m+m',0} \qquad \mbox{and} \qquad
\big[\alpha_m^1\,,\,\alpha_{m'}^2\big] = 0 \ .
\eeq
We denote $\alpha_m(u):=\alpha_m(u\,e^1)$ for $u\in
\IZ$ and $m\in\IZ$, so that
\beq
\big[\alpha_m(u)\,,\,\alpha_{m'}(u'\,)\big]=- \mbox{$\frac mp$} \ u\, u' \
\delta_{m+m',0} \ .
\eeq

This decomposition coincides with the isomorphism induced by the localization theorem
\beq
\cF_{M} \cong \big(\cF_{T_{v_1}M}\otimes \cF_{T_{v_2}M}\big)
\otimes_{\IC[\epsilon_1,\epsilon_2]} \IC(\epsilon_1,\epsilon_2) 
\eeq
and the parallel factorization stemming from
\beq
\alpha_m(\gamma)=\alpha_m\big(\gamma(v_1)\big)\otimes 1+ 1\otimes \alpha_m\big(\gamma(v_2)\big) \ ,
\eeq
where each factor corresponds to the instanton moduli space on
$T_{v_a}M\cong\IC^2$ and the one-dimensional oscillator algebra. This
map is easiest to describe in the fixed point basis: Fixed points of
$\Hilb_n(M)$ are labelled by assigning a Young diagram $\lambda_a$ to
each fixed point $v_a\in M$, describing the ideal sheaf $\cI_a$ of
$v_a$ for $a=1,2$, and the image of this point is $[\cI_1]\otimes
[\cI_2]$. In particular, the partition basis states can be written as
$|\lambda_1,\lambda_2\rangle= |\lambda_1\rangle\otimes
|\lambda_2\rangle$, where the natural basis for
$\cF_{\IC^2}=\bigoplus_{n\geq0}\, H^\bullet\big(
\Hilb_n(\IC^2)^\torus\,)$ is given by
\beq
|\lambda\rangle = \frac1{\mathfrak{z}(\lambda)}\, \prod_{i=1}^{\ell(\lambda)}\,
\alpha_{-\lambda_i}|0\rangle \qquad \mbox{with} \quad
\mathfrak{z}(\lambda) = \prod_{s\geq1}\, m_s(\lambda)! \ s^{m_s(\lambda)}
\eeq
and $m_s(\lambda)$ is the number of parts of the partition $\lambda= (\lambda_1,\dots,\lambda_{\ell(\lambda)})$
with $\lambda_i=s$; the \emph{Nakajima basis element} corresponding to
$|\lambda\rangle$ is the cohomological dual to the class of the
subvariety of $\Hilb_{|\lambda|}(\IC^2)$ with generic element a union
of subschemes of lengths $\lambda_1,\dots,\lambda_{\ell(\lambda)}$ supported at $\ell(\lambda)$ points
in $\IC^2$. Note that the standard inner product in $\torus$-equivariant
cohomology induces a \emph{non-standard} inner product on Fock space
after extension of scalars. The Fock space $\cF_{\IC^2}$ can be
naturally identified with the ring of symmetric polynomials in
infinitely many variables such that the Nakajima operators are
represented as multiplication by the power sum symmetric
functions. Under this correspondence
$|\lambda\rangle=J_\lambda^{(1/ \alpha)}\in\cF_{\IC^2} \otimes \IC[\epsilon_1,\epsilon_2]$
is the integral form of the Jack polynomial with parameter
$\alpha=-\epsilon_1/\epsilon_2$; at the Calabi-Yau locus
$\alpha=1$ the Jack polynomials specialize to Schur functions.
This factorization essentially reduces the description to
that of two copies of the $\IC^2$ results. In particular, using the
factorization (\ref{ZbifFp}) of the bifundamental weights on $M_{p,1}$
in terms of the bifundamental weights (\ref{ZbifC2}) on $\IC^2$, the
operator
\beq
\Phi^\mu(z):= z^{L_0} \ \Phi^\mu \ z^{-L_0} \ , \qquad z\in\IC
\eeq
defined by (\ref{bifundop}) can be written in terms of products of two $U(1)$
Carlsson-Okounkov free boson vertex operators~\cite{Carlsson} on
$\cF_{\IC^2}\otimes\cF_{\IC^2}$.

The tensor product with the group algebra $\IC[\Gamma]$ gives the Fock space $\cF_M^{\rm mid}$ a $\Gamma$-grading into subspaces with fixed $U(1)$ charge $u\in \Gamma$. The elements $|u\rangle\in\IC[\Gamma]$ are called ``zero-mode states'' and
they may be identified geometrically with the Chern characteristic
classes $\ch(\cO_{\IF_p}(u\,e^1))=\exp\big(c_1(\cO_{\IF_p}(u\,
e^1))\big)$; the group operation on $\IC[\Gamma]$ then corresponds to the
tensor product of line bundles in the Picard group ${\rm Pic}(M)$. The
grading operator $J_0$ introduced in (\ref{gradingops}) commutes with
all Nakajima operators $\alpha_m(\gamma)$, and
$\alpha_m^a|u\rangle=0$ for $a=1,2$. We may thus write the
$\torus$-equivariant cohomology of the instanton moduli space in terms
of its $\Gamma$-grading as
\beq
\calH= \bigoplus_{u\in \Gamma}\, \calH_{u} \qquad \mbox{with} \quad
\calH_{u}= \IC\big[\alpha_{-m}(\gamma) \ \big| \ m>0 \ , \ \gamma\in
H_c^\bullet(M,\IZ)\big]|u\rangle \ ,
\eeq
where $J_0\big|_{\calH_{u}}=u\ \Id_{\calH_{u}}$ is the $U(1)$
charge and $L_0=\frac1{2p}\,
J_0^2+ \sum_{m>0} \, \alpha_{-m}(\gamma)\, \alpha_m(\gamma)$ is the energy operator.

Thus the cohomology $\calH$ of the instanton moduli space carries an action of both the Heisenberg
Lie algebra $\widehat{u}(1)\oplus\widehat{u}(1) $ and the group
algebra $\IC[\Gamma]$. In fact, this implies that a much larger algebraic
entity acts on $\cF_{M}^{\rm mid}\otimes\IC[\Gamma]$, as it has the
structure of a \emph{vertex operator algebra}. For $p=2$, the space of vectors of
conformal dimension one in this algebra is naturally identified with the
simple Lie algebra $su(2)$, and $\cF_{M}^{\rm mid}\otimes\IC[\Gamma]$ is
the basic Fock space representation of $\widehat{su}(2)_1$; in this
case the
equivariant partition functions are matrix elements of operators in
this $\widehat{su}(2)_1$-module, and
this gives a geometric explanation for the appearence of affine
characters in \S\ref{subsec:affinechar}. The algebraic basis for this
equivalence is provided by Frenkel-Kac construction~\cite{nagaoFK}.

\bigskip

\section{Six-dimensional cohomological gauge theory\label{6dcohgt}}

\subsection{Instanton moduli spaces}

We now return to our original setting of Donaldson-Thomas theory from \S\ref{sec:curveCY3}. Donaldson-Thomas invariants have their geometric origin in the study
of the moduli space of holomorphic bundles (or better coherent torsion free
sheaves) on a Calabi-Yau threefold $X$. In this section we will see how
under this perspective the enumerative problem can be reformulated as
an instanton counting problem. The idea is to take
the perspective of the D6-branes and its topological effective field
theory. This effective field theory is described by a certain gauge
theory which sees the D2 and D0 branes bound to the D6-brane as
topologically non-trivial ground states of the worldvolume gauge
theory~\cite{Douglas:1995bn}. From this point of view Donaldson-Thomas
invariants count generalized instanton configurations of this gauge
theory: Computing Donaldson-Thomas invariants on a toric
Calabi-Yau manifold is precisely a higher-dimensional generalization
of instanton counting in four-dimensional supersymmetric Yang-Mills
theory that was discussed extensively in previous sections.

In the large radius phase, we can study bound states of $r$ D6-branes
with D2--D0 branes on $X$ using the Dirac-Born-Infeld theory on the
D6-brane worldvolume. For local K\"ahler threefolds $X$, the gauge theory
describing the low-energy excitations of the D6-branes is a
topological twist of maximally supersymmetric Yang-Mills theory in six
dimensions with gauge group
$U(r)$~\cite{Blau:1997pp,Acharya:1997gp,Hofman:2000yx}. Its bosonic
field content consists of a gauge field $A$ corresponding to a unitary connection
$\nabla_A=\dd+\ii A$ on a $U(r)$ vector bundle $\cE\to X$ whose curvature two-form $F_A=\dd A+A\wedge A$
has the complex decomposition $F_A=F_A^{2,0}+F_A^{1,1}+F_A^{0,2}$, a complex Higgs field
$\phi$ which is a local section of the adjoint bundle $\mathrm{ad}\,
\cE$ of $\cE$, and a
$(3,0)$-form $\omega\in \Omega^{3,0} ( X , \mathrm{ad}\, \cE)$, together with various other
fields which together define a six-dimensional gauge theory with $\cN=2$
supersymmetry. The gauge theory is parametrized by the complex
K{\"a}hler $(1,1)$-form $t=B+\ii J$ of $X$ and the six-dimensional
theta-angle which is identified with the topological string
coupling~$\lambda=g_s$. 

This gauge theory has a BRST symmetry and hence
localizes onto the moduli space $\scrM^{\rm inst}_r(X)$ of solutions of the fixed point
equations
\begin{eqnarray} F_A^{2,0} &=&
  \overline{\partial}_A{}^\dag\, \omega
  \ , \nonumber\\[4pt]
F_A^{1,1} \wedge t \wedge t + \omega\wedge\overline{\omega} &=& u_\cE \ t\wedge t\wedge t \ , \nonumber\\[4pt] \nabla_A \phi &=& 0 \ ,
\label{inste}\end{eqnarray}
where $u_\cE$ is proportional to the magnetic charge $\langle
c_1(\cE),t\wedge t\rangle_\Gamma$ of the gauge bundle $\cE\to X$.
The solutions of these equations yield minima of the gauge theory
and we will therefore call them generalized instantons or just
instantons. On a Calabi-Yau threefold we can consider minima where
$\omega=0$. Then the first two equations are the
Donaldson-Uhlenbeck-Yau equations which are conditions of
stability for holomorphic vector bundles $\cE$ over~$X$ with finite
characteristic classes. To compute Donaldson-Thomas invariants, we
restrict to bundles with $u_\cE=0$, which is
equivalent to excluding D4-branes from our counting of stable bound
states (this is automatic if $X$ has no compact divisors). Then these gauge
theory equations describe BPS bound states of D6--D2--D0 branes on $X$.

In a cohomological field theory the path integral localizes onto the
moduli space of solutions to the classical
field equations, which in our case is the generalized instanton moduli
space $\scrM^{\rm inst}_r(X)$ of holomorphic bundles (or torsion free coherent sheaves) $\cE$ on
$X$. We decompose $\scrM^{\rm inst}_r(X)$ into into its connected components $\scrM^{\rm inst}_{n,\beta;r}(X)$
which are labelled by the Chern characteristic classes
$(\ch_3(\cE),\ch_2(\cE)) = (n,-\beta)$. The path integral of the
topological gauge theory can be precisely defined as a
sum over the topologically distinct instanton sectors with an
appropriate measure factor, which arises from the ratio of fluctuation
determinants around each solution of the field equations. In
topological field theories this determinant generically has the form
of a particular characteristic class of a bundle over the
moduli space; in our case this is the Euler class $\eul(\scrN_{n,\beta;r})$ of the antighost or
obstruction bundle $\scrN_{n,\beta;r}\to \scrM^{\rm inst}_{n,\beta;r}(X)$. The partition function then
formally has the form
\begin{equation}
Z_{\rm gauge}^X(q,Q;r) = \sum_{n,\beta}\, q^n\, Q^\beta \
\int_{\scrM^{\rm inst}_{n,\beta;r}(X)}\, \eul(\scrN_{n,\beta;r}) \ .
\label{gaugepartfn}\end{equation}
{}From the gauge theory perspective we can understand the appearence
of the obstruction bundle as follows. The local geometry of the moduli
space $\scrM^{\rm inst}_r(X)$ can be characterized by the instanton deformation
complex~\cite{Iqbal:2003ds,Baulieu:1997jx}
\begin{equation} \label{defcomplex}
\xymatrix@1{\Omega^{0,0} ( X , \mathrm{ad}\, \cE) \quad
\ar[r]^{ C} & \quad {\begin{matrix}\Omega^{0,1} ( X ,
    \mathrm{ad}\, \cE) \\ \oplus \\ \Omega^{0,3} ( X , \mathrm{ad}\,
    \cE) \end{matrix}} \quad
\ar[r]^{ D_A} & \quad \Omega^{0,2} ( X , \mathrm{ad}\, \cE) } \ ,
\end{equation}
where $\Omega^{\bullet,\bullet} (X , \mathrm{ad}\, \cE)$ denotes
the bicomplex of $\IC$-differential forms taking values in the
adjoint gauge bundle over $X$, and the maps $C$ and $D_A$
represent a linearized complexified gauge transformation and the
linearization of the first equation in (\ref{inste}) respectively.
This complex is elliptic and
its first cohomology represents the holomorphic tangent space to
$\scrM^{\rm inst}_r(X)$ at a point corresponding to a holomorphic vector bundle
$\cE\to X$ with
connection one-form $A$. The
degree zero cohomology represents gauge fields $A$ that yield reducible connections, which we assume
vanishes. In general there is also a finite-dimensional
second cohomology that measures obstructions, which is the obstruction
or normal bundle $\scrN_r$ associated with
the kernel of the conjugate operator $D_A^\dag$. However, it is
difficult to give precise meaning to the integral $\int_{\scrM^{\rm inst}_r(X)}\,
\eul(\scrN_r)$. Below we will define it by using the formalism of
equivariant localization when $X$ is a toric manifold; in this case
the toric action lifts to the instanton moduli space and the
characteristic classes will involve the virtual
tangent bundle
$T^{\rm vir}\scrM^{\rm inst}_r(X)= T\scrM^{\rm inst}_r(X) \ominus \scrN_r$ rather than the stable tangent bundle
$T\scrM^{\rm inst}_r(X)$ (which is generally not well-defined here).

For rank $r=1$ this auxilliary gauge theory
reformulates Donaldson-Thomas theory as a (generalized) instanton counting
problem. The instanton multiplicities in the instanton expansion of
the gauge theory path integral represent the Donaldson-Thomas invariants. 
Note that in principle we can keep the rank $r$ arbitrary, since in
this framework it simply corresponds to studying an arbitrary number
$r$ of D6-branes with a nonabelian $U(r)$ worldvolume gauge
theory. Therefore this gauge
theory can in principle be used to also study higher rank
Donaldson-Thomas invariants, about which only a few results are currently
available. However, at present we only know how to make computational progress on the Coulomb branch of the gauge
theory where the gauge symmetry is completely broken down to the
maximal torus $U(1)^r$
by the Higgs field vacuum expectation values and the moduli space
essentially reduces to $r$ copies of the Hilbert scheme where
localization techniques have been successfully applied. This
is precisely the approach we used for instanton counting in
four-dimensional gauge theories.

An important issue which we shall not address here is that of
stability conditions. Strictly speaking, the set of gauge theory
equations (\ref{inste}) only describe stable D6--D2--D0 bound states
in the ``classical'' large radius region of the moduli space. Stable
BPS states of D-branes on the
entire Calabi-Yau moduli space should be properly understood as stable
objects in the bounded derived category $\frD(X)$ of coherent sheaves
on $X$. Different chambers of this moduli space should presumably be
accounted for by modifications of the gauge theory arising through a
noncommutative deformation of $X$ via a non-trivial $B$-field
background, through non-linear higher derivative corrections to the
field equations from the full Dirac-Born-Infeld theory on the branes, and through worldsheet instanton corrections.

\subsection{Singular instanton solutions}

We now explain how the instanton moduli space $\scrM^{\rm inst}_{n,0;r}(\IC^3)$
can be realized as the moduli scheme of (generalized) $n$-instanton
solutions in a six-dimensional
noncommutative $\cN=2$ gauge theory, or equivalently a particular
moduli space of torsion-free sheaves $\cE$ on $\mathbb{P}^3$ of rank
$r$ with $\ch_3(\cE)=-n$ which are framed on a plane $\bdiv$ at
infinity~\cite{Cirafici:2010bd}.
For rank $r=1$, the only non-trivial solutions of the
Donaldson-Uhlenbeck-Yau equations in (\ref{inste}) with $u_\cE=0$ are necessarily
singular. On $X=\IC^3$, we can make sense of such solutions by
passing to a noncommutative deformation $\IC_\theta^3$ defined by
Berezin-Toeplitz quantization of $\IC^3$ with respect to its canonical
K\"ahler form $\varpi= \sum_i\, \dd z_i\wedge\dd \bar z_i$, the
trivial prequantum line bundle $\calL\to X$, and holomorphic
polarization. Then the Hilbert space of geometric quantization
$\Hcal=H^0(X,\calL)=\ker\overline{\partial}_\vartheta$ is the space of
holomorphic sections of $\calL$ where $\varpi=\dd\vartheta$, and
holomorphic functions on $X$ are naturally realized as operators on
$\Hcal$. The Toeplitz quantization map
sends the local complex coordinates $z_i,\bar z_i$ of $X$ to operators with the
Heisenberg commutation relations
\beq
\big[ \bar z_i,z_j]=\delta_{ij} \qquad \mbox{and} \qquad
[z_i,z_j]=0=[\bar z_i,\bar z_j]
\label{Heisalg}\eeq
for $i,j=1,2,3$. The Hilbert space $\Hcal$ is isomorphic to the
unique irreducible representation of the algebra (\ref{Heisalg}) given by the Fock module
\beq
\calH:= \IC[\bar z_1,\bar z_2,\bar z_3]|0\rangle
=\bigoplus_{m_i\in\IZ_{\geq0}}\, \IC|m_1,m_2,m_3\rangle \ ,
\label{Fockspace}\eeq
where the vacuum vector $|0\rangle$ is a fixed section in
$\ker\overline{\partial}_\vartheta$; the dual vector space is
$\calH^*:=\langle0| \IC[z_1,z_2,z_3]$ and the pairing
is defined by $\langle0|0\rangle=1$.
This deformation of $X$ regulates the small instanton
singularities of the moduli space $\scrM^{\rm inst}(X):=\scrM^{\rm inst}_1(X)$. We then couple the
noncommutative gauge theory to Nekrasov's $\Omega$-background by
shifting the BRST supercharges by inner contraction with the vector
field generating the toric isometries of $\IC^3$ given by the torus
group $\mathbb{T}^3 = (t_1 = \e^{\ii
\epsilon_1} , t_2 = \e^{\ii \epsilon_2} , t_3 = \e^{\ii
\epsilon_3})$; this deformation provides a natural compactification of the instanton moduli
space $\scrM^{\rm inst}(X)$ by giving it a finite equivariant volume $\oint_{\IC^3}\, 1 = \frac1{\epsilon_1\,
  \epsilon_2\, \epsilon_3}$, and it localizes the instanton measure onto
point-like contributions which are $\torus^3$-invariant.

Using the Toeplitz quantization map, we replace all fields of the
six-dimensional cohomological gauge theory by operators acting on the
separable Hilbert space (\ref{Fockspace}).
The noncommutative deformation thus transforms the gauge theory into
an infinite-dimensional matrix model, and the instanton equations
(\ref{inste}) become algebraic operator equations for the
noncommutative fields which have the ``ADHM
form''
\begin{eqnarray}
\big[Z_{i}\,,\, Z_{j}\big] = 0  = \big[\,Z^\dag_{i}\,,\,
Z^\dag_{j}\,\big] \qquad \mbox{and} \qquad
\sum_{i=1}^3\, \big[\,Z^{i}\,,\, Z^{\dagger}_{i}\,\big] = 3~\Id_\calH \label{adhmform}
\end{eqnarray}
for $i,j=1,2,3$, where the operators $Z_i:=z_i+\ii A_{\bar z_i}$ are
called ``covariant coordinates''.

The vacuum solution of (\ref{adhmform}) has $A=0$ and can be
represented by harmonic oscillator algebra as $Z_i^{(0)}=z_i$. Generic
instanton solutions with non-trivial topological charges are given
by partial isometric transformations of the vacuum solution
$Z_i^{(0)}$: For each fixed integer $m\geq1$, let $U_m$ be a partial
isometry of the Hilbert space $\calH$ which removes all number basis states
$|m_1,m_2,m_3\rangle$ with $m_1+m_2+m_3<m$ from $\calH$. The instanton
charge
\beq
n:=\ch_3(\cE)=-\mbox{$\frac\ii6$}\, \Tr_\calH\big(F_A\wedge F_A\wedge
  F_A\big) = \mbox{$\frac16$}\, m\, (m+1)\, (m+2)
\eeq
is then the number of states removed by $U_m$ in $\calH$ with
$m_1+m_2+m_3<m$. The partial isometry $U_m$ isomorphically identifies
the Hilbert space $\calH$ with its subspace $\calH_{I_m}:=I_m(\bar z_1,\bar z_2,\bar
z_3)|0\rangle$, where
\beq
I_m(w_1,w_2,w_3)=\IC\big\langle w_1^{m_1}\, w_2^{m_2}\, w_3^{m_3} \
\big| \ m_1+m_2+m_3\geq m\big\rangle
\eeq
is a monomial ideal of codimension $n$ in the polynomial ring
$\IC[w_1,w_2,w_3]$. As in (\ref{pialpha}), this defines a plane partition
\beq
\pi_m=\big\{(m_1,m_2,m_3)\in\IZ_{\geq0}^3 \ \big| \ w_1^{m_1}\,
w_2^{m_2}\, w_3^{m_3} \notin I_m\big\}
\eeq
with $|\pi_m|=n=\ch_3(\cE)$ boxes.
The $\Omega$-background thus localizes the gauge theory partition
function (\ref{gaugepartfn}) onto $\torus^3$-invariant noncommutative
instantons which are parametrized by three-dimensional Young diagrams
$\pi$. One defines the integrals over the Euler classes of the
obstruction bundles via the virtual localization theorem in
equivariant Chow theory, which
extends the Atiyah-Bott localization theorem in equivariant
cohomology from smooth manifolds to schemes (see
e.g.~\cite[\S3.5]{Szabo:2009vw} for details in the present context). This gives
\beq
Z_{\rm gauge}^{\IC^3}(q) = \sum_{n=0}^\infty \, q^n \ \sum_{\pi\, :\, |\pi|=n}\
\frac{\eul(\scrN_{n,0})_\pi}{\eul\big(T_\pi\scrM^{\rm inst}_{n,0}(\IC^3)\big)} \ .
\label{ZNcombgen}\eeq
One shows that the ratio of
Euler classes evaluates to $(-1)^{|\pi|}$ at each fixed point $\pi$,
and the gauge theory path integral thus exactly reproduces the
anticipated MacMahon function
\beq
Z_{\rm gauge}^{\IC^3}(q) = M(-q) \ .
\eeq

For $r>1$, the instanton counting problem is also mathematically
well-posed for an arbitrary collection of D6-branes in the Coulomb
branch of the gauge theory. This branch is described by restricting the path
integral over Higgs field configurations $\phi$ whose eigenvalues
$a=(a_1,\dots,a_r)$ are all distinct; this breaks the gauge symmetry group
from $U(r)$ to its maximal torus $U(1)^r$ which acts by scaling the
trivialization of the instanton gauge bundle $\cE$ on $\bdiv$. In this case $U(1)^r$ noncommutative
instantons correspond to coloured partitions
$\vec\pi=(\pi_1,\dots,\pi_r)$, which are $r$-vectors of
three-dimensional Young diagrams $\pi_l$ labelled by $a_l$ with
$|\vec\pi|:=\sum_l\, |\pi_l|$ boxes. After
toric localization with respect to the torus $\torus^3\times U(1)^r$,
the gauge theory partition function (\ref{gaugepartfn}) becomes
\beq
Z_{\rm gauge}^{\IC^3}(q;r)=\sum_{\vec\pi}\, (-1)^{r\,|\vec \pi|}\,
q^{|\vec\pi|} = M\big((-1)^{r}\,q\big)^r \ .
\label{higherDT}\eeq
This is the generating function for
higher rank Coulomb branch Donaldson-Thomas invariants which was subsequently rigorously derived as a degenerate
central charge limit of Stoppa's higher rank Donaldson-Thomas
invariants for D6--D0 bound states~\cite{Stoppa}. The gauge theory in this branch
does not seem to be dual to topological string theory nor to even
enumerate holomorphic curves.

Finally, the construction just outlined carries through to the case of a general
toric manifold $X$ using gluing rules which are completely analogous
to those described in
\S\ref{subsec:gaugestringduality}. 
The relevant noncommutative deformations are described
in~\cite{Iqbal:2003ds}, while a rigorous treatment of the geometric quantization of the toric variety as a
K\"ahler manifold should deal with several subtleties including the
``half-form correction'', as explained in~\cite{mourao};
see~\cite{Cirio:2011eh,Szabo:2011mj} for an alternative approach.
The instantons sit on top of
each other at each vertex $v$ of the trivalent graph
$\Delta$ encoding the geometry of $X$, and along the edges
$e$ representing the local $\IC^3$ coordinate axes where they asymptote
to four-dimensional noncommutative instantons on the associated
rational curves. By
employing the localization formalism on the instanton moduli space, the gauge
theory path integral localizes onto a sum of contributions from
three-dimensional Young diagrams associated with each
vertex and a set of two-dimensional Young diagrams
that arise when gluing together two plane partitions as a section of a
common leg; the framing conditions map to a framing of the generalized
instanton gauge bundle on a compactification divisor at infinity. For the rank $r$ gauge theory in the
Coulomb branch one finds~\cite{Cirafici:2008sn}
\begin{equation}
{Z}_{\rm gauge}^{X}(q,Q;r) = \sum_{\vec\pi_v,\vec\lambda_e} \,
(-1)^{r\, D\{\vec\pi_v,\vec\lambda_e\}} \,
 q^{D\{\vec\pi_v,\vec\lambda_e\}} \ \prod_{ {\rm edges}\ e} \,
(-1)^{\sum_{l,l'=1}^r\, |\lambda_{e,l}|\, |\lambda_{e,l'}|\, m_{e,1}} \,
Q_e^{\sum_{l=1}^r \, |\lambda_{l,e}|} \ ,
\end{equation}
where 
\begin{equation}
D\{\vec\pi_v,\vec\lambda_e\}= \sum_{{\rm vertices}\ v} \ \sum_{l=1}^r \, |\pi_{v,l}|
+ \sum_{{\rm edges}\ e} \  \sum_{l=1}^r \ \sum_{(i,j) \in
  \lambda_{e,l}} \, \big(
m_{e,1}\,  (i-1) + m_{e,2} \, (j-1) + 1 \big)
\end{equation}
and the pairs of integers $(m_{e,1},m_{e,2})$ specify the normal bundles
over the rational curves corresponding to the edges $e$ of the graph
$\Delta$. For $r=1$ it is straightforward to check that this partition
function coincides with the large radius generating function
(\ref{ZDTcrystal}) for the Donaldson-Thomas invariants of $X$. Similar formulas for gauge theories on toric
surfaces are developed in~\cite{Nekrasov:2003vi}. As in the
four-dimensional case, possible descriptions of the partition
functions of D6--D2--D0 bound states as quasi-modular forms seem to be
rather subtle to deduce from the perspective of such combinatorial expansions.  

\subsection{Counting instantons}

We have provided a complete classification of the $\torus^3\times U(1)^r$ critical
points of the six-dimensional gauge theory in its Coulomb
branch, which are all isolated and parametrized by $r$-vectors of
three-dimensional Young diagrams $\vec\pi=(\pi_1,\dots,\pi_r)$. We will now sketch how to compute the quantum fluctuation determinants around each critical point. This
can be done explicitly in an ADHM-type formalism which provides a
concrete parametrization of the compactified instanton moduli space. 

It is customary in instanton computations to use collective
coordinates to study the local structure of the moduli space, as we
did in the four-dimensional case. This
corresponds to taking the point of view of the field theory on the D0-branes which characterize the instantons, in contrast to the point of view of the D6-brane gauge theory we have been considering so far. 
To compute the virtual equivariant characteristic classes in
(\ref{gaugepartfn}), we use the local model of the instanton moduli
space developed in~\cite{Cirafici:2008sn} from the instanton quantum
mechanics for $n$ D0-branes inside $r$ D6-branes on $\complex^3$; this is a rather powerful perspective since to apply toric localization we only need to understand the neighbourhood of each fixed point. We introduce two vector spaces $V$
and $W$ with $\dim V = n$ and $\dim W = r$, which represent
respectively the 
gas of $n$ D0-branes and the $r$ D6-branes. For abelian gauge theory one can
construct an explicit parametrization of the moduli space of ideal
sheaves via the Beilinson spectral sequence, whose first term is
\beq
E_1^{p,q} = \mathcal{O}_{{\mathbb P}^3}(p\,\bdiv) \otimes H^q \big( {\mathbb P}^3
\,,\, \mathcal{E}(-r\, \bdiv) \otimes \Omega^{-p}_{{\mathbb P}^3}(-p\,
\bdiv)
\big) 
\eeq
for any coherent sheaf $\mathcal{E}$ on $\mathbb P^3$. By an appropriate
choice of boundary conditions this spectral sequence degenerates at
the $E_2$-term, and the original sheaf $\cE$ can be
described as the only non-vanishing cohomology of a four-term
complex; the associated conditions yield a set of mutually commuting matrices
$B_i\in\End_\IC(V)$ for $i=1,2,3$ plus stability conditions. This
strategy is precisely a higher-dimensional generalization of the ADHM construction of the usual instanton moduli spaces
of four-dimensional gauge theories, as described in \S\ref{subsec:ALEmodsp}; in particular, in this case the vector
spaces $V$ and $W$ are explicitly realized as certain cohomology
groups of the original gauge sheaf $\cE$. The collection of commuting
matrices modulo the natural adjoint action of the gauge group $GL(n,\IC)$, together
with a particular stability condition, parametrize the moduli space of classical
solutions to the topological quantum mechanics describing the dynamics
of the collective instanton
degrees of freedom~\cite{Moore:1997dj,Moore:1998et}. The fields $B_i$
represent gauge fields on the D0-branes, whereas fields
$I\in\Hom_\IC(W,V)$ are associated with D6--D0 open strings whose role
is to label the colours of the three-dimensional Young diagrams
parametrizing the fixed points on the framed moduli space of coherent sheaves
on $\PP^3$. Other fields
are necessary to close the equivariant BRST algebra and localize
the matrix model on the generalized ADHM equations; see~\cite{Cirafici:2008sn} for a complete treatment.

Let $Q\cong\IC^3$ be the three-dimensional fundamental
$\mathbb{T}^3$-module with weight $(1,1,1)$. At the fixed points of the $\torus^3 \times
U(1)^r$ action on $\scrM^{\rm inst}_r(\IC^3)$, a gauge transformation is equivalent to
an equivariant rotation and we can decompose the vector spaces as
elements of the representation ring of $\torus^3 \times U(1)^r$ with
\begin{eqnarray}
V_{\vec\pi} = \sum_{l=1}^r \,e_l~ \sum_{(i,j,k)\in \pi_l}\,
t_1^{i-1} \,t_2^{j-1}\,t_3^{k-1} \qquad \mbox{and} \qquad W_{\vec\pi} =
\sum_{l=1}^r\,e_l
\label{decompos6d}\end{eqnarray}
regarded as polynomials in
$t_1$, $t_2$, $t_3$ and $e_l:=\e^{\ii a_l}$, $l=1,\dots,r$. Each term in the weight
decomposition of the vector space $V_{\vec\pi}$ corresponds to a box
in the collection of plane partitions~$\vec\pi$. 

Let us study the local
geometry of the instanton moduli space around a fixed
point $\vec\pi$ with corresponding commuting matrices
$(B_1,B_2,B_3)\in\End_\IC(V_{\vec\pi})\otimes Q$ and
$I\in\Hom_\IC(W_{\vec\pi}, V_{\vec\pi})$. The equivariant complex
\begin{equation} \label{adhmdefcomplexC3}
\xymatrix@1{
  \Hom_\IC(V_{\vec\pi} , V_{\vec\pi})
   \quad\ar[r] &\quad
   {\begin{matrix} \Hom_\IC(V_{\vec\pi} , V_{\vec\pi}) \otimes Q 
   \\ \oplus \\
   \Hom_\IC(W_{\vec\pi} , V_{\vec\pi}) \\ \oplus  \\ \Hom_\IC(V_{\vec\pi} ,
   V_{\vec\pi}) \otimes \bigwedge^3 
   Q \end{matrix}}\quad \ar[r] & \quad
   {\begin{matrix} \Hom_\IC(V_{\vec\pi} , V_{\vec\pi}) \otimes \bigwedge^2
       Q \\ \oplus \\ 
       \Hom_\IC(V_{\vec\pi},W_{\vec\pi}) \otimes \bigwedge^3 Q
   \end{matrix}}
}
\end{equation}
is the matrix quantum mechanics
analog of the instanton deformation complex (\ref{defcomplex}); the
first map is an infinitesimal (complex) gauge
transformation while the second map is the differential of the equations $[B_i, B_j] = 0$ that
define the moduli space. In a
similar way, its first cohomology is a local model of the Zariski
tangent space to the moduli space at the fixed point $\vec\pi$, while its second cohomology
parametrizes obstructions. The localization formula involves the ratio of the top Chern
class of the obstruction bundle over the weights coming from the
tangent bundle. The equivariant
index of the complex (\ref{adhmdefcomplexC3}) computes the virtual sum $\Ext_{\cO_{\PP^3}}^1\ominus \Ext_{\cO_{\PP^3}}^0\ominus \Ext_{\cO_{\PP^3}}^2$ of
cohomology groups. We assume that $\Ext_{\cO_{\PP^3}}^0$ vanishes, which is
equivalent to restricting attention to irreducible connections. The equivariant index is given in terms of
the characters of the representation evaluated at the fixed point
as
\begin{eqnarray} \label{character}
\Ch_\torus\big(T_{\vec\pi}^{\rm vir}\scrM^{\rm inst}_{n,0}(\IC^3)\big)=  W^*_{\vec\pi}
\otimes V_{\vec\pi} - 
\frac{ V^*_{\vec\pi} \otimes W_{\vec\pi}}{t_1\, t_2\, t_3} +
 V^*_{\vec\pi} 
\otimes V_{\vec\pi} ~\frac{(1-t_1)\, (1-t_2)\, (1-t_3)}{t_1\, t_2\,
  t_3} \ ,
\end{eqnarray}
and the inverse of the corresponding top Chern polynomial yields the desired
ratio of weights in (\ref{ZNcombgen})~\cite{Nekrasov:2003rj,Bruzzo:2002xf,Nekrasov:2003vi}; again the dual involution acts on the weights as
$t_i^*=t_i^{-1}$ and $e_l^*=e_l^{-1}$.
In~\cite{Cirafici:2008sn} it is shown that at the Calabi-Yau specialization
$\epsilon_1+\epsilon_2+\epsilon_3=0$ of
the $\Omega$-deformation, the Euler classes in (\ref{ZNcombgen}) coincide up to a
sign given by
\beq
\eul(\scrN_{n,0;r})_{\vec\pi}
 = (-1)^{r\,
  |\vec\pi|} \ \eul\big(T_{\vec\pi}\scrM^{\rm inst}_{n,0;r}(\IC^3)\big) \ .
\eeq

At an arbitrary point $\epsilon=(\epsilon_1,\epsilon_2,\epsilon_3)$ of
the $\Omega$-deformation, the classical part of the partition function
can be computed in the noncommutative gauge theory and is given
by~\cite[\S3.5]{Cirafici:2008sn}
\beq
 Z_{\rm class}^{\IC^3}(\epsilon,a;q;r) = \prod_{l=1}^r \,
 q^{-{a_l^3}/{6\, \epsilon_1\, \epsilon_2\, \epsilon_3} } \ ,
\eeq
while the expression for the vacuum contribution
in~\cite[eq.~(3.51)]{Cirafici:2008sn} holds at all points in parameter
space and gives the
perturbative partition function
\beq
Z_{\rm pert}^{\IC^3}(\epsilon,a;r) = \prod_{l,l'=1}^r \, 
\exp\Big(-\int_0^\infty\, \frac{\dd t}t \
\frac{\e^{t\,(a_l-a_{l'})}}{\big(1-\e^{t\, \epsilon_1}\big)\,
  \big(1-\e^{t\, \epsilon_2}\big)\, \big(1-\e^{t\,
    \epsilon_3}\big)} \, \Big) \ ,
\eeq
which should be properly defined using triple zeta-function
regularization similarly to the four-dimensional case. For $r=1$, the computation of the equivariant
instanton partition function is the content of~\cite[Thm.~1]{MNOPII}
which gives
\beq
Z_{\rm gauge}(\epsilon;q)=
M(-q)^{- \chi_{\torus^3}(\IC^3)} \ ,
\label{Z1instexpl}\eeq
where
\beq
\chi_{\torus^3}(X)=\int_{X}\, \ch_3(X)^{\torus^3} =
\frac{(\epsilon_1+\epsilon_2)\,
  (\epsilon_1+\epsilon_3)\,
  (\epsilon_2+\epsilon_3)}{\epsilon_1\, \epsilon_2\,
  \epsilon_3}
\eeq
is the $\torus^3$-equivariant Euler characteristic of $X=\IC^3$,
evaluated by the Bott residue formula.
This formula is proven using geometric arguments from relative
Donaldson-Thomas theory to develop an equivariant vertex formalism. An extension of this partition function to
the Coulomb branch of the rank $r$ gauge theory as a topological
matrix model is considered in~\cite{Awata:2009dd} and conjectured to
be independent of the Higgs parameters $a$, analogously to (\ref{higherDT}).

The
simplicity of the Coulomb branch invariants in this case may be
understood by rewriting them in terms of the more fundamental Joyce-Song generalized
Donaldson-Thomas invariants $\widehat{\DT}_k(X)$ which are completely
independent of the rank $r$ of the gauge theory, as explained
in~\cite{Cirafici:2010bd,Cirafici:2011cd}. In the present case they are defined
through
\beq
Z_{\rm gauge}^{\IC^3}(q;r) =:
\exp\Big(-\sum_{k=1}^\infty\, (-1)^{k\,r}\, k\, r \ \widehat{\DT}_k\big(\IC^3\big)\, (-q)^k\Big)
\ ,
\eeq
and they lead to the generalized Gopakumar-Vafa BPS invariants
$\BPS_k(X)$ defined by
\beq
\widehat{\DT}_k\big(\IC^3\big)=: \sum_{m|k}\, \frac1{m^2}\, \BPS_{k/m}\big(\IC^3\big) \ .
\eeq
The integers $\BPS_k(X)$ count M2 brane-antibrane bound states in
M-theory compactified on $X\times S^1$~\cite{Gopakumar:1998jq}. By
using the exponential representation (\ref{exprepMac}) with $Q=1$, we find explicitly
\beq
\widehat{\DT}_k\big(\IC^3\big) = \sum_{m|k}\, \frac1{m^2} \qquad \mbox{and} \qquad \BPS_k\big(\IC^3\big)=1 \
.
\eeq
The physical interpretation of these invariants in terms of D-brane
bound states is elucidated in~\cite{Cirafici:2011cd}.

\bigskip

\section{Stacky gauge theories\label{sec:Stacky}}

\subsection{Generalized McKay correspondence}

We will now describe the enumerative problem of noncommutative
Donaldson-Thomas invariants from \S\ref{sec:3CY} as an instanton counting problem. This
construction makes use of the generalized McKay correspondence for
Calabi-Yau threefolds, and it is inspired by the relation
between instanton moduli spaces on ALE varieties and the McKay quiver
which we discussed in \S\ref{subsec:ALEmodsp}. We will consider singular Calabi-Yau
orbifolds of the form $\complex^3 / G$ and work on the noncommutative
crepant resolution. We shall find that in order to construct the
noncommutative invariants we need to introduce a modification of the
six-dimensional cohomological gauge theory, which we call a stacky
gauge theory; we regard this gauge theory as the low-energy effective
field theory on the D6-branes, where certain stringy effects have been
added by hand. It turns out that these gauge theories are naturally
suited to the problem of constructing $ G$-equivariant instantons on
$\complex^3$, which will count $G$-equivariant closed subschemes of
$\IC^3$, or equivalently substacks of the quotient stack
$[\IC^3/G]$. As in the case of the ALE spaces, these instanton solutions depend sensitively on the boundary conditions at infinity.

A generalization of the ordinary McKay correspondence of
\S\ref{subsec:McKay} was given by
Ito-Nakajima~\cite{ItoNaka} for
three-dimensional orbifolds of the form $\complex^3 /  G$, where
$ G \subset SL(3,\complex)$ is a finite group, and their
natural smooth crepant Calabi-Yau
resolutions given by the Hilbert-Chow morphism $\pi : X  \rightarrow
\complex^3 \big/  G$, where 
$X=\mathrm{Hilb}_G(\complex^3)$ is the $ G$-Hilbert scheme consisting of $ G$-invariant
zero-dimensional subschemes $Z$ of $\complex^3$ of length $| G|$
such that $H^0(\cO_Z)$ is the regular representation of $ G$; for
simplicity we assume that $G$ is abelian. Roughly speaking, the McKay
correspondence in this setting is the statement that any well-posed
question about the geometry of the resolution $X$ should have a
$ G$-equivariant answer on $\complex^3$. 

Consider the universal scheme $\cZ \subset X\times
\complex^3$ with correspondence diagram
\begin{equation}
\xymatrix@=10mm{
  & \cZ \ar[ld]_{p_1}\ar[rd]^{p_2}& \\
  X & & \complex^3
}
\end{equation}
and define the tautological bundle
\begin{equation}
\calR := p_{1*} \cO_{\cZ} \ .
\label{tautdef}\end{equation}
Under the action of $ G$ on $\cZ$, the bundle $\calR$ transforms in the
regular representation. Its fibres are the $| G|$-dimensional
vector spaces $\IC[z_1,z_2,z_3]/I\cong H^0(\cO_Z)$ for the regular
representation of $ G$, where $I\subset \IC[z_1,z_2,z_3]$ is a
$ G$-invariant ideal corresponding to a zero-dimensional subscheme
$Z$ of $\IC^3$ of length $| G|$. 
Let $Q$ be the fundamental three-dimensional representation of
$ G\subset SL(3,\complex)$; if $G$ acts on $\IC^3$ with weights
$(a_1,a_2,a_3)$ obeying $a_1+a_2+a_3\equiv0$ and $\rho_{a_i}$ denotes the
irreducible one-dimensional representation of
$ G$ with weight $a_i$, then $Q = \rho_{a_1} \oplus
\rho_{a_2} \oplus \rho_{a_3}$.
The decomposition of the regular representation induces a
decomposition of the tautological bundle into
irreducible representations
\begin{equation}
\calR = \bigoplus_{a\in\widehat{G}}\, \calR_a \otimes \rho_a \ ,
\label{cRdecomp}\end{equation}
where $\{ {\rho_a} \}_{a\in\widehat{G}}$ is the set of irreducible
representations; we denote the trivial representation by $\rho_0$. The
tautological line bundles $\calR_a=\Hom_ G(\rho_a,\calR)$ form an integral basis for the Grothendieck group
$K(X)$ of vector bundles on $X$, where the bundle corresponding to the trivial representation
is the trivial line bundle $\calR_0 \cong \cO_X$. 

Similarly, one can introduce a dual basis $\cS_a$ of the
Grothendieck group $K^c(X)$ of 
coherent sheaves on the exceptional set $\pi^{-1}(0)$, or equivalently
of bounded complexes of vector bundles over $X$ which are
exact outside the exceptional locus $\pi^{-1} (0)$ given by
\begin{equation}
\cS_a \ : \ \xymatrix{
 \calR^{\vee}_a \ \ar[r] & \
 \displaystyle{\bigoplus_{b\in\widehat{G}}\, {\tt a}^{(2)}_{ab}\,
   \calR_b^{\vee} } \ \ar[r] & \ \displaystyle{
   \bigoplus_{b\in\widehat{G}}\, {\tt a}^{(1)}_{ab}\, \calR_b^{\vee} } \
 \ar[r] & \ 
 \calR_a^{\vee}
} \ ,
\end{equation}
where the arrows arise from the decomposition of the maps
$\bigwedge^{i-1} Q\otimes\calR\rightarrow\bigwedge^i Q
\otimes\calR$ for $i=1,2,3$ induced by multiplication with the coordinates
$(z_1,z_2,z_3)$ of $\IC^3$ and
\begin{equation} \label{tensordecomp}
\mbox{$\bigwedge^i$}\, Q \otimes \rho_a =
\bigoplus_{b\in\widehat{G}}\, {\tt a}^{(i)}_{ba}\, \rho_b \qquad
\mbox{with} \quad {\tt a}^{(i)}_{ba}=\dim\Hom_ G\big(\rho_b
\,,\,\mbox{$\bigwedge^i$}\, Q \otimes \rho_a \big) \ .
\end{equation}
Since $ G$ is a subgroup of $SL(3,\complex)$, one has
${\tt a}_{ab}^{(3)}=\delta_{ab}$ and ${\tt a}^{(2)}_{ba} = {\tt a}^{(1)}_{ab}$. These multiplicities can be computed explicitly from the decompositions
\beq
Q\otimes\rho_a=(\rho_{a_1} \oplus
\rho_{a_2} \oplus \rho_{a_3})\otimes\rho_a =\rho_{a_1+a} \oplus
\rho_{a_2+a} \oplus \rho_{a_3+a} \ ,
\eeq 
which comparing with (\ref{tensordecomp}) gives
\beq
{\tt a}_{ab}^{(1)}= \delta_{a,b+a_1}+\delta_{a,b+a_2}+\delta_{a,b+a_3} \qquad \mbox{and} \qquad {\tt a}_{ab}^{(2)}= \delta_{a,b-a_1}+\delta_{a,b-a_2}+\delta_{a,b-a_3} \ .
\label{ars12expl}\eeq 

This definition relates the representation theory
of $ G$ with the homology and K-theory of $X$, in particular the tensor
product decomposition (\ref{tensordecomp}) with the intersection
theory of~$X$. For this, define the collection of dual complexes $\{ \cS_a^{\vee} \,\}_{a\in\widehat{G}}$ by
\begin{equation}
\cS_a^{\vee} \ : \  - \Big[ \xymatrix{
 \calR_a \ \ar[r] & \ \displaystyle{\bigoplus_{b\in\widehat{G}}\,
   {\tt a}^{(1)}_{ab}\, \calR_b} \ \ar[r] & \
   \displaystyle{\bigoplus_{b\in\widehat{G}}\, {\tt a}^{(2)}_{ab}\, \calR_b} \
     \ar[r] & \ 
 \calR_a
} \Big] \ .
\end{equation}
As in \S\ref{subsec:McKay}, we define a perfect pairing on $K^c (X)$ by
\begin{equation}
(\cS , \cT  )_{K^c} = \big\langle \Xi (\cS) \,,\, \cT \big\rangle_K = \int_X\, \ch\big(\Xi
(\cS) \big)\wedge\ch( \cT)\w {\rm td}(X) \ ,
\end{equation}
which is a representation of the BPS intersection product
(\ref{DSZint}) on the K-theory lattice of fractional brane charges.
It follows that
\begin{equation}
\big(\cS_a^{\vee} \, , \, \cS_b \big)_{K^c} = \big\langle \Xi (\cS_a^{\vee}\,) \,,\,
\cS_b \big\rangle_K = \sum_{c\in\widehat{G}} \, \big( -  \delta_{ac} +
{\tt a}^{(2)}_{ac} - {\tt a}^{(1)}_{ac} + \delta_{ca}\big) \, \langle \calR_c , \cS_b \rangle_K = {\tt a}^{(2)}_{ab} -
{\tt a}^{(1)}_{ab} \ ,
\end{equation}
where we have used the fact that $\{ \calR_b \}_{b\in\widehat{G}}$ and $\{ \cS_a \}_{a\in\widehat{G}}$ are dual bases of $K(X)$ and $K^c (X)$. This result underlies the relation between the tensor
product decomposition (\ref{tensordecomp}) and the intersection theory
of~$X$, generalizing the pairing of \S\ref{subsec:McKay} in complex
dimension two which gave the extended Cartan matrix of an ADE singularity.

The dual bases $\{ \calR_a \}_{a\in \widehat{G}}$ and $\{ \cS_a
\}_{a\in \widehat{G}}$ of $K(X)$ and $K^c (X)$ correspond, via the McKay correspondence, with two
bases of $ G$-equivariant coherent sheaves on $\complex^3$~\cite{ItoNaka}. The Grothendieck groups of
$ G$-equivariant sheaves on $\complex^3$, $K_{ G}
(\complex^3)$ and $K^c_{ G}(\complex^3)$ (with coherent sheaves of compact
support), have respective bases $\{ \rho_a \otimes \cO_{\complex^3}
\}_{a\in \widehat{G}}$ and $\{ \rho_a \otimes \cO_0 \}_{a\in \widehat{G}}$ where
$\cO_0$ is the skyscraper sheaf at the origin; the latter basis is
naturally identified as the set of fractional $0$-branes. All of these groups are
isomorphic to the representation ring $R( G)$ of the orbifold
group $ G$. 

\subsection{Instanton moduli spaces}

We now introduce the concept of a stacky gauge theory as a gauge
theory on the quotient stack $[\IC^3/G]$, and study a moduli space of geometric objects which are naturally associated with the noncommutative Donaldson-Thomas enumerative problem. A stacky gauge theory is a  sequence of deformations of an ordinary gauge theory on $\complex^3$ whose observables are determined by $ G$-equivariant torsion free $\cO_{\complex^3}$-modules on $\complex^3$, i.e. $ G$-equivariant instantons.
We think of these gauge theories as describing the low-energy dynamics
of D-branes on orbifolds of the form $\complex^3 /  G$ in a
certain ``orbifold phase''. They are realized starting from the ordinary maximally supersymmetric Yang-Mills
theory on $\complex^3$ that was discussed in \S\ref{6dcohgt}; for the moment we discuss the $U(1)$ gauge theory, but
below we also consider the non-abelian $U(r)$ gauge theory in its Coulomb
branch. 
One then considers the orbifold action of $ G$ which is a diagonal
subgroup of the torus group $\torus^3\subset SL(3,\IC)$. Under this
action, the Fock space of the noncommutative gauge theory is a $ G$-module which decomposes as  
\beq
\calH=\complex[\bar z_1,\bar z_2,\bar
z_3]|0\rangle = \bigoplus_{a\in\widehat{G}}\, \calH_{a} \qquad \mbox{with} \quad
\calH_{a} = \bigoplus_{\sum_i\, m_i\, a_i \equiv a} \, {\IC}|m_1,m_2,m_3\rangle \ .
\label{Hisotop}\eeq
As a result the covariant coordinate operators $Z_{i}$ decompose as
\beq
Z_i =\bigoplus_{a\in \widehat{G}}\, Z_i^{(a)} \qquad \mbox{with} \quad
Z_i^{(a)} \in\Hom_\IC\big(\calH_{a} \,,\, \calH_{a+a_i}\big)
\label{orbcovcoord}\eeq
and the first of the instanton equations (\ref{adhmform}) becomes
\beq
Z_i^{(a+a_j)}\, Z_j^{(a)}= Z_j^{(a+a_i)}\, Z_i^{(a)} \ .
\label{orbcomm}\eeq
Partial isometries $U_m$ decompose accordingly and the resulting
noncommutative instanton solutions are parametrized by
$\widehat{G}$-coloured plane partitions
$\pi=(\pi_a)_{a\in\widehat{G}}$, where
$(m_1,m_2,m_3)\in\pi_a$ if and only if $m_1\, a_1+m_2\,a_2+m_3\,
a_3\equiv a$.

These solutions are associated with a certain framed moduli space
of torsion free sheaves $\cE$ of rank $r$ and topological charge
$\ch_3(\cE)=n$ on the compact toric orbifold $\PP^3 /  G$ by an
application of Beilinson's theorem. This describes the original sheaf
$\cE$ as the single non-vanishing cohomology of a complex which is
characterized by two vector spaces $V$ and $W$ of dimensions $n$ and
$r$ which are $ G$-modules, along with the set of tautological
bundles constructed from the representation
theory of $ G$ via the McKay correspondence that characterize the
homology of the resolved space $X=\mathrm{Hilb}_{ G} (\complex^3)$. In particular the
framing $ G$-module $W$ is associated with the fibre of $\cE$ at
infinity. One discovers that the
relevant moduli spaces can be described in terms of representations
$(V,W,B,I)$ of the \textit{framed} McKay quiver $\hat\sfQ_G$ associated
with the orbifold singularity $\complex^3 /  G$, where $B\in
\Hom_G(V,Q\otimes V)$ and $I\in\Hom_G(W,V)$. As before, the nodes of the
quiver $\sfQ_G$ are the vector spaces $V_a$ in the isotopical decomposition of
$V$ into irreducible representations $\rho_a$ of $ G$, and there are
${\tt a}_{ab}^{(1)}$ arrows $B$ between the nodes labelled by
$a,b\in\widehat{G}$ satisfying ${\tt a}_{ab}^{(2)}$ relations; the dimensions $n_a=\dim V_a$ are associated with multi-instantons which transform in the irreducible representation $\rho_a$. 
The new ingredients are the framing nodes which arise from the
isotopical decomposition of the vector space
$W=\bigoplus_{a\in\widehat{G}}\, W_a\otimes \rho_a^*$ into
irreducible representations. The framing nodes label boundary
conditions on the Higgs fields at infinity where the gauge fields are
required to approach a flat connection; whence the gauge sheaf is associated with a representation $\rho$ of the orbifold group $ G$ and the
dimensions $\dim W_a = r_a$ label the multiplicities of
the decomposition of $\rho$ into irreducible representations, with
$\sum_{a\in\widehat{G}}\, r_a=r$.
The arrows from the framing nodes correspond to equivariant maps
$I\in\Hom_ G(W,V)$, which by Schur's lemma decompose into
linear maps $I^{(a)}\in\Hom_\complex(W_a,V_a)$.

This construction thus gives a correspondence between a certain class
of sheaves $\cE$ and representations of a framed McKay quiver.
Moreover, from the complex derived via Beilinson's theorem one can express the Chern character of the original torsion free sheaf $\cE$ in terms of data associated with the representation theory of the orbifold group via the McKay correspondence as
\begin{eqnarray}
\mathrm{ch} (\cE) &=& - \mathrm{ch} \Big( \big( V \otimes \mathcal{R} (-2\bdiv) \big)^{ G} \Big) +
 \mathrm{ch} \Big(
 \big(\mbox{$V \otimes \bigwedge^2 Q^{*}$} \otimes \mathcal{R} (-\bdiv) \big)^{ G} \Big)
 \cr & & -\, \mathrm{ch} \Big(
\big(( {V \otimes Q^{*} \oplus W}) \otimes \mathcal{R} \big)^{ G}
 \Big) + \mathrm{ch} \Big( \big(  {V} \otimes \mathcal{R} (\bdiv) \big)^{ G} \Big) \ , \label{chE}
\end{eqnarray}
where the Chern classes $c_1(\calR_a)$ of the set of tautological bundles (\ref{cRdecomp})
give a basis of $H^2 (X , \zed)$ dual to the basis of exceptional
curves in the crepant resolution $X$. In the algebraic framework the tautological bundles map to projective objects in the category of quiver representations.

\subsection{Counting instantons and orbifold BPS invariants\label{subsec:orbifoldBPS}}

To study the local structure of the instanton moduli space of the
stacky gauge theory, let us now consider the instanton quantum
mechanics which corresponds to taking the point of view of the
fractional D0-branes which characterize the instantons. For this, we
will linearize the complex obtained via Beilinson's theorem to
construct a local model for the instanton moduli space. As before, the dynamics
of the collective coordinates is described by a cohomological matrix
model whose classical field equations are given by the orbifold generalized ADHM equations
\begin{equation} \label{ADHMorb}
B_i^{(a + a_j)} \ B_j^{(a)} = B_j^{(a+a_i)} \ B_i^{(a)}
\end{equation}
together with a suitable stability condition. The set of equations (\ref{ADHMorb}) arises as an ideal of relations in the path
algebra $\sfA_G$ of the McKay quiver $\sfQ_G$. This algebra is the
noncommutative crepant resolution which ``desingularizes'' the
orbifold singularity $\IC^3/G$, in the sense that the centre of $\sfA_G$
is isomorphic to the coordinate algebra $\IC[z_1,z_2,z_3]\rtimes\IC [G]$
of the quotient stack $[\IC^3/G]$. Then the generalized McKay
correspondence dictates that the bounded derived category of coherent
sheaves on the resolution $X$ is functorially equivalent to the
bounded derived category of (stable) representations of the McKay
quiver.

In the Coulomb branch of the topological matrix model, the BRST fixed points are parametrized by
$r$-vectors of $G$-coloured plane partitions $\vec \pi = \left( \pi_1 , \dots , \pi_r
\right)$ with $|\vec\pi|=\sum_l\, |\pi_l|=n$ boxes, where
$\pi_l=(\pi_{l,a})_{a\in \widehat{G}}$ with $\sum_l\,
|\pi_{l,a}|=\dim(V_a)$. Since the orbifold group $ G$ is a subgroup of the torus group $\torus^3$, the fixed points onto which the matrix quantum mechanics localizes are the same as in the case of the affine space $\complex^3$, the only difference being that one now has to keep track of the $ G$-action.
A local model for the moduli space near a fixed point of the action of
the torus $\torus^3\times U(1)^r$ is realized by a $G$-equivariant version of the instanton deformation complex
\begin{equation} \label{equivdefcomplex}
\xymatrix{
  \Hom_{ G}  (V_{\vec\pi} , V_{\vec\pi})
   \quad\ar[r] &\quad
   {\begin{matrix} \Hom_{ G} (V_{\vec\pi} , V_{\vec\pi} \otimes Q )
   \\ \oplus \\
   \Hom_{ G} (W_{\vec\pi} , V_{\vec\pi}) \\ \oplus  \\ \Hom_{ G} (V_{\vec\pi} ,
   V_{\vec\pi}  \otimes \bigwedge^3 
   Q) \end{matrix}}\quad \ar[r] & \quad
   {\begin{matrix} \Hom_{ G} (V_{\vec\pi} , V_{\vec\pi}  \otimes \bigwedge^2
       Q) \\ \oplus \\ 
       \Hom_{ G} (V_{\vec\pi},W_{\vec\pi} \otimes \bigwedge^3 Q)
   \end{matrix}}
}
\end{equation}
from which we can extract the character at the fixed points
\begin{equation} \label{orbcharacter}
\Ch_{\torus}\big(T_{\vec\pi}^{\rm vir}\scrM^{\rm inst}_{n,0}(\IC^3)\big)^G = \Big(\, W^*_{\vec\pi}
\otimes V_{\vec\pi} - 
\frac{ V^*_{\vec\pi} \otimes W_{\vec\pi}}{t_1\, t_2\, t_3} +
 V^*_{\vec\pi} 
\otimes V_{\vec\pi} ~\frac{(1-t_1)\, (1-t_2)\, (1-t_3)}{t_1\, t_2\,
  t_3} \, \Big)^{ G} \ ,
\end{equation}
where $t_i=\e^{\ii\eps_i}$ for $i=1,2,3$. This yields all the data we
need for the construction of noncommutative Donaldson-Thomas
invariants which enumerate
$ G$-equivariant torsion free sheaves on $\complex^3$ via the McKay
correspondence.

We can construct a partition function for these invariants from the local structure of the instanton moduli space. Neglecting the $ G$-action, the two vector spaces $V$ and $W$ can
be decomposed at a
fixed point $\vec\pi=(\pi_1,\dots,\pi_r)$ of the $\torus^3\times U(1)^r$ action on the
instanton moduli space as in (\ref{decompos6d}). Each partition $\pi_l$
carries an action of $ G$. However this action is offset by the $
G$-action of the factor $e_l=\e^{\ii a_l}$ for $l=1,\dots,r$ which corresponds to the choice of a boundary condition on the gauge field at infinity. One still has to
specify in which superselection sector one is working which is
characterized by choosing which of the eigenvalues $e_l$ are in a particular
irreducible representation of $ G$. Following~\cite{Cirafici:2010bd}, we define a
boundary function ${\tt b}:\{1,\dots,r\}\to\widehat{G}$ which to each
sector $l=1,\dots,r$ associates the weight ${\tt b}(l)$ of the
$G$-module generated by the eigenvalue $e_l$. The relation between the
instanton numbers and the number of boxes in a partition associated
with a given irreducible representation is then given by $n_a = \sum_{l=1}^r \, |\pi_{l,a-{\tt b}(l)}|$.
The contribution of an instanton to the gauge theory fluctuation
determinant can be now derived from the local character
(\ref{orbcharacter}) of the moduli space near a fixed point at the
Calabi-Yau specialization $t_1\,t_2\,t_3=1$ of the $\Omega$-background; it is
given by $(-1)^{{\mathcal K}_G (\vec\pi;\mbf r,{\tt b})} $, with~\cite{Cirafici:2010bd}
\begin{eqnarray}
\cK_G(\vec\pi;\mbf r) &=& \sum_{l=1}^r ~ \sum_{a\in\widehat{G}}\, |\pi_{l,a}| \ r_{a+{\tt b}(l)} - \sum_{l,l'=1}^r \ \sum_{a\in\widehat{G}}\, |\pi_{l,a}|\, \Big( |\pi_{l',a+{\tt b}(l)-{\tt b}(l'\,)-a_1-a_2}| - |\pi_{l',a+{\tt b}(l)-{\tt b}(l'\,)-a_1}| \nonumber \\ && \hspace{6cm} -\, |\pi_{l',a+{\tt b}(l)-{\tt b}(l'\,)-a_2}| + |\pi_{l',a+{\tt b}(l)-{\tt b}(l'\,)}| \Big)
\label{instmeasure}\end{eqnarray}
where the
$|G|$-vector $\mbf
r=(r_a)_{a\in\widehat{G}}= (\dim W_a)_{a\in\widehat{G}}$ parametrizes
the number of eigenvalues
of the Higgs fields $e_l$ which correspond to a particular irreducible
representation $\rho_a$ of $G$. For rank $r=1$ and the trivial
framing, the sign factor (\ref{instmeasure}) coincides with that
of~\cite[Ex.~23]{bryan} which was computed geometrically using techniques of orbifold
Donaldson-Thomas theory. From (\ref{chE}) we can read off the fixed
point values of the pertinent Chern characteristic classes
\bea
\ch_2(\cE_{\vec\pi})&=& \sum_{a,b\in\widehat{G}}\, \bigg(
\Big( r_b\, \delta_{ab} - \big( {\tt a}^{(2)}_{ab} - {\tt
  a}^{(1)}_{ab} \big) \, \sum_{l=1}^r \, |\pi_{l,b-{\tt b}(l)}| \Big)
\ \mathrm{ch}_2 ({\calR_a}) \nonumber \\ && \qquad \qquad +\,\big(  {\tt a}_{ab}^{(2)}  - 3 \delta_{ab} \big) \, \sum_{l=1}^r\, |\pi_{l,b-{\tt b}(l)}| \ c_1 \big(\cO_{\overline{X}}(\bdiv)\big) \wedge c_1(\calR_a)\bigg) \ , \label{ch3Naction} \\[4pt] \ch_3(\cE_{\vec\pi}) &=&
 -\sum_{a,b\in\widehat{G}}\, \bigg(\Big( r_b\,\delta_{ab} - \big( {\tt
   a}^{(2)}_{ab} - {\tt a}^{(1)}_{ab} \big) \, \sum_{l=1}^r\,
 |\pi_{l,b-{\tt b}(l)}| \Big) \ \mathrm{ch}_3 ({\calR_a})
 +\frac{\delta_{ab}}{|G|}\, |\pi_{l,b-{\tt b}(l)}| \
 \frac{t\wedge t\wedge t}6 \cr && \qquad \qquad
  +\,\big( {\tt a}_{ab}^{(2)} - 3 \delta_{ab} \big) \, \sum_{l=1}^r\,
  |\pi_{l,b-{\tt b}(l)}| \ \Big( c_1 \big(\cO_{\overline{X}}(\bdiv)\big) \wedge \ch_2 (\calR_a)
  \nonumber \\ & & \qquad \qquad \qquad \qquad\qquad \qquad\qquad \qquad \qquad
  \qquad +\,  c_1
  (\calR_a) \wedge \ch_2 \big(\cO_{\overline{X}}
  (\bdiv)\big)\Big) \bigg) \ , \nonumber
\end{eqnarray}
where the choice of boundary condition ${\tt b}$ enters not only
explicitly in the dimensions $r_a$, but also implicitly in the plane
partitions.

Finally the instanton partition function for noncommutative Donaldson-Thomas invariants of type $\mbf r$ is given by
\begin{equation}
Z_{\rm gauge}^{[\complex^3 /  G]}(q,Q;\mbf r) = \sum_{\vec \pi} \,
(-1)^{\cK_G(\vec\pi;\mbf r)} \, q^{\ch_3(\cE_{\vec\pi})} \,
Q^{\ch_2(\cE_{\vec\pi})} \ ,
\end{equation}
where the counting weights are naturally expressed via (\ref{ch3Naction}) in terms of intersection
indices on the homology of the crepant resolution $X= \mathrm{Hilb}_{
  G} (\complex^3)$, via the McKay correspondence, which 
can be determined from the toric graph $\Delta$ of $X$. However it is
computed via the $G$-equivariant instanton charges that characterize
the noncommutative Donaldson-Thomas invariants, which are the relevant
variables in the noncommutative crepant resolution chamber: As shown
in~\cite{Cirafici:2010bd,Cirafici:2011cd}, there exists a simple
change of variables from the large radius parameters $(q,Q)$ to
orbifold parameters $p=(p_a)_{a\in\widehat{G}}$ with
$\prod_{a\in\widehat{G}}\, p_a=q$ such that the gauge theory partition
function assumes the form
\beq
Z_{\rm gauge}^{[\complex^3 /  G]}(p;\mbf r ) = \sum_{\vec \pi} \,
(-1)^{\cK(\vec\pi;\mbf r)} \ \prod_{a\in\widehat{G}} \,
  p_a^{\sum_{l=1}^r \, |\pi_{l,a-{\tt b}(l)}|} \ ,
\eeq
which should be compared with the BPS partition function
(\ref{NCZBPS}) associated with the McKay quiver~$\sfQ_G$.

In this way the instanton counting problem yields the orbifold
Donaldson-Thomas invariants defined in~\cite{youngbryan} for ideal sheaves
and more generally in~\cite{joycesong}. For this, we associate to our framed quiver $\hat\sfQ_G$ the representation space
\begin{equation}
{\sf Rep}_G (\mbf n , \mbf r) =  \Hom_{G} (V , Q \otimes V) \ 
 \oplus \ \Hom_{G} (V , \mbox{$\bigwedge^3$} Q \otimes V) \ 
\oplus \ \Hom_{G} (W , V) \ ,
\end{equation}
and let ${\sf Rep}_G (\mbf n , \mbf r;B)$ be the subvariety cut out by the $G$-equivariant decomposition of the matrix equations
(\ref{ADHMorb}), which generate the ideal of relations in
the instanton quiver path algebra $\sfA_G$. This allows us to define
the BPS quiver moduli space as the quotient stack
\begin{equation}
\scrM_G (\mbf n ,\mbf r) = \Big[ {\sf Rep}_G (\mbf n , \mbf r;B) \,
\Big/ \, \prod_{a\in\widehat G}\, GL(n_a,\IC) \Big]
\end{equation}
by the gauge group which acts as basis change automorphisms of the
$G$-module $V$; we regard this stack as a moduli space of stable framed
representations, where every object in the category of quiver representations with relations is $0$-semistable~\cite[\S7.4]{joycesong}. Noncommutative
Donaldson-Thomas invariants may now be defined using Behrend's
weighted topological Euler characteristic which can here be identified
explicitly as
\begin{equation}
\DT_{\mbf n , \mbf r}(\sfA_G) = \chi \big( \scrM_G (\mbf n ,\mbf r) \,
, \, \nu_{\sfA_G} \big) = \sum_{\vec\pi\,:\, \sum_l\,|\pi_{l,a-{\tt b}(l)}|=n_a}\, (-1)^{\cK_G(\vec\pi;\mbf r)} \ ,
\label{BehrendNC}\end{equation}
where $\nu_{\sfA_G}: \scrM_G (\mbf n ,\mbf r) \to\zed$ is an invariant
constructible function. In the rank one case $r=1$, these invariants coincide with the
invariants given in (\ref{DTfantechi}). 

Note that the collective coordinate dynamics is determined in terms of
\emph{cyclic} $\sfA_G$-modules $V$, i.e. $V=\sfA_G v_a$ is generated by
the action of the path algebra $\sfA_G$ of the quiver on a reference node
$v_a\in V_a$. Hence the moduli space of $0$-semistable $\hat
\sfA_G$-modules parametrizes the $a$-cyclic modules, or equivalently
finite-dimensional quotients of the projective $\hat\sfA_G$-module
$\sfP_a=\sfe_a\hat\sfA_G$ of dimension vector $\mbf n$~\cite{reineke}.
The choice of boundary
function $\tt b$ in (\ref{BehrendNC}) labels a superselection sector in the space of states
of the worldvolume gauge theory. In the framework
of~\cite{Ooguri:2008yb}, it determines how cyclic modules of the
framed
McKay quiver are based and therefore the particular enumerative
problem; in this setting, a choice of reference vertex $v_a$ for the
counting is simply a choice of asymptotic boundary condition on the instanton gauge fields. However, all invariants $\DT_{\mbf n , \mbf r}(\sfA_G)$
are {equivalent}, as they can all be expressed in terms of the same
set of quiver invariants which are independent of the boundary
conditions~\cite{joycesong}; this feature nicely agrees with physical
expectations of the noncommutative BPS
invariants. See~\cite{Cirafici:2010bd,Cirafici:2011cd} for further
details of the properties of these invariants. 

\subsection{Closed topological vertex geometry}

Let us consider the explicit example $\IC^3/\IZ_2\times\IZ_2$
following~\cite{Cirafici:2010bd}. The orbifold group
$G=\IZ_2\times\IZ_2$ contains the identity $g_0$ plus three elements
$g_1,g_2,g_3$ acting on $\IC^3$ with respective weights
\beq
a_1=(1,1,0) \ , \qquad a_2=(1,0,1) \qquad \mbox{and} \qquad
a_3=a_1+a_2=(0,1,1) \ .
\eeq
The framed McKay quiver $\hat\sfQ_{\IZ_2\times\IZ_2}$ is given by
\beq
\vspace{4pt}
\begin{xy}
\xymatrix@C=20mm{
& & \ {W_0 \ar@{.>}@//[d]^< < < < < < < < {I^{(0)}}} & & \\
& & \ V_0 \ \ar@/^/[ddr] \ar@/_1pc/[ddl]  \ar@/^/[d] & & \\
& & \ V_3 \ \ar@/^/[u] \ar@/^/[dl] \ar@/^/[dr] & & \\
{W_1 \ \ar@{.>}@//[r]^< < < < < < < <{\!\!\!\!\!\!\!\!I^{(1)}}} & \ V_1 \ \ar@/^/[uur] \ar@/^/[ur] \ar@//[rr] & &
\ V_2  \ \ar@/_1pc/[uul] \ar@/^/[ul] \ar@/^/[ll] & \ \ar@{.>}@//[l]_<
< < < < < < <{ \ \ \ \ \ \ \ \ I^{(2)}} \ W_2 \\
& & {W_3 \ar@{.>}@//[uu]_< < < {I^{(3)}}} & & 
}
\end{xy}
\vspace{4pt}
\eeq
and the natural crepant resolution $X={\rm
  Hilb}_{\IZ_2\times\IZ_2}(\IC^3)$ is the closed topological vertex
geometry whose toric diagram $\Delta$ consists of four vertices joined
pairwise by three edges with corresponding K\"ahler parameters denoted
$Q_1$, $Q_2$ and $Q_3$. Then the rank one instanton partition function
of the quotient stack $[\IC^3/\IZ_2\times\IZ_2]$ for the trivial
boundary condition $\mbf r=(1,0,0,0)$ is given by
\begin{equation}
Z_{\rm gauge}^{[\IC^3/\zed_2 \times \zed_2]}(p,p_1,p_2,p_3) = \sum_{\pi} \, (-1)^{|\pi_1| + |\pi_2| + |\pi_3|} \ p^{|\pi|}\, p_1^{|\pi_1|}\, p_2^{|\pi_2|}\, p_3^{|\pi_3|} \ ,
\end{equation}
where the change of variables given by
\begin{eqnarray}
p  \ = \ p_0\, p_1\, p_2\, p_3 &=& q^{5/8}\, Q_1 \, Q_2 \, Q_3 \ , \nonumber\\[4pt]
p_1 &=& q^{-1/2}\,Q^{-2}_2 \, Q^{-2}_3 \ , \nonumber\\[4pt]
p_2 &=& q^{-1/2}\, Q^{-2}_1 \, Q^{-2}_3 \ , \nonumber\\[4pt]
p_3 &=& q^{-1/2}\,Q^{-2}_1 \, Q^{-2}_2 
\end{eqnarray}
is the mapping between fractional D0-brane
charges, corresponding to each configuration represented by a
4-coloured plane
partition, and the D2--D0 charges on $X$. In this case, the Donaldson-Thomas
partition functions of $[\IC^3/\IZ_2\times \IZ_2]$ and its natural crepant resolution
$X={\rm Hilb}_{\IZ_2\times \IZ_2}(\complex^3)$ are related through
\beq
Z_{\rm gauge}^{[\IC^3/\zed_2 \times \zed_2]} (p , p_1 , p_2 , p_3) =
M(-p)^{-4} \ Z^X_{\rm top}(p,p_1,p_2,p_3)\, Z^X_{\rm
  top}(p,p^{-1}_1,p^{-1}_2,p^{-1}_3)
\eeq
where the topological string partition function is given by
\beq
Z^X_{\rm top}(p,p_1,p_2,p_3)= M(-p)^4 \ \frac{M(p_1\,p_2,-p)\,
  M(p_1\,
  p_3,-p)\,M(p_2\,p_3,-p)}{M(p_1,-p)\,
  M(p_2,-p)\, M(p_3,-p)\,
  M(p_1\,p_2\, p_3,-p)} \ .
\eeq
Here the variables $p_1$, $p_2$ and $p_3$ correspond to the basis
of curve classes $Y$ (D2-branes) in $X$ and $p$ to the Euler number
$\chi(\cO_Y)$ (D0-branes). The stacky gauge theory on
$[\IC^3/\IZ_2\times \IZ_2]$ thus realizes the anticipated
wall-crossing behaviour of
the BPS partition function $\cZ_{\rm BPS}^{\sfA_{\IZ_2\times\IZ_2}}(p)$, connecting
the orbifold point with the large radius point in the
K\"ahler moduli space through collapsings of two-cycles in the
resolved geometry.

\subsection{Noncommutative mirror symmetry}

Recall~\cite{Hori:2000kt} that the mirror manifold $\widetilde{X}$ of a toric Calabi-Yau
threefold $X$ is defined by an equation of the form
\begin{equation}
u \, v + P_\Delta (z,w;t) = 0 
\end{equation}
where $u,v\in \complex$, $z,w \in \complex^\times$, and $P_\Delta(z,w;t)$ is
the Newton polynomial whose monomial terms $z^n\, w^m$ are
constructed from the vertices $(n,m)\in\IZ_{\geq0}^2$ of the toric diagram
$\Delta$ of $X$. The mirror B-model topological string theory is described by a
Landau-Ginzburg model whose superpotential $W:\widetilde{X}\to\IC$ is a functional of the
Newton polynomial $P_\Delta$.

On the other hand, the formalism we have outlined in this section computes BPS states in the noncommutative
crepant resolution chamber of the Calabi-Yau moduli space. In this region the geometry is replaced by the
path algebra of a certain framed quiver, where the framing nodes are
associated with non-compact D6-branes wrapping the full Calabi-Yau
threefold. It is natural to ask how this picture behaves under mirror
symmetry. This question was partly addressed
in~\cite{Ooguri:2009ri}, where it was shown that the $q\to1$
limit of the BPS partition function is captured by the genus zero topological string amplitude of the mirror manifold. It would be interesting however to examine
directly
what is the analog of the noncommutative crepant resolution on the
mirror side, and if one can set up the BPS state counting problem
there.

A possible avenue of investigation is the proposal of
\cite{auroux,bocklandt}, where the proper object to consider on the
mirror side would be the \textit{wrapped} Fukaya category constructed
in \cite{seidel}, whose objects also include non-compact Lagrangian
submanifolds which are mirror to the non-compact D6-branes. The
proposal of \cite{bocklandt} consists in looking for an appropriate full
subcategory of the wrapped Fukaya category, precisely as here we are
considering the category of quiver representations which can be
regarded as a subcategory of the derived category of coherent
sheaves. This subcategory is constructed using a certain dimer
model, i.e. a quiver with superpotential on a Riemann surface whose Jacobi algebra is a
noncommutative 3-Calabi-Yau algebra. It would be interesting to understand the relation between this
dimer model and our framed quivers. If one neglects the framing, we
expect the two pictures to be related by the dimer duality of
\cite{Feng:2005gw}. The framing nodes introduce additional complications;
in particular it is not clear what would be the analog of the
noncommutative invariants~$\DT_{{\mbf n},{\mbf r}}(\sfA_G) $. 

\bigskip

\section{Instantons and the special McKay correspondence\label{sec:specialmckay}}

\subsection{Special McKay quivers}

So far we have shown how (generalized) instanton moduli spaces are
deeply related with moduli spaces of quiver representations. This is
true for toric Calabi-Yau twofolds and threefolds, where we have set
up the instanton counting problem. It is natural to wonder how this
picture can be generalized. In this final section we will discuss a
possibility in this direction by considering more general
singularities of the form $\mathbb{C}^2 /  G$ where $ G \subset GL(2
,\mathbb{C})$. The resolutions of these singular orbifolds are just
the Hirzebruch-Jung surfaces from \S\ref{subsec:HJsurfaces}. The
construction of instanton moduli spaces on generic Hirzebruch-Jung
surfaces is an open problem. The extension of the formalism discussed in
\S\ref{subsec:N2quiver} only holds for the resolutions of $A_{p,1}$
singularities; in fact this family only includes the $A_1$ ALE space,
whose instanton moduli space and partition functions were singled out
with distinctive special properties,
whereas all ALE spaces can be treated using the formalism of quiver
varieties from \S\ref{subsec:ALEmodsp}, which was also the approach
taken in \S\ref{subsec:orbifoldBPS} to construct instanton moduli
spaces on singular orbifolds in six dimensions; a monadic
parametrization of framed moduli spaces on $A_{p,1}$ resolutions is
developed in~\cite{rava} which may lead to a suitable reformulation as
a topological matrix model. Hence we suggest that
the problem should be addressed within the framework of the
\textit{special McKay correspondence}, which we shall now describe.

The main issue in the general case is that there are more irreducible
representations of the orbifold group $ G$ than there are exceptional
curves (fractional instantons) in the resolution $M$, as is witnessed by the $A_{p,1}$
singularities with $p>2$ for example; hence we lose one of the main features that
tautological bundles in crepant resolutions form an
integral basis of $K(M)$ dual to the exceptional divisors.
There is however a canonical set of
tautological sheaves which are dual to the exceptional set with
respect to the intersection pairing (\ref{intersection}) on Chow theory, and which are
called \textit{special}; similarly, we will define
special representations to be those representations of $ G$ whose
associated tautological sheaf is dual to a rational curve in the
exceptional set. This suggests that the construction of the instanton moduli space
should be rephrased in terms of special representations alone, in
order to correctly obtain the appropriate mapping between fractional
0-brane charges on the orbifold and D2--D0
brane charges on the resolution; the \emph{special McKay quiver} is constructed out of the special sheaves.

Consider the universal scheme $\cZ \subset M_{p,p'} \times
\complex^2$ with correspondence diagram
\begin{equation}
\xymatrix@=10mm{
  & \cZ \ar[ld]_{p_1}\ar[rd]^{p_2}& \\
  M_{p,p'} & & \complex^2
}
\end{equation}
and define the tautological bundle
\begin{equation}
\calR := p_{1*} \cO_{\cZ} \ .
\end{equation}
Under the action of $ G=G_{p,p'}$ on $\cZ$, the bundle $\calR$ transforms in the
regular representation and can thus be decomposed into
irreducible representations as
\begin{equation}
\calR = \bigoplus_{a\in \widehat{G}} \, \calR_a \otimes \rho_a
\end{equation}
where $\{ {\rho}_a \}_{a\in \widehat{G}}$ is the set of irreducible
representations of $G\cong\IZ_p$. 

Being a pullback of reflexive sheaves, the tautological sheaf satisfies
the cohomological condition $H^1(M,\calR^\vee\otimes K_M)=0$. 
The special sheaves are the tautological line bundles $\calR_a$ which
obey $H^1 (M , \mathcal{R}_a^{\vee}\, ) =
0$; we will reserve the notation $\calL_i$ for the special
sheaves. Note that when $M$ is Calabi-Yau all tautological bundles
$\calR_a$ are special. The
line bundles $\calL_i$, together the trivial line bundle $\calL_0=\cO_M$, form
an integral basis of the K-theory group $K(M)$ of vector bundles on
$M=M_{p,p'}$ which are dual to the exceptional set~\cite{wunram},
i.e. $\int_{D_i}\, c_1(\calL_j)=\delta_{ij}$. A stronger result
actually holds: The direct sum
\beq
\cW =
\bigoplus_{i= 0}^m \, \calL_i
\eeq
is a tilting bundle \cite{vandenbergh} and
consequently the set $\{\calL_i\}_{i\geq0}$ forms a full strongly exceptional collection~\cite[Prop.~6.6]{crawreview}. The associated functor $\Hom_{\cO_M} (\cW
, - )$ induces an equivalence between the bounded derived category
$\frD(M)$ of
coherent sheaves on $M_{p,p'}$ and the bounded derived category
$\frD(\sfA)$ of
representations of the tilting algebra $\sfA= \mathrm{End}_{\cO_M}(\cW)$.

To rephrase the special McKay correspondence as an
equivalence of derived categories, Craw introduces in \cite{craw}
the bound special McKay quiver $(\tilde\sfQ_{p,p'},\tilde\sfR)$ as
the complete bound quiver of the exceptional collection $\calL_i$;
its path algebra coincides with the tilting algebra
$\sfA$. In particular this means that the nodes of the quiver
correspond to the special sheaves, while the arrows and
relations can be read off from the cohomology groups of tensor
products of the bundles $\calL_i$. Craw proves in~\cite{craw} that the resolution $M_{p,p'}$ is isomorphic to the fine moduli space of stable representations of the special McKay quiver with a particular dimension vector; whence this quiver plays the same role that the ordinary McKay quiver played for $A_p$ singularities.

The relation between the special McKay quiver and geometry is
even more clear in its explicit construction, due to Wemyss
\cite{wemyss}, which generalizes that of the ordinary McKay correspondence. The
derived equivalence between left $\sfA$-modules and coherent sheaves on
$M$ has a stronger form, proven by van den Bergh \cite{vandenbergh},
which is encoded in the commutative diagram
\begin{equation}
\begin{xy}
\xymatrix@C=10mm{ \frD \left( M\right)  \ \ar[r] & \ \frD\left( \sfA
  \right) \\  \mathfrak{P}(M) \ \ar[u] \ar[r] & \ \frM od(\sfA) \ar[u]
}
\end{xy}
\end{equation}
where the vertical arrows are inclusions of subcategories, and the
subcategory $\mathfrak{P}(M)$ of perverse coherent sheaves on $M$
consists of objects satisfying certain conditions on their cohomology
sheaves; for our purposes we can regard $\mathfrak{P}(M)$ as the
subcategory into which the module subcategory $\frM od(\sfA) $ is mapped
by the derived equivalence. In particular, the fractional 0-brane simple
modules $\sfD_i$, $i=0,1,\dots,m$ in $\frM od(\sfA) $ are mapped to the objects
\begin{equation}
\cO_{D_0} \ , \ \cO_{D_1} (-1) [1] \ , \ \dots \ , \ \cO_{D_m} (-1)
[1] \ ,
\end{equation}
where $D_i$, $i=1,\dots,m$ are the exceptional curves corresponding to
the special representations and the fundamental cycle
$D_0=\sum_{i=1}^m \, D_i$ corresponds to the trivial representation of
$G$; note that $n=\chi(\cO_{D_0})=0$ by (\ref{chiOD}).
This relation translates algebraic problems into geometric ones. In
particular, we can reduce the problem of finding the arrows and
relations of the special McKay quiver, i.e. of computing the
dimensions of the groups $\Ext_{\sfA}^p(\sfD_i ,\sfD_j)$ between the
simple representations $\sfD_i$, to a problem in algebraic
geometry. The numbers of arrows and relations are given by the
dimensions of the cohomology groups ${\tt b}_{ij}^{(1)}:= \dim
\Ext^1_{\sfA} ( \sfD_i , \sfD_j)$ and ${\tt b}_{ij}^{(2)}:= \dim
\Ext^2_{\sfA} ( \sfD_i , \sfD_j)$ respectively, with~\cite{wemyss}
\begin{eqnarray}
\Ext^1_{\sfA} \left( \sfD_i , \sfD_j \right) &=& \Ext_{\cO_M}^1 \big(
\cO_{D_i} (-1)  \,,\, \cO_{D_j} (-1) \big) \ , \nonumber \\[4pt]
\Ext^2_{\sfA} \left( \sfD_i , \sfD_j \right) &=& \Ext_{\cO_M}^2 \big(
\cO_{D_i} (-1)  \,,\, \cO_{D_j} (-1) \big) \ , \nonumber \\[4pt]
\Ext^1_{\sfA} \left( \sfD_0 , \sfD_0 \right) &=& \Ext_{\cO_M}^1 (
\cO_{D_0} , \cO_{D_0} ) \ , \nonumber \\[4pt]
\Ext^2_{\sfA} \left( \sfD_0 , \sfD_0 \right) &=& \Ext_{\cO_M}^2
(\cO_{D_0} , \cO_{D_0} ) 
\end{eqnarray}
for maps between vertices $i,j\neq 0$ or between vertex $0$ and
itself. These are the maps between objects of the same degree. The
remaining arrows and relations occur between objects of different
degree with
\begin{eqnarray}
\Ext^1_{\sfA} \big( \sfD_i , \sfD_0 \big) &=& \Ext_{\cO_M}^1 \big(
  \cO_{D_0} \,,\,  \cO_{D_i} (-1)[1]  \big) \ = \ \Ext_{\cO_M}^2
\big( \cO_{D_0}\,,\,  \cO_{D_i} (-1) \big) \ , \nonumber \\[4pt]
\Ext^2_{\sfA} \big( \sfD_i , \sfD_0 \big) &=& \Ext_{\cO_M}^2 \big(
  \cO_{D_0} \,,\,  \cO_{D_i} (-1) [1]  \big) \ = \ \Ext_{\cO_M}^3
\big( \cO_{D_0} \,,\,  \cO_{D_i} (-1) \big) \ , \nonumber \\[4pt]
\Ext^1_{\sfA} \big( \sfD_0 , \sfD_i \big) &=& \Ext_{\cO_M}^1 \big(
  \cO_{D_i} (-1) [1] \,,\, \cO_{D_0} \big) \ = \ \Hom_{\cO_M} \big(
  \cO_{D_i} (-1) \,,\, \cO_{D_0} \big)\ , \nonumber \\[4pt]
\Ext^2_{\sfA} \big( \sfD_0 , \sfD_i \big) &=& \Ext_{\cO_M}^2 \big(
  \cO_{D_i} (-1) [1] \,,\, \cO_{D_0} \big) \ = \ \Ext_{\cO_M}^1
\big(  \cO_{D_i} (-1) \,,\, \cO_{D_0} \big) \ .
\end{eqnarray}

To construct the quiver $\tilde\sfQ_{p,p'}$, one draws a node for each summand of the
tilting bundle $\cW$, corresponding to the divisors $D_0,D_1,\dots,D_m$.
There is also a
canonical cycle $Y_K$ defined by
\begin{equation}
\langle Y_K , D_i\rangle_\Gamma := - \langle K_M, D_i \rangle_\Gamma =
\langle D_i,D_i\rangle_\Gamma + 2 \ .
\end{equation}
Note that this cycle is trivial in the Calabi-Yau case. Equipped with
these definitions we can build the special McKay quiver whose data is
summarised in Table~\ref{special}; for $x\in\IR$, we use the notation $[x]_\pm=
\pm\, x$ for $\pm\,x>0$ and $[x]_\pm=0$ for $\pm\, x\leq0$.
\begin{table}[htdp]
\begin{center}
\begin{tabular}{|c|c|c|} \hline
 $v_1 \longrightarrow v_2$ & arrows ${\tt b}^{(1)}$ & relations ${\tt b}^{(2)}$  \cr \hline \hline
  $i \longrightarrow j$ & $ \big[\langle D_i ,
  D_j\rangle_\Gamma\big]_+$ &  $\big[-1-\langle D_i , D_j\rangle_\Gamma\big]_+$  \cr \hline
  $ 0 \longrightarrow 0 $ & $0$ & $-1-\langle D_0, D_0\rangle_\Gamma$ \cr \hline
  $ i \longrightarrow 0 $ &  $-\langle D_i , D_0\rangle_\Gamma$ & $0$ \cr \hline
  $ 0 \longrightarrow i $ & $\big[\langle Y_K - D_0 ,
  D_i\rangle_\Gamma \big]_+$ & $\big[\langle Y_K - D_0, D_i\rangle_\Gamma \big]_-$ \cr  \hline
  \end{tabular}
\end{center}
\caption{\footnotesize{The numbers of arrows in $\tilde\sfQ_1$ and
    relations in
    $\tilde\sfR$ for the special McKay quiver $\tilde\sfQ_{p,p'}$.}}
\label{special}
\end{table}

Throughout this section we will consider the explicit example of the
$A_{7,2}$ singularity with $(p,p'\,)=(7,2)$ for illustration. Then $M=M_{7,2}$ is the minimal resolution of
$\complex^2/ G$ where $G=G_{7,2}$ acts on $\IC^2$ by $(z_1,z_2) \mapsto (\zeta\, z_1, \zeta^2\, z_2)$ with $\zeta^7=1$. The resolved geometry has two exceptional curves $D_1$ and $D_2$ characterized by the intersection matrix
\begin{equation}
C=\left( \begin{matrix} - 4 & 1\\ 1 & -2 \end{matrix} \right) \ ,
\end{equation}
as the continued fraction expansion (\ref{contfrac}) in this instance
reads $\frac{7}{2} = [4,{2}]$. From this matrix one can easily compute
the requisite intersection products in Table~\ref{special} corresponding
to the matrices of the number of arrows ${\tt b}_{ij}^{(1)}$ and
relations ${\tt b}_{ij}^{(2)}$ for $i,j=0,1,2$ with
\begin{equation}
{\tt b}^{(1)} = \left( \begin{matrix} 0 & 3 & 1 \\ 1 & 0 & 1 \\ 1 & 1 & 0 \end{matrix} \right)
\qquad \mbox{and} \qquad {\tt b}^{(2)} = \left( \begin{matrix} 3 & 0 &
    0 \\ 0 & 3 & 0 \\ 0 & 0 & 1 \end{matrix} \right) \ .
\end{equation}
This construction gives the special McKay quiver $\tilde\sfQ_{7,2}$
with diagram 
\begin{equation}
\begin{xy}
\xymatrix@C=30mm{
& \ 2  \ \ar@/^/[dr]|{ \ a_2 \ }  \ar@/^/[dl]|{ \ a_3 \ } & \\
  1  \ \ar@/^/[ur]|{ \ a_4 \ }   \ar@/_4pc/[rr]|{ \ a_6 \ }
  \ar@/_3pc/[rr]|{ \ a_7 \ }  \ar@/_2pc/[rr]|{ \ a_8 \ } & & \ar[ll]|{
    \ a_5 \ }  \ar@/^/[lu]|{ \ a_1 \ } \ 0
}
\end{xy}
\label{72quiver}\end{equation}
whose ideal of relations is generated by the set
\begin{eqnarray} \label{rels1712}
\tilde{\sf{R}} &=& \Big\{ \ a_1\, a_2 - a_5\, a_6  \ , \  a_5\, a_7 -
a_1\, a_3\, a_6  \ , \  a_5\, a_8 - a_1\, a_3\, a_7 \  , \  a_4\, a_3
- a_6\, a_5 \ , \ \cr
&&  \qquad \ a_7\, a_5 -a _6\, a_1\, a_3 \  ,  \  a_8 \,a_5 - a_7\,
a_1\, a_3 \  , \   a_2\, a_1 - a_3 \,a_4  \ \Big\} \ .
\end{eqnarray}

\subsection{From McKay to special McKay quivers}

The special McKay quiver can also be obtained from the ordinary McKay
quiver of $ G$, by deleting nodes which do not correspond to special representations and
modifying the ideal of relations suitably to set to zero any
composition of arrows which passes through the removed node. Starting from the
resolution $M$ we can construct the McKay quiver in the usual way;
this is the quiver whose vertices are labelled by the irreducible
representations of $ G\subset GL(2,\IC)$. The main difference from the
ordinary McKay quiver is that the matrix of coefficients ${\sf
  a}_{ab}$, while still traceless, is no longer symmetric in general. For $ G=G_{p,p'}$ the fundamental
representation is $Q = \rho_1 \oplus \rho_{p'}$ (recall that $p'=p-1$ recovers the $A_{p-1}$ du Val singularity). The tensor product representation again decomposes as
\begin{equation}
Q \otimes \rho_a = \rho_{a+1} \oplus \rho_{a+p'} \ ,
\end{equation}
while the determinant representation gives
\begin{equation}
\mbox{$\bigwedge^2$} Q\otimes \rho_a=  \rho_{1+p'}\otimes \rho_a =
\rho_{1+a+p'} \ .
\label{decomporeps}\end{equation}
Each vertex labelled by a representation $\rho$ has two incoming arrows, $a_1^\rho$ from the vertex $\rho \otimes \rho_1$ and $a_2^{\rho}$ from the vertex $\rho \otimes \rho_{p'}$.
From this data we can construct the McKay quiver $(\sfQ_{p,p'},\sfR)$ where
the ideal of relations is generated by the set
\begin{equation}
{\sf R} = \Big\{ \, c_2^{\rho \otimes \rho_1}\, c_1^{\rho} - c_1^{\rho
  \otimes \rho_{p'}} \, c_2^{\rho}  \ \Big| \ \rho \in \widehat{G} \,
\Big\} \ .
\end{equation}
This construction
realizes $M=M_{p,p'}$ as the moduli space of stable representations of
the bound McKay quiver. One can prove~\cite{craw}
that $(\sfQ_{p,p'},\sfR)$ is the complete bound quiver
generated by the sections of the tautological bundles $\calR_a$
labelled by
\textit{all} irreducible representations of $G$, and that the
corresponding path algebra is isomorphic to $\End_{\cO_M}
\big(\bigoplus_{a\in\widehat{G}}\, \calR_a\big)$.

Let us consider again the example $(p,p'\,)=(7,2)$. The irreducible
representations ${\rho}_a$ correspond to the characters $\zeta^a$ for $a=1,\dots,6$, plus the trivial representation ${\rho}_0$. Then the tensor product decomposition is
\begin{eqnarray}
Q \otimes \rho_a = \rho_{a+1} \oplus \rho_{a+2} \ .
\end{eqnarray}
Note that the matrix encoding the decomposition between
representations and the intersection matrix have different rank, since
the underlying resolution has only two exceptional curves. The McKay
quiver $\sfQ_{{7,2}}$ takes the form
\begin{equation}
\begin{xy}
\xymatrix@C=5mm{ &  & \ar@/^/[rrd]|{ \ c_1 \ }  \ \rho_0  \ \ar@/_/[rrdd]|{ \ c_2 \ } &  & \\
  \ \rho_1  \  \ar@/^/[urr]|{ \ c_1 \ }  \ar@/^/[rrrr]|{ \ c_1 \ } &  &  & & \ar@/_/[ddl]|{ \ c_2 \ }  \  \rho_6  \  \ar@/^/[d]|{ \ c_1 \ } \\
 \ar@/_/[uurr]|{ \ c_2 \ }  \ \rho_2  \  \ar@/^/[u]|{ \ c_1 \ } &   &
 & &  \ \rho_5 \   \ar@/^/[ld]|{ \ c_1 \ }   \ar@/_/[dlll]|{ \ c_2 \ }
 \\  &  \ar@/_/[uul]|{ \ c_2 \ }  \  \rho_3  \  \ar@/^/[ul]|{ \ c_1 \
 }  & &  \ar@/^/[ll]|{ \ c_1 \ }  \ \rho_4  \  \ar@/_/[ulll]| { \ c_2 \ } & 
}
\end{xy}
\end{equation}
With the conventions of \cite{craw}, to each representation $\rho$ we
can associate the tautological bundle $\calR_{\rho^*}$ which
corresponds to the module $( \complex^2 \otimes \rho^{*}
)^{ G}$. To derive the special McKay quiver we keep only the
special tautological sheaves, which in this case are $\calR_{\rho_1^*}=:\calL_2
$ and $\calR_{\rho_2^*}=:\calL_1$ corresponding
respectively to the two exceptional curves $D_2$ and $D_1$, together
with the trivial sheaf $\calL_0 := \cO_{M}$. Removing the rest of the
vertices gives rise to another sequence of arrows which come from the
paths connecting the remaining vertices via the vertices that have
been removed. For example, removing the node corresponding to $\rho_4$
yields the new quiver
\begin{equation}
\begin{xy}
\xymatrix@C=5mm{ &  & \ar@/^/[rrd]|{ \ c_1 \ }  \ \rho_0 \  \ar@/_/[rrdd]|{ \ c_2 \ } &  & \\
  \ \rho_1  \  \ar@/^/[urr]|{ \ c_1 \ }  \ar@/^/[rrrr]|{ \ c_1 \ } &  &  & & \ar@/_/[dllll]|{ \ c_2\, c_2 \ }   \ \rho_6  \  \ar@/^/[d]|{ \ c_1 \ } \\
 \ar@/_/[uurr]|{ \ c_2 \ }  \ \rho_2  \  \ar@/^/[u]|{ \ c_1 \ } &   &
 & &  \ \rho_5 \   \ar@/^/[llld]|{ \ c_1 \, c_1 \ }   \ar@/_/[dlll]|{ \ c_2 \ } \\  &  \ar@/_/[uul]|{ \ c_2 \ }  \  \rho_3  \  \ar@/^/[ul]|{ \ c_1 \ }  & & &
}
\end{xy}
\end{equation}
and so on. The net result of this process is the special quiver
$\tilde\sfQ_{{7,2}}$ generated by the sections of the special
tautological sheaves $\cO_M , \calL_1 , \calL_2$; it coincides with
(\ref{72quiver}) with the same ideal of relations~(\ref{rels1712}). 

We can now construct the fine moduli space of
$\tilde{\theta}$-stable representations of the special McKay quiver
$\scrM_{\tilde{\theta}} (\tilde{\sfQ}_{p,p'} , \tilde{\sfR})$; it is
isomorphic to the minimal resolution of the $\complex^2 /  G$
singularity, i.e. the moduli scheme $M=\mathrm{Hilb}_G
(\complex^2)$ of $G$-clusters. Its path algebra $\sfA_{p,p'} = \complex \tilde{\sf{Q}}_{p,p'} /
\langle \tilde{\sf{R}} \rangle$ is isomorphic to the endomorphism
algebra $\End_{\cO_M} \big(  \bigoplus_{i \in \widehat{G}_s}\, \calL_i
\big)$, where the sum runs over the special representations
$\tilde{\rho}_i \in \widehat{G}_s $. This algebra is isomorphic to a
particular subalgebra of the path algebra of the McKay quiver,
obtained by keeping only the special and trivial representations. We
turn next to the study of this subalgebra.

\subsection{Reconstruction algebras}

To the special McKay quiver one can associate its path algebra; we
wish to base the construction of the instanton moduli spaces on this algebra, which can be studied on its own to provide the necessary homological ingredients. This
algebra was extensively studied in \cite{wemyssA} for cyclic
singularities and is called the {\it reconstruction algebra}
$\sfA=\sfA_{p,p'}$. The reconstruction algebra has an abstract
definition rooted
in the representation theory of the discrete group $ G=G_{p,p'}$; we will not need this abstract machinery and will regard
$\sfA$ as the path algebra of the special McKay quiver. 

One of the ingredients we need is the global dimension of the reconstruction algebra. The geometric data of the resolution are encoded in the continued
fraction expansion (\ref{contfrac}), which can be depicted in a
decorated Dynkin diagram of type A as
\begin{equation} 
\xymatrix@1{
 \alpha_1 \  \ar@{-}[r] 
  & \ \alpha_2 \ \ar@{-}[r] & \ \cdots \ \ar@{-}[r] & \ \alpha_m
  }
\end{equation}
As usual this diagram needs to be extended by adding a further node,
labelled by $0$, in order to obtain a diagram of affine type. To the
affine diagram we associate its double quiver, obtained by adding a
pair of arrows in opposite directions for each link. For
every $\alpha_i > 2$ we then add $\alpha_i -2$ additional arrows from
vertex $i$ to vertex $0$. This algebraic procedure gives precisely the
special McKay quiver $\tilde\sfQ_{p,p'}$. In particular, the usual
A-type McKay quivers are
correctly obtained when all $\alpha_i =2$; for the example
$(p,p'\,)=(7,2)$ we recover again the quiver (\ref{72quiver}). 
The relations of this quiver can also be determined algebraically in
an algorithmic way, which is however somewhat more involved; we will take them as given by the ideal of relations of the special McKay quiver. The corresponding reconstruction algebra $\sfA=\sfA_{p,p'}$ has finite global dimension given by
\begin{equation}
\mathrm{gldim} \, \sfA_{p,p'} =  \left\{ \begin{matrix}  \ 2 &  \qquad
  \text{for $p'=p-1$} \ , \\  \ 3 &  \qquad \text{otherwise} \ .
\end{matrix} \right . 
\end{equation}

We will also need the explicit forms of the projective resolutions of the
simple $\sfA$-modules associated with the vertices labelled by
$\alpha_i$ for $i=0,1,\dots,m$. By~\cite[Thm.~6.17]{wemyssA} it
follows that if $1 \le i \le m$ then the simple module $\sfD_i$ has the projective resolution
\begin{equation} \label{simplet}
\begin{xy}
\xymatrix@C=8mm{  0 \ \ar[r] &  \ \sfP_i{}^{\oplus(\alpha_i -1)}  \ \ar[r] &
  \ \sfP_{i-1} \oplus  \sfP_0{}^{\oplus(\alpha_i -2)}  \oplus  \sfP_{i+1}  \ \ar[r] &  \ \sfP_i  \  \ar[r] &  \ \sfD_i \ar[r] \   & \  0
}
\end{xy}
\end{equation}
where $\sfP_i=\sfe_i\sfA$ and throughout we use the convention $i+m+1\equiv i$. On the other hand, for the
affine node $i=0$ one has
\begin{equation} \label{simple0}
\begin{xy}
\xymatrix@C=8mm{  0 \ \ar[r] & \ \displaystyle{\bigoplus_{i=1}^m \,
    \sfP_i {}^{\oplus(\alpha_i -2)} } \ \ar[r] & \  
    \sfP_0 {}^{\oplus(1 + \sum_{i=1}^m \, (\alpha_i - 2) )} \  \ar[r]
  & \ \sfP_m \oplus \sfP_1 \ \ar[r] &  \  \sfP_0  \
  \ar[r] & \  \sfD_0 \  \ar[r]  & \  0 \ .
}
\end{xy}
\end{equation}
Note that the projective dimension of $\sfD_i$ is always two for $i
\neq 0$, while $\sfD_0$ generically has projective dimension three
except when the singularity is of A-type, as expected. The explicit
form of the maps between the modules can be found in
\cite{wemyssA}. Each term of these resolutions implicitly defines the
matrices ${\tt b}^{(p)}_{ij}$ for $p=1,2$, since the projective
resolutions of the simple modules already tell us the numbers of
arrows and relations by reading off the dimensions using
(\ref{projressimple})--(\ref{dpwvdimExt}). For the non-zero numbers of
arrows $\ttb^{(1)}_{ij}=d^1_{j,i}$ we find
\beq
\ttb^{(1)}_{i,i\pm1}=1=\ttb^{(1)}_{10}=\ttb^{(1)}_{m0} \qquad
\mbox{and} \qquad \ttb_{0i}^{(1)}=\alpha_i-1 \ ,
\eeq
while the non-zero numbers of relations $\ttb^{(2)}_{ij}=d^2_{j,i}$ are given by
\beq
\ttb_{00}^{(2)}=1 + \sum_{i=1}^m\, \left( \alpha_i - 2 \right) \qquad
\mbox{and} \qquad \ttb_{ii}^{(2)}= \alpha_i - 1
\eeq
for $i\neq0$. It is straightforward to check that these dimensions
agree with those of Table~\ref{special} which were computed
geometrically. However, from the resolution
(\ref{simple0}) we also find the non-vanishing dimension
\begin{equation}
d^3_{i,0}=\dim \Ext^3_{\sf{A}} \left( \sfD_0 , \sfD_i \right) = \alpha_i - 2
\end{equation}
which implies that the special quiver $\tilde{\sfQ}_{p,p'}$ has a non-trivial set of relations between relations. One can indeed check that
\begin{eqnarray}
\dim \Ext^3_{\cO_M} \big( \cO_{D_i} (-1) [1]\,,\, \cO_{D_0} \big) &=&
\dim \Ext^2_{\cO_M}\big( \cO_{D_i} (-1)\,,\, \cO_{D_0} \big)
\\[4pt] &=& \dim \Hom_{\cO_M}\big( \cO_{D_0} \,,\, \cO_{D_i}
(-1 - \langle D_0,D_i\rangle_\Gamma)\big) \ = \ \alpha_i - 2 \ , \nonumber 
\end{eqnarray}
where we have used Serre duality and borrowed some results from
\cite{wemyss}. From the point of view of representation theory, the
existence of non-trivial relations among relations is a consequence of the fact that the
determinant representation $\bigwedge^2 Q$ is generally non-trivial
for $G \subset GL(2 , \complex)$, and hence a non-zero set of integers $\ttb_{ij}^{(3)}=d_{j,i}^3$ arises via (\ref{decomporeps}).

It is instructive at this point to consider the case of the cyclic du Val
singularities $A_{p-1}$. In this case all Dynkin labels are $\alpha_i
= 2$, the reconstruction algebra $\sfA_{p,p-1}$ has global dimension two, and every
representation is special; one has $\ttb_{ij}^{(3)}=0$ and $\ttb_{ij}^{(2)}=\delta_{ij}$. The projective resolutions
should in this instance be compared
with the basis ${\cS_a}$ of $K^c (M)$ consisting of complexes of
holomorphic bundles which are exact outside the exceptional locus,
given in (\ref{McKaycSbasis}) with
$\tta_{ab}=\delta_{a,b+1}+\delta_{a,b-1}$. Each of these complexes
follows from the projective resolutions
(\ref{simplet})--(\ref{simple0}) with $\alpha_i-2=0$. Being resolutions they are
exact at each arrow except the rightmost one, which
represents the corresponding exceptional curve $D_i$; but outside this
locus the sequence is exact at every arrow. In this case the
reconstruction algebra reduces to the preprojective algebra of the
McKay quiver with zero deformation parameter, which is just the path algebra with the usual ideal of relations.

Let us now go back to our previous example of the $A_{7,2}$
singularity. Then we have the sequence
\begin{equation}
\begin{xy}
\xymatrix@C=8mm{  0 \ \ar[r] &  \ \sfP_1 {}^{\oplus
    3} \ \ar[r]^{\!\!\!\!\!\!\!\!{\sf p}_2} & \ \sfP_2 \oplus \sfP_0
  {}^{\oplus 3} \ \ar[r]^{ \ \ \ \ \ {\sf p}_1} & \ \sfP_1
  \ \ar[r] & \ \sfD_1 \ \ar[r]  & \ 0
}
\end{xy}
\end{equation}
where the map ${\sf p}_1$ is the scalar product with $(a_4 , a_6, a_7 ,
a_8)$, while the image of the injective map ${\sf p}_2$ is the kernel
\begin{equation}
\ker {\sf p}_1 = (a_3 , - a_5 , 0,0) \sfP_1+ (0, a_1\, a_3 , - a_5, 0
)\sfP_1 + (0,0,a_1 \, a_3 , a_5) \sfP_1 \ .
\end{equation}
Similarly we have
\begin{equation}
\begin{xy}
\xymatrix@C=8mm{  0 \ \ar[r] & \ \sfP_1{}^{ \oplus
    2} \ \ar[r]^{{\sf q}_3} & \ \sfP_0 {}^{\oplus 3} \
  \ar[r]^{\!\!\!{\sf q}_2} & \ \sfP_2 \oplus \sfP_1 \ \ar[r]^{ \ \
    \ \ {\sf q}_1} & \ \sfP_0 \ \ar[r] & \ \sfD_0 \ \ar[r]  & \ 0 \\ 
}
\end{xy}
\end{equation}
where the map ${\sf q}_1$ is the scalar product with $(a_1 , a_5)$, while
\begin{equation}
{\sf q}_2 = \left( \begin{matrix} a_3 \, a_7 & a_3\, a_6 & a_2 \\  - a_8 &
    -a_7 &  -a_6 \end{matrix} \right) \qquad \mbox{and} \qquad {\sf q}_3 =
\left( \begin{matrix} a_5 & 0 \\ - a_1\, a_3 & a_5 \\ 0 & -a_1\,
    a_3 \end{matrix} \right) \ .
\end{equation}
It follows that
\begin{equation} \label{q1q2}
{\sf q}_1 \, {\sf q}_2 = 0 = \left( \begin{matrix} a_1 \,a_3\, a_7 - a_5\, a_8 \\
    a_1\, a_3\, a_6 - a_5\, a_7 \\ a_1\, a_2 - a_5\, a_6 \end{matrix}
\right)
\end{equation}
gives a subset of the relations (\ref{rels1712}). Similarly
\begin{equation}
{\sf q}_2 \, {\sf q}_3 = 0 = \left( \begin{matrix} a_3\, a_7\, a_5 - a_3\, a_6\, a_1\, a_3 & a_3\, a_6\, a_5 - a_2\, a_1\, a_3 \\ -a_8\, a_5 + a_7\, a_1\, a_3 & -a_7\, a_5 + a_6\, a_1\, a_3 \end{matrix} \right)
\end{equation}
gives a set of conditions which are independent from (\ref{q1q2}), but
are {automatically satisfied} if one takes into account the full set of relations (\ref{rels1712}).

\subsection{Counting instantons} 

We would like to propose that the homological structure that we have
outlined should play a prominent role in the construction of the
instanton moduli space on the varieties $M=M_{p,p'}$. Although a
direct parametrization is plagued by some technical difficulties, we
would nevertheless like to sketch now some arguments in support of
such putative ADHM-type constructions on generic Hirzebruch-Jung surfaces. 

We have described a basis of $K(M)$ given by the special
sheaves $\calL_i$ (including the trivial sheaf $\calL_0= \cO_{M}$). We can now construct a dual basis for $K^c (M)$, the Grothendieck group of bounded complexes
of vector bundles over $M$ which are exact outside the exceptional
locus $\pi^{-1}
(0)$; in fact we have already done so in our discussion of the reconstruction
algebras: The resolutions of the simple modules essentially represent
the sheaves supported on the exceptional set that we are looking
for. Define the complexes
\begin{equation}
\cT_k \ : \ \xymatrix@1{ \displaystyle{  \bigoplus_{j=0}^m\,
  \ttb_{kj}^{(3)} \, \calL_j^{\vee} } \  \ar[r] & \ \displaystyle{ 
  \bigoplus_{j=0}^m\,  \ttb_{kj}^{(2)} \, \calL_j^{\vee} } \ \ar[r] &
\ \displaystyle{ \bigoplus_{j=0}^m\, \ttb_{kj}^{(1)}\, \calL_j^{\vee} } \ \ar[r] & \ \calL_k^{\vee}
  }
\end{equation}
for $k=0,1,\dots,m$,
which give (conjecturally) a basis of $K^c (M)$ dual to the basis of
bundles $\calL_i$ in the sense that
\begin{equation}
\langle \calL_i , \cT_j \rangle_K := \int_M\, \mathrm{ch}(\calL_i) \wedge
\mathrm{ch}(\cT_j) \wedge \mathrm{td} (M) = \delta_{ij} \ .
\end{equation}
There is a natural pairing on $K^c (M)$ given as before by
\begin{equation}
(\cS , \cT  )_{K^c} = \big\langle \Xi(\cS) \,,\, \cT \big\rangle_K \ ,
\end{equation}
such that the pairing between two generators of $K^c (M)$ gives
\begin{equation}
\big(\cT_k^{\vee} \, , \, \cT_j\big)_{K^c} = \big\langle
\Xi(\cT_k^{\vee}\,) \,,\, \cT_j \big\rangle_K = \sum_{i=0}^m\, \big(
\delta_{ki} - \ttb^{(1)}_{ki} + \ttb^{(2)}_{ki} - \ttb^{(3)}_{ki}
\big) \, \langle \calL_i , \cT_j \rangle_K =: \tilde{C}_{kj} \ .
\end{equation}
Remarkably, the matrix $\tilde{C}= (\tilde C_{kj})$ is precisely an
``affine'' extension of the intersection matrix of the $A_{p,p'}$
singularity given as
\begin{equation} \label{affintersection}
\tilde{C} = \left( 
\begin{matrix}
- 2 - \displaystyle{\sum_{i=1}^{m} \, \left( \alpha_i - 2 \right)} & \alpha_1 -1 & \alpha_2 -2 & \dots & \alpha_{m-1} -2 & \alpha_m -1 
\\
 \alpha_1 -1 & - \alpha_1 & 1 & 0 & \dots & 0
\\
 \alpha_2 -2 & 1 & - \alpha_2 & 1 & \dots & 0
\\
\vdots & \vdots & \ddots & \ddots & \ddots & \vdots & 
\\
 \alpha_{m-1} -2 & 0 & \cdots & 1 & - \alpha_{m-1} & 1
\\
 \alpha_m -1 & 0 & \dots &  0 & 1 & - \alpha_{m} 
\end{matrix}
\right)
\end{equation} 
where as usual $\alpha_i$ is the self-intersection number of the
exceptional curve $D_i$. When all $\alpha_i = 2$, i.e. the
resolution is crepant, we recover (minus) the affine Cartan matrix of the
A-type singularities. In particular, the matrix
(\ref{affintersection}) has zero determinant, while the intersection
matrix (\ref{int_mat}) of the $A_{p,p'}$ singularity has determinant
equal to $p$. Therefore the two bases $\calL_i$ and $\cT_i$
play essentially the same role that the bases of sheaves $\calR_a$ and
$\cS_a$ played for ALE spaces in \S\ref{subsec:McKay}. This is further
exemplified by the fact that $\calL_i$ form a basis of $K (M)$ which
generate the full derived category $\frD(M)$. One can therefore try to
adapt the standard Beilinson monad construction to the problem at hand. The object
\begin{equation}
\cW^{\vee} \stackrel{L} \boxtimes \cW \ \longrightarrow \ \cO_{\Delta}
\label{derivedres}\end{equation}
is a resolution of the diagonal sheaf~\cite{king}. In the following we will attempt to
use it to generalize Beilinson's theorem to arbitrary resolutions
$M=M_{p,p'}$.

To start, we need a practical way to compute the derived tensor
product in (\ref{derivedres}); this is provided by~\cite{king,king2}. The idea is to consider the
path algebra $\sfA =
\End_{\cO_M}(\cW)$ of the special McKay quiver as a bimodule over
itself and construct a projective resolution $
\sfS^{\bullet}\rightarrow \sfA$ of the form
$\sfS^k = \bigoplus_{i,j}\, \sfA \sfe_i \otimes V^k_{ij} \otimes
\sfe_j \sfA$, 
where $V_{ij}^k = \Tor_\sfA^k
(\mathsf{M}_i , \mathsf{M}_j)$ and $\mathsf{M}_{i}$ are
$\sfA$-modules. The form of the projective resolution in our case is
derived directly from the individual projective resolutions of the
simple modules of the path algebra $\sfA$ and of its opposite algebra $\sfA^{\mathrm{op}}$. 
For this, let us introduce
some notation. We define maps ${\sf h},{\sf
  t}:\tilde\sfQ_1\rightrightarrows\tilde\sfQ_0$ which identify the head and tail
vertices in $\tilde\sfQ_0$
of each arrow $a:{\sf t}(a)\longrightarrow{\sf h}(a)$ in
$\tilde\sfQ_1$. The set of functional relations written as paths in
the quiver is denoted $\tilde\sfR$; as the reconstruction algebra
$\sfA$ has global dimension three, there is also a set of ``relations
among relations'' $\tilde{\mathsf R}\tilde{\sfR}\subset\sfA$. Then the
projective resolution of the algebra $\sfA$ is~\cite{king,king2}
\begin{equation}
\xymatrix@1{
  \sfS^3 \ \ar[r] & \ \sfS^2 \ \ar[r] &  \
  \sfS^1 \ \ar[r]
  & \ \sfS^0 \ \ar[r] & \ \sfA
  }
\end{equation}
where the individual terms are given by
\begin{eqnarray}
\sfS^3 &=&   \bigoplus_{rr \in \tilde{\mathsf R}\tilde{\sfR}}\, \sfA
\sfe_{\sft({rr})} \otimes [rr] \otimes \sfe_{\sfh({rr})} \sfA \ ,
\nonumber \\[4pt]
\sfS^2 &=& 
 \bigoplus_{r \in \tilde{\mathsf R}}\, \sfA \sfe_{\sft (r)} \otimes [r]
 \otimes \sfe_{\sfh(r)} \sfA \ , \nonumber \\[4pt] \sfS^1 &=&
 \bigoplus_{a \in \tilde\sfQ_1}\, \sfA \sfe_{\sft(a)} \otimes [a]
 \otimes \sfe_{\sfh(a)} \sfA  \ , \nonumber \\[4pt]  \sfS^0 &=&
 \bigoplus_{i \in \sfQ_0}\, \sfA \sfe_{i} \otimes [i] \otimes \sfe_{i}
 \sfA \ ,
\label{sfScomplex}\end{eqnarray}
and $[rr]$, $[r]$, $[a]$ and $[i]$ are one-dimensional vector spaces of
``labels''. After expanding the sums, these vector spaces act as
multiplicity spaces encoding the numbers of relations and arrows
between two vertices of the quiver. To go from the projective
resolution of $\sfA$ to the resolution of $\cO_{\Delta}$, one only
has to pick the multiplicity spaces and replace the projective modules
at vertices by
special tautological bundles. Note that the sums in (\ref{sfScomplex})
are over relations, arrows, and vertices, and hence need to be
re-expressed as sums over the line bundles.

From this we infer that
\begin{equation}
\bigoplus_{i,j=0}^m\, p_1^* \calL_i^{\vee} \otimes \ttb^{(k)}_{ij}
\otimes p_2^* \calL_j
\end{equation}
is a locally free resolution of $\cO_{\Delta}$. It is given explicitly by
\begin{equation} \label{specialres}
\xymatrix@C=10mm{
&  \displaystyle{ \calL_0 \ \boxtimes \ \bigoplus_{i=1}^m \,
    \big(\calL_i^{\vee}\,\big)^{\oplus (\alpha_i-2)} } \
  \ar[r] & \
   {\begin{matrix}
\calL_0  \boxtimes \big(\calL_0^{\vee}\, \big)^{\oplus (1+\sum_i\, (\alpha_i -2))}
\\  \oplus \\ \displaystyle{ \bigoplus_{i=1}^m \, \calL_i \boxtimes
  \big(\calL_i^{\vee}\, \big)^{\oplus (\alpha_i-1)} }  \end{matrix}} \ \ar[r] & 
  \\ \ar[r] & \
 {\begin{matrix}
\calL_0 \boxtimes \big(\calL_m^{\vee} \oplus \calL_1^{\vee} \, \big)\\
\oplus \\ \displaystyle{ \bigoplus_{i=1}^m\, \calL_i \boxtimes \left(
    \calL_{i-1}^{\vee} \oplus \big(\calL_0^{\vee}\, \big)^{\oplus (\alpha_i-2)}
    \oplus \calL_{i+1}^{\vee}  \right) } \end{matrix}} \ 
   \ar[r] & \
    {\begin{matrix}
\calL_0 \boxtimes \calL_0^{\vee} \\  \oplus \\  \displaystyle{
  \bigoplus_{i=1}^m\, \calL_i \boxtimes \calL_i^{\vee} }
  \end{matrix}} \
   \ar[r] & \
\cO_{\Delta}  \ . }
\end{equation}

In the case of du Val singularities, when all $\alpha_i=2$,
this complex collapses to the expected resolution
\begin{equation}
\xymatrix@1{
  \big( \calR \boxtimes \calR^{\vee} \otimes
  \bigwedge^2 Q^{*} \big)^{ G} \ \ar[r] & \ \big( \calR \boxtimes
  \calR^{\vee} \otimes Q^{*} \big)^{ G} \ \ar[r] & \ \big( \calR
  \boxtimes \calR^{\vee} \, \big)^{ G} \ \ar[r] & \ 
\cO_{\Delta}   }
\end{equation}
from \S\ref{subsec:ALEmodsp}, where $Q \cong \complex^2$ is the trivial bundle on which the regular
representation of $G=G_{p,p-1}$ acts. This complex can be extended to
the orbifold compactifictation of $M$ by replacing the trivial bundle
$Q$ with the sheaf of differential forms $\cQ$ which enters the Koszul
complex, as explained in~\S\ref{subsec:ALEmodsp}.

For our previous example of the $A_{7,2}$ singularity, if we write
$\cW = \cO_M \oplus \calL_2 \oplus \calL_1$ then the object $
\cW \stackrel{L} \boxtimes \cW^{\vee} $ is~\cite{craw}
\begin{equation}
\xymatrix@1{
 \cO_M \boxtimes \big(\calL_1^{\oplus 2}\big)^{ \vee} \ \ar[r]^{\!\!\!\dd_3} & \ 
 {\begin{matrix}   
 \cO_M \boxtimes \big(\cO_M^{\oplus 3}\big)^{ \vee} \\ \oplus \\ 
  \calL_2 \boxtimes \calL_2^{\vee} 
 \\ \oplus \\  \calL_1 \boxtimes \big(\calL_1^{\oplus 3 }\big)^{ \vee}    \end{matrix}}
   \quad\ar[r]^{\dd_2} &\quad
   {\begin{matrix} 
    \cO_M \boxtimes \calL_2^{\vee}  \\ \oplus \\
  \calL_2 \boxtimes \cO_M^{\vee}  \\ \oplus \\
     \calL_2 \boxtimes \calL_1^{\vee}  \\ \oplus \\
    \calL_1 \boxtimes \calL_2^{\vee}  \\ \oplus \\
  \cO_M \boxtimes \cO_M^{\vee}
\\ \oplus \\
   \calL_1 \boxtimes \big(\cO_M^{\oplus 3}\big)^{ \vee} 
   \end{matrix}} \quad \ar[r]^{ \ \ \ \dd_1} & \quad
   {\begin{matrix} 
    \cO_M \boxtimes \cO_M^{\vee} \\ \oplus \\
    \calL_2 \boxtimes \calL_2^{\vee}  \\ \oplus \\
    \calL_1 \boxtimes \calL_1^{\vee}  
   \end{matrix}}
}
\end{equation}
where
\begin{equation}
\dd_1 = \left(
\begin{matrix}
-a_1^{[1]} & a_2^{[2]} & 0 & 0 & -a_5^{[1]} & a_6^{[2]} & a_7^{[2]} & a_8^{[2]} \\
a_1^{[2]} & -a_2^{[1]} & -a_3^{[1]} & a_4^{[2]} & 0 & 0 & 0 & 0 \\
0 & 0 & a_3^{[2]} & -a_4^{[1]} & a_5^{[2]} & -a_6^{[1]} & -a_7^{[1]} & -a_8^{[1]}
\end{matrix}
\right)
\end{equation}
and
\begin{equation}
\dd_2 = \left(
\begin{matrix}
a_2^{[1]}  &  -a_3^{[1]} \, a_6^{[1]} & -a_3^{[1]}\, a_7^{[1]} & 0 & -a_6^{[2]}\,\, a_3^{[1]} & -a_7^{[2]}\, a_3^{[1]} & a_2^{[2]} \\
a_1^{[2]} & 0 & 0 & 0 & 0 & 0 & a_1^{[1]} \\
0 & -a_1^{[2]}\, a_6^{[1]} & -a_1^{[2]}\, a_7^{[1]} & a_4^{[2]} & -a_6^{[2]}\, a_1^{[2]} & -a_7^{[2]}\, a_1^{[2]} & -a_4^{[1]} \\
0 & 0 & 0 &  a_3^{[1]} & 0 & 0  & -a_3^{[2]} \\
-a_6^{[1]} & a_7^{[1]} & a_8^{[1]} & -a_6^{[2]} & a_7^{[2]} & a_8^{[2]} & 0 \\
-a_5^{[2]} & -a_1^{[2]}\, a_3^{[2]} & 0 & -a_5^{[1]} & -a_1^{[1]}\, a_3^{[1]} & 0 & 0  \\
0 & a_5^{[2]} & -a_1^{[2]}\, a_3^{[2]} & 0 & a_5^{[1]} & -a_1^{[1]}\, a_3^{[1]} & 0  \\
0 & 0 & a_5^{[2]} & 0 & 0 & a_5^{[1]} & 0
\end{matrix}
\right) \ ,
\end{equation}
while $\dd_3$ is the kernel map; here we use the notation $a_s^{[1]} =
a_s \boxtimes 1$ and $a_s^{[2]} = 1 \boxtimes a_s$. Note that the maps $a_s^{[2]}$ are elements of the opposite quiver and their composition goes in the opposite sense.

The problem encountered now is that one ends up with a resolution of
the diagonal sheaf of $M_{p,p'} \times M_{p,p'}$, while what is really
required is a resolution of the diagonal sheaf of $\overline{M_{p,p'}}
\times \overline{M_{p,p'}}$, where $\overline{M_{p,p'}}$ is an
``appropriate'' compactification of $M_{p,p'}$. Presumably this
compactification could be obtained by adding a certain stacky divisor
at infinity of the form $\PP^1 /  G$ with the $ G$-action suitably modified
so that only special representations
occur; this restriction on the boundary conditions of the instanton
gauge field is possible since any irreducible representation of $G$
can be expressed as tensor products of special representations. Furthermore, the reconstruction algebra $\sfA$ is generally not
homogeneous and therefore not a Koszul
algebra, hence the resolution (\ref{specialres}) cannot be extended at infinity by gluing it with
a Koszul resolution e.g. of $\PP^2\times\PP^2$, as we did in the case of du Val singularities
in \S\ref{subsec:ALEmodsp}. In other words, we do not know how to
impose boundary conditions on the gauge theory in the general
case. Such gluing problems do not arise in one-point compactifications
of Hirzebruch-Jung spaces, see e.g.~\cite{calderbank}. Carrying such
an ADHM-type construction through would produce a description
of the instanton moduli space analogous to that of Nakajima's quiver
varieties, which could lead to interesting representation theoretic
interpretations of the decomposition into regular and fractional
instantons in terms of the special representations of the orbifold
group $G$. Further impetus into the problems alluded to above could
come from comparing the constructions of this section with the
alternative monadic
description of framed torsion free sheaves on Hirzebruch surfaces
given in~\cite{rava}; however, even in these cases, a
parametrization in terms of explicit ADHM matrix data is still lacking.

\bigskip

\section{Enumerative invariants\label{EIsummary}}

We end by summarizing the various enumerative invariants we have encountered in the course of this survey, their associated moduli spaces and their mutual relations, emphasizing the geometrical and algebraic aspects. 

All the invariants we have discussed have a physical origin in field theory or in string theory. The problem we are interested in is the structure of the BPS space of states. The associated enumerative problem is to compute the BPS degeneracies $\Omega_X \left( \gamma  \right)$ defined in (\ref{Windex}) as the Witten indices of the BPS Hilbert spaces for a D-brane bound state with total charge $\gamma$. These degeneracies are identified with the generalized Donaldson-Thomas invariants, and in the large radius limit of the Calabi-Yau correspond to the intersection theory of the moduli space of stable sheaves. One can be more precise in the case of ordinary Donaldson-Thomas invariants, which correspond to the particular case of charge vector $\gamma = (1 , 0 , - \beta , n)$. States with these charges are geometrically represented by ideal sheaves and the ordinary Donaldson-Thomas invariants $\Omega_X(n,\beta)=\DT_{n,\beta}(X)$ can be rigorously defined via virtual integration over the moduli space of ideal sheaves $\scrM^{\rm BPS}_{n,\beta} (X)$. These invariants are conjectured, and in certain cases proved, to be equivalent to the Gromov-Witten invariants $
\GW_{g , \beta} (X)$ which correspond to worldsheet instantons and are defined via integration over the compactified moduli space of stable curves $\overline{\scrM}_{g,\beta} (X)$. The equivalence means that the generating functions for the two kind of invariants are equal, after a rather non-trivial parameter identification.

These invariants, whether they are associated with curves or ideal sheaves, are \textit{geometrical} in nature. Other have a more \textit{algebraic} definition, such as the noncommutative Donaldson-Thomas invariants $\Omega_{\sfA,v_0}(\mbf n)= \DT_{\mbf n , {v_0}}(\sfA) := \chi \big(\scrM_{\mbf n}(\hat\sfQ,v_0)
\,,\, \nu_\sfA
\big)$ whose associated moduli space characterizes stable representations of a certain quiver, with fixed dimension vector $\mbf n$. These invariants correspond to a situation where the geometry of the Calabi-Yau becomes singular and is replaced by the notion of a noncommutative crepant resolution via the path algebra of a quiver. Physically they still correspond to D-brane bound states and can be in principle obtained from the $\DT_{n,\beta}(X)$ via an infinite series of wall-crossings.

The Donaldson-Thomas invariants can also be given an equivalent but more physical definition via the gauge theory perspective. According to this point of view they corresponds to integrals over the generalized instanton moduli space of a cohomological Yang-Mills theory in six dimensions, with a specific measure given by the Euler class of the obstruction bundle $\int_{\scrM^{\rm inst}_r(X)} \, \eul(\scrN_r)$. In the case of ordinary Donaldson-Thomas invariants, as well as in the case of higher rank invariants but in the Coulomb branch, this integrals can be rigorously defined and evaluated via virtual localization. One advantage of this perspective is that also the noncommutative invariants can be understood as generalized instantons, via the McKay correspondence. The net result is that the relevant moduli space is now identified with a certain sub variety of the framed McKay quiver representation space and the noncommutative invariants (\ref{BehrendNC}) defined as weighted Euler characteristics $\DT_{\mbf n , \mbf r}(\sfA_G) = \chi \big( \scrM_G (\mbf n ,\mbf r) \,
, \, \nu_{\sfA_G} \big)$ which again can be computed explicitly via virtual localization. These invariants also depend on the framing label $\mbf r$ and generalize the noncommutative invariants described above. Physically this generalization corresponds to fixing different boundary conditions for the gauge fields. However generic invariants of type  $\mbf r$ can be shown to be equivalent to the rank one noncommutative invariants. Note however that this is just a computational limitation; if one could describe the $\mbf r$ invariants outside of the Coulomb branch, they would be genuinely new, and related to the rank one case only via wall-crossing.

The relation between instanton invariants and geometry is not limited to six dimensions. It holds also in four dimensions. On a toric surface $M$ one can define a BPS counting problem corresponding to subschemes of dimension zero and one with compact support in $M$. The associated invariants (\ref{BPSEulerchar}) are the topological Euler characteristics of the BPS moduli spaces $\chi\big(\scrM^{\rm BPS}_{n,\beta}(M)\big)$. In this case the BPS moduli space  is just the Hilbert scheme $\scrM^{\rm BPS}_{n,\beta}(M)=\Hilb_{n,\beta}(M)$. On the gauge theory side one considers Vafa-Witten theory whose partition function is the generating function of the Euler characteristics (\ref{chiMPinst}) of the instanton moduli spaces $\chi \big(\mathscr{M}^{\rm inst}_\cP(M) \big)$. The two counting problem are quite similar but not precisely identical due to a different treatment of the sum over line bundles. In the case of ALE spaces, the instanton moduli space has an algebraic nature and is constructed via the McKay correspondence; as a consequence the instanton generating functions have interesting modular properties. For other geometries, such as generic Hirzebruch-Jung surfaces, not much is known and the problem is still open.

Some of these characteristics are present non trivially in the much less studied case of $\cN=2$ gauge theories on toric surfaces. In this case the appropriate invariants correspond to equivariant integrals over the instanton moduli spaces. In this case the integrand depends sensitively on the specific theory, in particular on its matter content. Yet in many cases the appropriate generating functions can be expressed in terms of the representation theory of affine algebras.

\bigskip

\section*{Acknowledgments}

In the course of this project we have
benefited from discussions with several colleagues; we thank in
particular Amir Kashani-Poor and Ani Sinkovics for the many discussions
and collaborations on some of the topics we have presented here. We also wish
to thank Ali Craw, Anatoly Konechny, Elizabeth Gasparim, Nikita Nekrasov and
Balasz Szendr\H{o}i for helpful discussions. Some parts of this work were presented by M.C. at the thematic period on  
``Matrix Models and Geometry'' at the Instituto Superior T\'ecnico in
the fall of 2009, by R.J.S. at the workshop ``Noncommutative Algebraic
Geometry and D-Branes'' at the Simons Center for Geometry
and Physics in December 2011, and by R.J.S at the programme ``Mathematics and Applications of Branes in String and
M-Theory'' at the Isaac Newton Institute for Mathematical Sciences in
March 2012; we are thankful to the participants of these programmes for their comments  
and discussions. In particular, R.J.S. expresses his gratitude to Charlie
Beil for the invitation to the Simons Center, and to David Berman, Neil Lambert and
Sunil Mukhi for the invitation to the Newton Institute. Finally, we thank Ugo
Bruzzo, Gunther Cornelissen, Giovanni Landi and Vladimir Rubtsov for
the invitation to contribute this article to the special issue, which
is based on themes from the workshop ``Noncommutative Algebraic
Geometry and its Applications to Physics'' which was held at the
Lorentz Center in March 2012. The work of M.C. is partially supported by the Funda\c{c}\~{a}o para a Ci\^{e}ncia e a
Tecnologia (FCT/Portugal). The work of R.J.S. is supported in part by the
Consolidated Grant ST/J000310/1 from the UK Science and Technology
Facilities Council, and by Grant RPG-404 from the Leverhulme Trust.

\end{document}